\documentclass[12pt]{article}

\usepackage{graphicx} 
\usepackage[utf8]{inputenc}
\usepackage{pythontex}
\usepackage{algpseudocode}
\usepackage{amsmath}
\usepackage{hyperref}
\usepackage{amssymb}
\usepackage{physics}
\usepackage{amsfonts}
\usepackage{nicematrix}
\usepackage{soul}
\usepackage{array}   
\usepackage{booktabs} 
\usepackage[table]{xcolor}
\usepackage{colortbl}
\usepackage{multirow}      
\usepackage{lscape}
\usepackage{float} 

\usepackage{multicol}

\usepackage{caption}
\usepackage{subcaption}
\usepackage{arydshln}
\usepackage{natbib}
\newtheorem{definition}{Definition}
\usepackage[dvipsnames]{xcolor}
\hypersetup{colorlinks,linkcolor={blue},citecolor={blue},urlcolor={blue}} 
\usepackage[margin=1in]{geometry}
\usepackage{tikz-cd}
\usepackage{accents}
\usepackage{tabularx}
\usepackage{rotating}   
\usepackage{setspace}
\usepackage[english]{babel}
\newtheorem{theorem}{Theorem}
\usepackage{enumitem}   

\newtheorem{lemma}[theorem]{Lemma}

\title{Understanding Geopolitical Alignments Through Covariate Augmented Spectral Clustering of Heterogeneous UNGA Voting Data}
\author{Chirayata Kusari \footnote{Statistical Sciences Division, Indian Statistical Institute, Kolkata, India}, 
Souvik Roy\footnote{Statistical Sciences Division, Indian Statistical Institute, Kolkata, India}, and
Srijan Sengupta\footnote{Department of Statistics, North Carolina State University, Raleigh, NC, 27606. ssengup2@ncsu.edu}
}
\date{}
\begin{document}

\maketitle

\begin{abstract}
Community detection is a fundamental problem in network analysis. While many existing methods focus on homogeneous networks, real world networks are often heterogeneous, involving multiple node types and interaction mechanisms. In addition, node specific covariates frequently provide valuable information about the underlying community structure. Existing methodologies typically account for either network heterogeneity or covariate information, but seldom both simultaneously.

In this paper, we propose a covariate assisted spectral clustering framework for heterogeneous networks that jointly utilizes network connectivity {in a heterogeneous setting} and node level covariates. The proposed method extends covariate assisted spectral clustering to heterogeneous settings and operates directly on the heterogeneous network without relying on projection based simplifications. Under a heterogeneous node contextualized stochastic blockmodel, we establish theoretical guarantees for the proposed procedure, including concentration results, eigenspace perturbation bounds, and an explicit upper bound on the misclustering rate.

Simulation studies demonstrate that incorporating covariate information substantially improves community recovery and consistently outperforms several benchmark methods. We further apply the proposed framework to United Nations General Assembly voting data, where it reveals meaningful geopolitical structures by combining voting interactions with auxiliary covariate information.
\end{abstract}

\section{Introduction}

Community detection is one of the central problems in statistical network analysis, where the objective is to identify latent groups of nodes exhibiting similar connectivity patterns. Such communities often reveal meaningful hidden structures in complex systems and have important applications in areas such as social networks, biological systems, recommendation platforms, and international relations. More broadly, network analysis has become an indispensable tool for studying complex relational data, with applications ranging from community detection and diffusion analysis to recommendation systems and anomaly detection. Comprehensive surveys of statistical network models and methodologies may be found in \cite{goldenberg_survey} and \cite{Sengupta_survey}.

\subsection{Background and Related Work}

Several probabilistic frameworks have been proposed for modelling community structure in networks. One important class consists of latent variable models. For example, \cite{Hoff_02} introduced the latent space model, where edge probabilities depend on distances between latent node positions, and \cite{Handcock_07} subsequently incorporated model based clustering through finite Gaussian mixtures. Another influential class of models is the stochastic blockmodel (SBM) \citep{sbm_holland,sbm_nowicki}, where nodes are partitioned into latent communities and edge probabilities depend only on block memberships. Important extensions include the degree corrected blockmodel \citep{dcbm}, which accommodates degree heterogeneity, and the popularity adjusted blockmodel \citep{pabm}, which allows more flexible connectivity patterns.

Among the many algorithmic approaches for community detection, spectral clustering has emerged as one of the most widely used methodologies because of its computational efficiency, scalability, and strong theoretical guarantees \citep{Luxburg_sc_2007,jordan_sc}. The basic idea is to construct a low dimensional embedding of the network using eigenvectors of a graph based matrix and subsequently apply a clustering algorithm such as $K$ means or {$K$ mediods} to the embedded points. Weak consistency of spectral clustering under the stochastic blockmodel, characterized through vanishing misclustering error rates, was first established by \cite{rohe_chatterjee_2010}. Since then, extensive research has expanded our understanding of its statistical properties.

In many modern applications, network topology alone may not provide sufficient information for accurately recovering latent communities. Nodes are often accompanied by auxiliary attributes that carry additional information about the underlying community structure. Motivated by this observation, several authors have developed methods that incorporate covariate information into community detection procedures. Examples include the probabilistic framework of \cite{cov_newman}, the semidefinite programming approach of \cite{Yan_sarkar_21}, graphon estimation approach integrating graph information and node covariate information [\cite{graphon_cov}] and the methodologies developed in \cite{zhang_2016,pairwise_cov_18,yang_2021,grdpg_cov,cov_dim_reduction_21,sc_sbm_cov}. Very recently \cite{profile_likelihood_cov} proposes a model using profile least squares to estimate heterogeneous network interactions by separating the effects of observed edge attributes from unobserved latent factors.
Among these, spectral methods have attracted particular attention because of their favourable computational and theoretical properties. In particular, \cite{cov_sc_bink} proposed a covariate assisted spectral clustering framework and established weak consistency under the SBM. More recently, covariate assisted spectral clustering has been extended to multilayer network settings by \cite{mult_cov_1,mult_cov_2}, while \cite{net_adj_cov_24} established strong consistency results in multiscale network settings with nodal covariates.

Most of the existing literature on covariate assisted community detection focuses on homogeneous networks, where all nodes belong to a single category and all interactions are treated uniformly. However, many real world networks are inherently heterogeneous and involve multiple types of nodes and interactions. 

Examples of heterogeneous networks include author--paper networks, user--item networks, social media platforms, and country--resolution voting networks. Such networks consist of multiple types of nodes and interactions, leading to structural heterogeneity that cannot be adequately captured by models designed for homogeneous networks.
For instance, a social media platform such as Facebook may contain users, pages, groups, and events, each representing a distinct node type. Interactions may occur both within and across these categories; users may form friendship links with other users, join groups, follow pages, or participate in events. Furthermore, the attributes associated with these entities are inherently different, resulting in varying connectivity patterns and interaction mechanisms across node types.
In such settings, both the interaction mechanisms and the associated covariates may vary substantially across node types {may contain additional information}. Despite their practical importance, statistical methodologies for heterogeneous network clustering remain relatively limited. An important contribution in this direction is the heterogeneous spectral clustering framework of \cite{het_sengupta}, which developed spectral and regularized spectral clustering methods under heterogeneous stochastic blockmodel {and degree corrected stochastic blockmodel respectively. Alongside they} established theoretical guarantees for weak recovery {of misclustered nodes.}

Although covariate assisted community detection and heterogeneous network clustering have each received considerable attention, the intersection of these two research directions remains largely unexplored. Existing covariate assisted methods are primarily designed for homogeneous networks, whereas existing heterogeneous spectral clustering procedures typically rely only on network connectivity. Consequently, there is a need for a unified framework that can simultaneously exploit heterogeneous network structure and node specific covariate information.

\subsection{Our Contributions}

In this paper, we develop a covariate assisted spectral clustering framework for heterogeneous networks. Our methodology combines the covariate assisted spectral clustering approach of \cite{cov_sc_bink} with the heterogeneous stochastic blockmodel framework of \cite{het_sengupta}. The resulting procedure jointly incorporates network connectivity and node level covariates while explicitly accounting for multiple node types and heterogeneous interaction structures.

The proposed framework possesses several features that distinguish it from existing approaches. First, it extends covariate assisted spectral clustering from homogeneous networks to heterogeneous settings involving multiple node types and interaction mechanisms. Second, the method operates directly on the heterogeneous network itself and therefore avoids the one mode projection techniques commonly used in bipartite and multipartite networks. Such projections may discard important structural information and introduce biases into subsequent analyses. By working directly with the heterogeneous network, the proposed procedure preserves the full relational structure of the data.

Beyond the methodological development, we establish a comprehensive theoretical framework for the proposed procedure under a heterogeneous node contextualized stochastic blockmodel. We characterize the population eigenspace associated with the proposed covariate assisted operator. We then derive concentration bounds for the empirical operator, establish eigenspace perturbation results through a Davis--Kahan type analysis, and obtain explicit upper bounds on the node misclustering rate. These results yield weak consistency of the proposed clustering procedure and provide rigorous statistical guarantees for community recovery in heterogeneous networks with nodal covariates.

Through extensive simulation studies, we investigate the effects of network heterogeneity and block specific covariate separation on clustering performance. The results demonstrate that incorporating covariate information can substantially reduce misclustering error rates and improve community recovery. We further compare the proposed method with several benchmark procedures, including methods that account only for heterogeneity and methods that account only for covariates. The proposed framework consistently achieves superior performance, highlighting the benefits of simultaneously utilizing heterogeneous network structure and auxiliary covariate information.

\subsection{Application to UNGA Voting Data}

To illustrate the practical utility of the proposed methodology, we analyse voting records from the United Nations General Assembly (UNGA) \citep{voeten_2009}. The UNGA provides a particularly compelling application because it naturally gives rise to a heterogeneous, {particularly} bipartite network consisting of {two types of nodes, namely} countries and resolutions. Moreover, meaningful covariate information is available for both node types. Economic indicator such as GDP per capita provide additional information about countries, while issue based classifications provide contextual information for resolutions.

The UNGA dataset has been widely used for studying geopolitical alignments and international voting behaviour. Existing approaches include dynamic ordinal spatial models \citep{voeten_2017}, Bayesian Dynamic Dirichlet Process Mixture Models \citep{unga_ddpm}, and network based analyses of voting similarity patterns \citep{Gallop_2021}. In contrast to similarity based approaches that first project the data onto a country only network, our framework directly models the country resolution bipartite network and incorporates auxiliary covariates for both node types. This allows us to investigate latent geopolitical structures while preserving the heterogeneous nature of the underlying data.

As an empirical illustration, we apply the proposed methodology to the UNGA voting data during the 1990s, a period of profound geopolitical transformation following the collapse of the Cold War order.
The analysis demonstrates how combining heterogeneous network structure with auxiliary covariates can provide deeper insights into international voting coalitions and evolving geopolitical relationships.

\vspace{2mm}

The remainder of the paper is organized as follows. Section 2 introduces the proposed heterogeneous node contextualized stochastic blockmodel. Section 3 presents the covariate assisted spectral clustering methodology and the associated algorithm. Section 4 develops the theoretical properties of the proposed procedure. Section 5 reports simulation results. Section 6 contains the analysis of the UNGA voting data and the resulting geopolitical insights. Technical proofs are deferred to the Appendix.

\section{Preliminaries on Graph Structure and Nodal Covariates}
\label{sec2}

Let $G = (V, E)$ be an undirected graph with $N$ nodes, and let $\boldsymbol{A} \in \{0,1\}^{N \times N}$ denote the adjacency matrix of $G$. The degree $d_v$ of a node $v \in V$ is defined as the number of edges incident to $v$. The \emph{degree matrix} of $G$ is the diagonal matrix
\[
\boldsymbol{D} := \mathrm{diag}(d_1, d_2, \ldots, d_N) \in \mathbb{R}^{N \times N}.
\]

The \emph{normalized graph Laplacian} is defined as
\[
\boldsymbol{L} := \boldsymbol{D}^{-1/2} \boldsymbol{A} \boldsymbol{D}^{-1/2}.
\]
Intuitively, the normalization by $\boldsymbol{D}^{-1/2}$ adjusts for differences in node degrees so that highly connected nodes do not dominate the spectral structure of the graph. Consequently, $\boldsymbol{L}$ captures the connectivity pattern of the network in a more balanced manner and forms the basis of many spectral clustering methods.

To further stabilize the spectral behaviour, particularly in sparse networks, one often considers a regularized version of the graph Laplacian. The \emph{regularized graph Laplacian} is defined as
\[
\boldsymbol{L}_\tau := (\boldsymbol{D} + \tau \boldsymbol{I})^{-1/2} \boldsymbol{A} (\boldsymbol{D} + \tau \boldsymbol{I})^{-1/2}
= \boldsymbol{D}_\tau^{-1/2} \boldsymbol{A} \boldsymbol{D}_\tau^{-1/2},
\]
where
\[
\boldsymbol{D}_\tau := \boldsymbol{D} + \tau \boldsymbol{I},
\]
and $\tau > 0$ is a regularization parameter.

The parameter $\tau$ is introduced to improve the spectral behaviour of $\boldsymbol{L}$, particularly in the sparse graph regime where low-degree nodes may introduce instability in the eigenspace structure. It has been shown that an appropriate choice of $\tau$ can significantly enhance the performance of spectral clustering methods in sparse networks (\cite{chaudhuri12}). Following the prior work of \cite{qin_rohe_2013}, we consider the average degree as a suitable choice of $\tau$:
\[
\tau = \frac{1}{N} \sum_{i=1}^N d_i.
\]
This choice helps smooth degree heterogeneity and improves the spectral gap, thereby facilitating more accurate recovery of the underlying communities (\cite{qin_rohe_2013}, \cite{chaudhuri12}).

Each node $v \in V$ is associated with a random covariate vector $\boldsymbol{X}_v \in [-J, J]^R$, for some fixed $J > 0$, where $R$ denotes the dimension of the covariate space. Let $\boldsymbol{X} \in [-J, J]^{N \times R}$ denote the covariate matrix, whose $v$-th row corresponds to the covariate vector associated with node $v$, and whose columns represent the different covariate features. Such covariates provide auxiliary node-level information that complements the structural information contained in the network.

Throughout the paper, for a matrix $\boldsymbol{M} \in \mathbb{R}^{m \times n}$, let $\| \boldsymbol{M} \|$ and $\| \boldsymbol{M} \|_F$ denote its spectral norm and Frobenius norm, respectively. These are defined as
\begin{equation*}
\| \boldsymbol{M} \|
:= \max_{\boldsymbol{u} \in \mathbb{R}^n : \| \boldsymbol{u} \|_2 = 1}
\| \boldsymbol{M} \boldsymbol{u} \|_2
= \sigma_{\max}(\boldsymbol{M}),
\end{equation*}
and
\begin{equation*}
\| \boldsymbol{M} \|_F
:= \left(
\sum_{i=1}^{m}
\sum_{j=1}^{n}
\boldsymbol{M}_{i,j}^2
\right)^{1/2},
\end{equation*}
where $\sigma_{\max}(\boldsymbol{M})$ denotes the largest singular value of $\boldsymbol{M}$. Intuitively, the spectral norm measures the maximum stretching effect of the matrix, whereas the Frobenius norm measures its overall magnitude aggregated across all entries.


\section{Methodology}

In this section, we first review the heterogeneous stochastic blockmodel framework that forms the probabilistic foundation of our analysis. We then introduce the proposed covariate-assisted spectral clustering methodology for heterogeneous networks, followed by its algorithmic implementation. Finally, we describe the procedure for selecting the tuning parameter, adopting the approach proposed by \cite{cov_sc_bink}.

\subsection{The Heterogeneous Stochastic Blockmodel Framework}

The stochastic blockmodel (SBM) is one of the most widely used probabilistic models for studying community structure in random networks (\cite{sbm_holland}, \cite{sbm_nowicki}). Under SBM, the network consists of $N$ nodes, each belonging to one of $K$ blocks or communities. The probability of an edge between two nodes depends only on their respective block memberships. Consequently, nodes within the same block behave similarly from a probabilistic perspective and are often referred to as \emph{stochastically equivalent}.

Most classical SBM formulations focus on homogeneous networks, where all nodes and edges are of the same type. However, many real-world networks are heterogeneous in nature and involve multiple categories of nodes together with different forms of interactions. In such settings, the edge formation mechanism depends not only on community memberships but also on node types.

To model such networks, we adopt the Heterogeneous stochastic blockmodel (Het-SBM) proposed in \cite{het_sengupta}. Consider a network with $N$ nodes, where each node belongs to one of $T$ node types and one of $K$ latent underlying communities. To simultaneously capture type-based and community-based heterogeneity, each community is further divided according to node types. As a result, there are effectively $TK$ sub-blocks in the network, indexed by node type and community label.

Let $\boldsymbol{Z} \in \{0,1\}^{N \times TK}$ denote the sub-block membership matrix, where
\[
\boldsymbol{Z}_{i,(t,k)} =
\begin{cases}
1 & \text{if node } i \text{ is of type } t \text{ and belongs to block } k, \\
0 & \text{otherwise},
\end{cases}
\]
for all $t = 1,\ldots,T$ and $k = 1,\ldots,K$.

Let $\boldsymbol{P} \in [0,1]^{TK \times TK}$ denote the probability matrix governing the connectivity structure among the sub-blocks. This matrix may be viewed as a $T \times T$ block matrix:
\[
\boldsymbol{P} =
\begin{bmatrix}
\boldsymbol{P}_{11} & \boldsymbol{P}_{12} & \cdots & \boldsymbol{P}_{1T} \\
\boldsymbol{P}_{21} & \boldsymbol{P}_{22} & \cdots & \boldsymbol{P}_{2T} \\
\vdots & \vdots & \ddots & \vdots \\
\boldsymbol{P}_{T1} & \boldsymbol{P}_{T2} & \cdots & \boldsymbol{P}_{TT}
\end{bmatrix},
\]
where each block $\boldsymbol{P}_{tt'} \in [0,1]^{K \times K}$ specifies the connection probabilities between nodes of type $t$ and type $t'$.
In other words, $\boldsymbol{P}_{tt'}(a,b)$ denotes the probability that a node of type $t$ belonging to community $a$ is connected to a node of type $t'$ belonging to community $b$. Thus, the model allows edge probabilities to vary jointly across node types and latent communities.

Under this framework, for every pair of nodes $1 \leq i < j \leq N$, the adjacency matrix entry $\boldsymbol{A}_{ij}$ is generated independently as
\[
\boldsymbol{A}_{ij} \mid \boldsymbol{Z}
\sim
\mathrm{Bernoulli}
\left(
\boldsymbol{Z}_{i\cdot}
\boldsymbol{P}
\boldsymbol{Z}_{j\cdot}^{\top}
\right),
\]
where $\boldsymbol{Z}_{i\cdot}$ denotes the $i^{th}$ row of $\boldsymbol{Z}$.

This framework generalizes the classical SBM by explicitly incorporating node-type heterogeneity. In the special case $T=1$, the model reduces to the standard homogeneous stochastic blockmodel.

\subsection{Covariate-Assisted Spectral Clustering for Heterogeneous Networks}

We now introduce the proposed covariate-assisted spectral clustering framework for heterogeneous networks. We first briefly review covariate-assisted spectral clustering and then extend it to the heterogeneous network setting.

\subsubsection{Covariate-Assisted Spectral Clustering}

Spectral clustering is one of the most widely used clustering methodologies because of its computational efficiency, scalability, and strong theoretical guarantees (\cite{Luxburg_sc_2007}). Unlike many classical clustering methods, spectral clustering can effectively detect non-convex or nonlinearly separable communities by exploiting the spectral structure of the network.

The standard spectral clustering procedure involves constructing a graph Laplacian from the observed adjacency matrix, computing a suitable eigenspace representation, and subsequently applying $K$-means clustering to the leading eigenvectors.
Before introducing covariate-assisted spectral clustering, we briefly define assortative networks.

\medskip

\noindent
\textit{Assortative Network:}
Consider a network generated from a $K$-block stochastic blockmodel with block probability matrix $\Pi \in [0,1]^{K \times K}$. The network is said to be \emph{assortative} if
\[
\min_{k \in [K]} \Pi_{kk}
>
\max_{j \neq l} \Pi_{jl}.
\]
Intuitively, this means that within-community connections are denser than between-community connections. In disassortative networks, the reverse phenomenon occurs.

\medskip

To incorporate node-level covariate information into spectral clustering, we adopt the Covariate-Assisted Spectral Clustering (CASC) framework proposed by \cite{cov_sc_bink}. The main idea is to combine structural information from the graph with similarity information derived from node covariates.
The resulting matrices are defined as
\begin{align*}
\Bar{\boldsymbol{L}}(\alpha)
& :=
\boldsymbol{L}_{\tau}
+
\alpha \boldsymbol{X}\boldsymbol{X}^{\top}
\qquad
\text{(for assortative graphs)},
\\
\Tilde{\boldsymbol{L}}(\alpha)
& :=
\boldsymbol{L}_{\tau}\boldsymbol{L}_{\tau}
+
\alpha \boldsymbol{X}\boldsymbol{X}^{\top}
\qquad
\text{(general case)},
\end{align*}
where $\alpha \in [0,\infty)$ is a tuning parameter.

Here, $\boldsymbol{L}_{\tau}$ captures the network connectivity structure, whereas $\boldsymbol{X}\boldsymbol{X}^{\top}$ captures similarity induced by node covariates. The parameter $\alpha$ controls the relative contribution of these two information sources. Intuitively, larger values of $\alpha$ place greater emphasis on covariate similarity, while smaller values rely more heavily on the graph structure.

Categorical covariates are converted into binary dummy variables, while continuous covariates are standardized by centering and scaling each feature column. Since $\boldsymbol{L}_{\tau}$ is typically sparse and $\boldsymbol{X}\boldsymbol{X}^{\top}$ is often low-rank, the resulting matrices retain computational tractability for spectral decomposition.
Clustering is then performed using the leading eigenvectors of $\Bar{\boldsymbol{L}}(\alpha)$ or $\Tilde{\boldsymbol{L}}(\alpha)$. Unless stated otherwise, we use $\Tilde{\boldsymbol{L}}(\alpha)$ in the subsequent analysis.

\subsubsection{Heterogeneous Covariate-Assisted Spectral Clustering Algorithm (Het-Cov-SC)}

In a heterogeneous network with $T$ node types and $K$ communities per type, the effective number of clusters becomes $TK$. However, since node types are assumed to be known a priori, the clustering problem naturally decomposes into $T$ separate clustering tasks, one for each node type.
More precisely, nodes belonging to the same type are clustered into $K$ communities independently of nodes from other types. This decomposition considerably simplifies the clustering procedure while preserving the heterogeneous structure of the network.
The proposed Heterogeneous Covariate-Assisted Spectral Clustering (Het-Cov-SC) algorithm is described below.

\medskip

\noindent
\textbf{Algorithm 1: Het-Cov-SC}

\medskip

\noindent
\textbf{Input:}
Adjacency matrix $\boldsymbol{A}$, covariate matrix $\boldsymbol{X}$, tuning parameter $\alpha$, regularization parameter $\tau \geq 0$, number of communities $K$, and number of node types $T$.

\begin{enumerate}

\item Compute the covariate-assisted regularized Laplacian:
\[
\Tilde{\boldsymbol{L}}_{\tau}(\alpha)
:=
\boldsymbol{L}_{\tau}\boldsymbol{L}_{\tau}
+
\alpha \boldsymbol{X}\boldsymbol{X}^{\top},
\]
where $\boldsymbol{L}_{\tau}$ denotes the regularized graph Laplacian defined in Section~\ref{sec2}.

\item Since $\Tilde{\boldsymbol{L}}_{\tau}(\alpha)$ is symmetric, compute the $TK$ eigenvectors corresponding to the eigenvalues with largest absolute values. Denote these eigenvectors by
\[
\boldsymbol{U}_1,\ldots,\boldsymbol{U}_{TK}.
\]
Construct the matrix
\[
\boldsymbol{U}
=
[\boldsymbol{U}_1,\ldots,\boldsymbol{U}_{TK}]
\in \mathbb{R}^{N \times TK}.
\]

\item (Optional) Normalize each row of $\boldsymbol{U}$ to have unit Euclidean norm:
\[
\boldsymbol{U}^{*}(i,j)
:=
\frac{\boldsymbol{U}(i,j)}
{\sqrt{\sum_j \boldsymbol{U}(i,j)^2}}.
\]
This normalization step reduces scale variation across rows and is commonly used in spectral clustering.

\item For each node type $t=1,\ldots,T$, extract the rows of $\boldsymbol{U}^{*}/\boldsymbol{U}$ corresponding to nodes of type $t$ and apply the $K$-means algorithm separately within that subset. Since node types are known beforehand, this step clusters nodes only among comparable node categories.

\end{enumerate}

\noindent
\textbf{Output:}
A partition of the nodes into $TK$ communities respecting the heterogeneous network structure.

\medskip

Note that the construction of $\Tilde{\boldsymbol{L}}_{\tau}(\alpha)$ requires an a priori choice of the tuning parameter $\alpha$. The procedure for selecting $\alpha$ is discussed next.

\subsubsection{Selection of the Tuning Parameter \texorpdfstring{$\alpha$}{alpha}}
\label{select tuning param}

The tuning parameter $\alpha$ controls the balance between graph-based structural information and covariate-based similarity information. Therefore, selecting an appropriate value of $\alpha$ is crucial for effective clustering performance.

We adopt and extend the tuning strategy proposed by \cite{cov_sc_bink}. The idea is to select the value of $\alpha$ that minimizes the within-cluster variability in the spectral embedding space.
Formally, let
\[
\Phi(\alpha)
=
\sum_{i=1}^{TK}
\sum_{\boldsymbol{u}_j \in F_i}
\left\|
\boldsymbol{u}_j(\alpha)
-
\boldsymbol{C}_i(\alpha)
\right\|^2,
\]
denote the $K$-means objective function, where $\boldsymbol{u}_j(\alpha)$ is the $j$-th row of the eigenvector matrix $\boldsymbol{U}(\alpha)$, $\boldsymbol{C}_i(\alpha)$ denotes the centroid of cluster $i$, and $F_i$ is the set of points assigned to cluster $i$.
The optimal tuning parameter is then chosen as
\[
\alpha^{*}
=
\underset{\alpha \in [\alpha_{\min},\alpha_{\max}]}
{\arg\min}
\,
\Phi(\alpha).
\]

The search interval $[\alpha_{\min},\alpha_{\max}]$ is determined from the spectral properties of $\boldsymbol{L}_{\tau}\boldsymbol{L}_{\tau}$ and $\boldsymbol{X}\boldsymbol{X}^{\top}$:
\begin{align*}
\alpha_{\min}
&=
\frac{
\lambda_{TK}(\boldsymbol{L}_{\tau}\boldsymbol{L}_{\tau})
-
\lambda_{TK+1}(\boldsymbol{L}_{\tau}\boldsymbol{L}_{\tau})
}
{
\lambda_1(\boldsymbol{X}\boldsymbol{X}^{\top})
},
\\
\alpha_{\max}
&=
\frac{
\lambda_1(\boldsymbol{L}_{\tau}\boldsymbol{L}_{\tau})
}
{
\lambda_R(\boldsymbol{X}\boldsymbol{X}^{\top}) \cdot \mathbb{I}(R \leq TK)
+
\left[
\lambda_{TK}(\boldsymbol{X}\boldsymbol{X}^{\top})
-
\lambda_{TK+1}(\boldsymbol{X}\boldsymbol{X}^{\top})
\right]
\cdot
\mathbb{I}(R > TK)
},
\end{align*}
where $\lambda_m(\cdot)$ denotes the $m$-th largest eigenvalue of the corresponding matrix.

Intuitively, these bounds define a reasonable range within which neither the graph structure nor the covariate information overwhelmingly dominates the clustering procedure.
Both $\alpha_{\min}$ and $\alpha_{\max}$ are directly computable from the observed data matrices. The resulting optimization strategy ensures that the final clustering effectively combines structural and attribute-based information.

\section{Theoretical Results: Block Model on Heterogeneous Networks in the Presence of Nodal Covariates}

In this section, we present the theoretical foundations of the proposed covariate-assisted spectral clustering procedure for heterogeneous networks. We first introduce the probabilistic model assumptions governing the network and nodal covariates. Subsequently, we establish consistency results for the proposed clustering procedure under the heterogeneous node-contextualized stochastic blockmodel framework.

The overall proof strategy closely follows the spectral perturbation framework commonly employed in the spectral clustering literature. Broadly speaking, the argument proceeds in three stages. First, we establish concentration bounds for the empirical covariate-assisted Laplacian around its population counterpart. Next, we apply the Davis--Kahan perturbation theorem to control the deviation between the sample and population eigenvector matrices, up to an orthogonal transformation. Finally, these results are combined to derive bounds on the node-wise misclustering error rates for each node type.


\subsection{Model Assumptions}

We consider a joint probabilistic model for the graph and the nodal covariates. The framework extends the heterogeneous stochastic blockmodel by incorporating node-level covariate information whose distribution depends on the latent block memberships.

Suppose the network consists of $N$ nodes, each belonging to one of $T$ node types and one of $K$ communities (or blocks). Consequently, each node belongs to exactly one of the $TK$ type-block combinations.
Let $\boldsymbol{Z} \in \{0,1\}^{N \times TK}$ denote the membership matrix, where
\[
\boldsymbol{Z}_{i,(t,k)} =
\begin{cases}
1, & \text{if node } i \text{ belongs to type } t \text{ and block } k,\\
0, & \text{otherwise}.
\end{cases}
\]

As in the standard SBM framework, edges are assumed to be generated independently conditional on the latent memberships. However, unlike homogeneous SBMs, the edge probabilities now depend jointly on both node types and community memberships.

Throughout this section, population counterparts of sample-level quantities are denoted by calligraphic symbols; for example, $\boldsymbol{A}$ and $\boldsymbol{\mathcal{A}}$.
Let 
$\boldsymbol{P} \in [0,1]^{TK \times TK}$
be the symmetric full-rank matrix of connection probabilities. The population adjacency matrix is defined as
\[
\boldsymbol{\mathcal{A}}^{N \times N}
:=
\mathbb{E}(\boldsymbol{A}\mid \boldsymbol{Z})
=
\underbrace{\boldsymbol{Z}}_{N \times TK}
\underbrace{\boldsymbol{P}}_{TK \times TK}
\underbrace{\boldsymbol{Z}^{\mathsf T}}_{TK \times N}.
\] 
Thus, conditional on $\boldsymbol{Z}$,
\[
\boldsymbol{A}_{ij}
\sim
\mathrm{Bernoulli}
\left(
\boldsymbol{Z}_{i\cdot}\boldsymbol{P}\boldsymbol{Z}_{j\cdot}^{\mathsf T}
\right),
\qquad 1\le i<j\le N,
\]
independently across node pairs.

Intuitively, the matrix $\boldsymbol{\mathcal{A}}$ represents the ideal noiseless connectivity structure of the graph representing the network, while the observed adjacency matrix $\boldsymbol{A}$ corresponds to a random realization around this population structure.

\vspace{0.2cm}

In addition to the graph structure, each node is associated with an $R$-dimensional covariate vector. We allow the covariate dimension $R$ to grow with the number of nodes.
Let
$\boldsymbol{X}\in[-J,J]^{N\times R}$
denote the bounded covariate matrix for some fixed constant $J>0$. The distribution of the covariates depends only on the underlying type-block combination of the node.
Define 
$\boldsymbol{M}\in[-J,J]^{TK\times R}$,
where $\boldsymbol{M}_{ij}$ represents the expectation of the $j^{\text{th}}$ covariate for a node belonging to the $i^{\text{th}}$ type-block combination. Then the population covariate matrix is given by
\[
\boldsymbol{\mathcal{X}}
:=
\mathbb{E}(\boldsymbol{X}\mid \boldsymbol{Z})
=
\underbrace{\boldsymbol{Z}}_{N\times TK}
\underbrace{\boldsymbol{M}}_{TK\times R}.
\]
Conditional on $\boldsymbol{Z}$, the covariates are assumed to be independent across nodes and follow the above block-dependent structure.

The above formulation naturally captures the intuition that nodes belonging to the same latent type-block combination should exhibit similar connectivity behaviour as well as similar covariate characteristics.


\subsection{Consistency of Covariate-Assisted Spectral Clustering}

We now establish statistical consistency of the proposed covariate-assisted spectral clustering procedure under the heterogeneous node-contextualized stochastic blockmodel.

The key idea is to view spectral clustering as an eigenspace estimation problem. The empirical clustering algorithm operates on the eigenvectors of a sample-level covariate-assisted Laplacian, while the population eigenvectors encode the true latent community structure. Therefore, consistency reduces to showing that the sample eigenspace concentrates around its population counterpart.


\subsubsection{Population Covariate-Assisted Laplacian}

Recall that the empirical covariate-assisted Laplacian is defined as
\[
\Tilde{\boldsymbol{L}}(\alpha)
=
\boldsymbol{L}_{\tau}\boldsymbol{L}_{\tau}
+
\alpha \boldsymbol{X}\boldsymbol{X}^{\mathsf T}.
\]
The corresponding population version is defined as
\[
\widetilde{\boldsymbol{\mathcal L}}(\alpha)
=
(\boldsymbol{\mathcal D}+\tau \boldsymbol I)^{-1/2}
\boldsymbol{\mathcal A}
(\boldsymbol{\mathcal D}+\tau \boldsymbol I)^{-1}
\boldsymbol{\mathcal A}
(\boldsymbol{\mathcal D}+\tau \boldsymbol I)^{-1/2}
+
\alpha\,\mathbb E(\boldsymbol X\boldsymbol X^{\mathsf T}\mid \boldsymbol Z).
\]
where $\boldsymbol{\mathcal{D}}$ denotes the population degree matrix corresponding to $\boldsymbol{\mathcal{A}}$.

Intuitively, $\boldsymbol{\tilde{\mathcal{L}}}(\alpha)$ represents the ideal population-level object that combines structural information from the graph and information from the nodal covariates. The observed matrix $\Tilde{\boldsymbol{L}}(\alpha)$ may therefore be viewed as a noisy empirical approximation of this quantity.

Let $\boldsymbol{U}$ and $\boldsymbol{\mathcal{U}}$ denote the matrices containing the leading $TK$ eigenvectors of $\Tilde{\boldsymbol{L}}$ and $\Tilde{\boldsymbol{\mathcal{L}}}$, respectively.
Since both matrices are symmetric positive semidefinite, their nonzero eigenvalues are non negative and the corresponding eigenvectors are orthogonal.


\subsubsection{Structure of the Population Eigenspace}

We first establish that the population eigenspace exactly encodes the underlying latent type-block structure. Intuitively, this result shows that, at the population level, the proposed covariate-assisted spectral embedding perfectly separates the communities. Consequently, any clustering error arises only due to random fluctuations in the observed network and covariates.

Before formally stating the lemma, we introduce several quantities that characterize the population-level structure of the model.
Under the node-contextualized heterogeneous stochastic blockmodel, define
\[
\boldsymbol{\mathcal{D}_{P}
=
\mathrm{diag}(PZ^{T}\mathbf{1}_{N} + \tau)},
\qquad 
\boldsymbol{\Tilde{W}
=
Z^{T}Z},
\]
and
\[
\boldsymbol{\Tilde{P}
=
\mathcal{D}_{P}^{-1/2}
PZ^{T}\mathcal{D}_{\tau}^{-1}
ZP
\mathcal{D}_{P}^{-1/2}}
+
\alpha \boldsymbol{MM^{T}}.
\]

Here, $\boldsymbol{\Tilde{W}}$ records the sizes of the type-block combinations through the membership structure captured in $\boldsymbol{Z}$. The matrix $\boldsymbol{\Tilde{P}}$ may be viewed as the effective population connectivity matrix after incorporating both graph structure and covariate information. 

Next, define
$\mathfrak{X}
=
\underset{l}{\max}
\, |c_{l} - \Bar{c}|$, 
where
\[
c_{l}
=
\sum_{i}
\mathrm{var}(X_{il}\mid Z_i=l),
\qquad
\Bar{c}
=
\frac{1}{TK}\sum_{l} c_l.
\]
The quantity $\mathfrak{X}$ measures the maximum deviation of the within type-block covariate variability from its overall average level. Thus, smaller values of $\mathfrak{X}$ correspond to more homogeneous covariate variability across the latent type-block combinations.

We now state the following structural result for the population eigenspace.

\begin{lemma}\label{lemma1}

Let
$\boldsymbol{\mathcal{U}} \in \mathbb{R}^{N \times TK}$
denote the matrix whose columns are the eigenvectors corresponding to the top $TK$ eigenvalues of the population covariate-assisted Laplacian matrix $\Tilde{\boldsymbol{\mathcal{L}}}$. Suppose that
\[
\lambda_{TK}(\Tilde{\boldsymbol{P}}\Tilde{\boldsymbol{W}})
>
2\alpha \mathfrak{X},
\]
where $\lambda_{TK}(\cdot)$ denotes the $TK^{\text{th}}$ largest eigenvalue.
Then there exists an orthogonal matrix
$V \in \mathbb{R}^{TK \times TK}$
such that
\[
\boldsymbol{\mathcal{U}}
=
\boldsymbol{Z(Z^{T}Z)^{-1/2}V}.
\]

Moreover, for any nodes $i$ and $j$,
\[
\boldsymbol{Z_i(Z^{T}Z)^{-1/2}V}
=
\boldsymbol{Z_j(Z^{T}Z)^{-1/2}V}
\]
if and only if
$\boldsymbol{Z_i = Z_j},$
where $\boldsymbol{Z_i}$ denotes the $i^{\text{th}}$ row of the membership matrix $\boldsymbol{Z}$.

\end{lemma}

The above lemma shows that the rows of the population eigenvector matrix are completely determined by the latent type-block memberships. In particular, nodes belonging to the same type-block combination are mapped to exactly the same point in the population eigenspace, whereas nodes belonging to different type-block combinations are mapped to distinct points.

Consequently, the latent communities become perfectly separable at the population level. This property forms the fundamental basis for the consistency of the proposed spectral clustering procedure.


\subsubsection{Concentration of the Empirical Laplacian}
The next theorem establishes concentration of the empirical covariate-assisted Laplacian around its population counterpart. Intuitively, the result shows that the random matrix
$\Tilde{\boldsymbol{L}}$ 
constructed from the observed network and covariates remains close, in spectral norm, to its population analogue 
$\Tilde{\boldsymbol{\mathcal{L}}},$
provided the graph is not excessively sparse and the covariate signal is sufficiently informative. Such concentration results form the key technical ingredient for establishing consistency of spectral clustering procedures.

Before stating the theorem formally, we introduce several quantities that will appear throughout the subsequent analysis. Let 
$d
=
\underset{i}{\min}\,
\mathcal{D}_{ii},$
denote the minimum expected degree of the {population} network. This quantity plays an important role in controlling the sparsity level of the graph.
Further, define
$\mathcal{X}_{ik}^{(p)}
=
E(X_{ik}^{p}),$
to be the $p^{\text{th}}$ moment of the $k^{\text{th}}$ covariate corresponding to node $i$.

Next, define
\[
\Bar{\omega}
=
\sum_{k}
\left\{
\sum_{i}
\mathcal{X}_{ik}^{(2)}
\sum_{l}
\left(
\mathcal{X}_{lk}^{(2)}
-
\mathcal{X}_{lk}^{2}
\right)
+
\mathcal{X}_{ik}^{(4)}
\right\}.
\]
The quantity $\Bar{\omega}$ serves as an upper bound on the aggregate variability contributed by the covariates and quantifies the extent to which covariate fluctuations influence the spectral concentration behavior of the proposed matrix.\footnote{More precisely,
$||\sum_{k} var(\alpha X_k X_k^T)|| \leq \Bar{\omega}$.}
We also define the quantity 
\[
S
=
3\alpha N J^{2},
\]
which depends on the tuning parameter $\alpha$, the network size $N$, and the covariate bound $J$.
Finally, let
\[
\delta
=
12(d+\tau)^{-1/2}
+
\Bar{\omega}^{1/2}.
\]
Intuitively, $\delta$ captures the combined effects of graph-level and covariate-level variability on the deviation between the empirical and population covariate-assisted Laplacian matrices. The first term in $\delta$ reflects the contribution of graph noise arising from network sparsity and edge randomness, whereas the second term quantifies the contribution of covariate-induced noise. Together, these two components determine the overall magnitude of the perturbation between the sample and population matrices.

\begin{theorem}\label{thm2}

Suppose $\epsilon>0$ satisfies
\begin{enumerate}[label=(\roman*)]
\item \label{thm2_cond_i}
$d+\tau
>
3\log(8N/\epsilon),$ and 

\item \label{thm2_cond_ii}
$\Bar{\omega}/S^{2}
>
3\log(8N/\epsilon).$
\end{enumerate}

Then, with probability at least $(1-\epsilon)$,
\[
\left\|
\Tilde{L}
-
\Tilde{\mathcal{L}}
\right\|
\le
\delta
\left\{
3\log(8N/\epsilon)
\right\}^{1/2}.
\]

\end{theorem}

Condition~\ref{thm2_cond_i} controls the sparsity level of the graph and ensures that the network is sufficiently dense for spectral concentration to hold. In sparse graphs, random fluctuations in the adjacency structure can dominate the underlying signal, making spectral recovery unstable. The regularization parameter $\tau$ helps alleviate this issue by stabilizing the degree structure.
Condition~\ref{thm2_cond_ii} imposes a growth condition on the covariate contribution relative to the network size. Intuitively, this assumption ensures that the covariate information carries sufficient signal strength and does not vanish asymptotically. In particular, the covariate signal remains strong enough to dominate the random fluctuations arising from the covariate space, thereby preventing the informative structure from being overwhelmed by noise. Consequently, the covariates contribute meaningfully to the spectral structure of the combined matrix and can effectively assist the clustering procedure.


\subsubsection{Eigenvector Perturbation Bound}

Having established concentration of the empirical covariate-assisted Laplacian around its population counterpart, we next study the behaviour of the associated eigenspaces. The main objective here is to show that the leading empirical eigenvectors closely approximate the corresponding population eigenvectors. This step is important because spectral clustering is ultimately performed on these eigenvectors.

The result follows by combining the concentration bound obtained in Theorem~\ref{thm2} with the classical Davis--Kahan perturbation theorem (\cite{davis_kahan_1970}), which relates perturbations of matrices to perturbations of their eigenspaces.

Before stating the theorem formally, we introduce the following notation. Let $\lambda_{TK}$ denote the $TK^{\text{th}}$ largest eigenvalue of the population matrix $\Tilde{\boldsymbol{\mathcal{L}}}$, and let $\boldsymbol{\mathcal{O}}$ be an orthogonal matrix accounting for the rotational non-identifiability of eigenspaces.
Further, let $\boldsymbol{U}$ and $\boldsymbol{\mathcal{U}}$ denote the matrices containing the leading $TK$ eigenvectors of $\Tilde{\boldsymbol{L}}$ and $\Tilde{\boldsymbol{\mathcal{L}}}$, respectively.
\begin{theorem}\label{thm3}

Under the assumptions of Lemma~\ref{lemma1}, conditions~\ref{thm2_cond_i} and~\ref{thm2_cond_ii} of Theorem~\ref{thm2}, and further assuming that
\begin{enumerate}[label=(\roman*)]
\item \label{cond_thm_3}
\[
\delta
\{3\log(8N/\epsilon)\}^{1/2}
\le
\frac{\lambda_{TK}}{2},
\]
\end{enumerate}
we have, with probability at least $(1-\epsilon)$,
\[
\|
\boldsymbol{U}
-
\boldsymbol{\mathcal{U}}\mathcal{O}
\|_{F}
\le
\frac{
8\delta
\{3TK\log(8N/\epsilon)\}^{1/2}
}{
\lambda_{TK}
}.
\]

\end{theorem}

The above theorem shows that the empirical eigenspace consistently estimates the population eigenspace whenever the signal strength, represented by $\lambda_{TK}$, dominates the perturbation noise level. In particular, a larger eigengap leads to greater stability of the spectral embedding under random fluctuations in the observed network and covariates.

Condition~\ref{cond_thm_3} ensures that the perturbation level is sufficiently small relative to the population spectral signal. Under this regime, the leading empirical eigenvectors remain close to their population counterparts, thereby enabling accurate recovery of the latent community structure through spectral clustering.


\subsubsection{Misclustering Error Bound and Consistency}

We now translate the eigenspace perturbation result into bounds on the node misclustering rate.
Recall that spectral clustering applies the $K$-means algorithm to the rows of the eigenvector matrix $\boldsymbol{U}$. Let $\boldsymbol{C_i^{(t)}}$ and $\boldsymbol{\mathcal{C}_i^{(t)}}$ denote the empirical and population cluster centroids corresponding to the cluster containing the $i^{\text{th}}$ node of type-$t$, respectively.

A node is considered correctly clustered if its empirical centroid is closer to the corresponding population centroid than to any centroid associated with a different community of the same node type.
Since eigenspaces are identifiable only up to orthogonal transformations, the clustering definition is stated modulo an orthogonal rotation matrix.

\begin{definition}
Let $\boldsymbol{\mathcal{O}}$ be the orthogonal matrix minimizing
\[
||
\boldsymbol{U\mathcal{O}^{T} - \mathcal{U}}
||_{F}.
\]
Define the set of misclustered nodes of type-$t$ as
\[
\mathcal{M}_{t}
=
\left\{
i\in \text{type-}t
:
\exists j\in\text{type-}t,\ j\neq i
\text{ such that }
||
\boldsymbol{C_i^{(t)}\mathcal{O}^{T}-\mathcal{C}_i^{(t)}}
||_2
>
||
\boldsymbol{C_i^{(t)}\mathcal{O}^{T}-\mathcal{C}_j^{(t)}}
||_2
\right\}.
\]
\end{definition}

We now state the main consistency result for the proposed covariate-assisted spectral clustering procedure on heterogeneous networks. The theorem provides an upper bound on the number of misclustered nodes for each node type and establishes weak consistency of the method under suitable asymptotic conditions.

Before stating the theorem formally, we introduce the following notation. Consider a $T$-type, $K$-block heterogeneous stochastic blockmodel with
\[
N=\sum_{t=1}^{T} N_t,
\]
where $N_t$ denotes the number of nodes of type $t$.

Further, let
\[
n_{\max}
=
\underset{i}{\max}
\,
(\boldsymbol{Z^{T}Z})_{ii}
\]
denote the size of the largest type-block cluster. Intuitively, $n_{\max}$ measures the largest community size among all type-block combinations and appears naturally in the misclustering bound.

\begin{theorem}\label{thm4}

Under all assumptions of Lemma~\ref{lemma1}, conditions~\ref{thm2_cond_i} and~\ref{thm2_cond_ii} of Theorem~\ref{thm2}, and condition~\ref{cond_thm_3} of Theorem~\ref{thm3}, we have, with probability at least $(1-\epsilon)$,
\[
|\mathcal{M}_t|
\le
\frac{
c\,TK\,n_{\max}\,\delta^{2}\log(8N/\epsilon)
}{
\lambda_{TK}^{2}
}.
\]

\end{theorem} where $|S|$ denoting the cardinality of the set $S.$

The theorem establishes weak consistency of the proposed covariate-assisted spectral clustering procedure for heterogeneous networks. In particular, whenever the population signal strength, represented through $\lambda_{TK}$, grows sufficiently faster than the perturbation error, the proportion of misclustered nodes converges to zero asymptotically.

The result also illustrates the interplay between graph structure, covariate information, and spectral separation. Stronger community signal and informative covariates improve clustering accuracy, whereas higher perturbation noise or weaker eigengap leads to larger misclustering error.


\section{Simulation Studies}

In this section, we investigate the empirical performance of the proposed heterogeneous covariate assisted spectral clustering framework through a series of simulation experiments. The primary objective of these experiments is to understand how heterogeneity across node types and the strength of covariate information jointly influence clustering accuracy.

We consider several simulation settings to study the effect of the heterogeneity parameter, the separation between block specific covariate means, and their combined influence on the misclustering error rate. In addition, we compare the proposed method with several existing benchmark clustering procedures commonly used in network analysis.

\subsection{Description of the Simulation Settings}

In this section, we describe the simulation framework used to evaluate the empirical performance of the proposed heterogeneous covariate assisted spectral clustering method. The primary objective of these experiments is to understand how heterogeneity across node types and the strength of covariate information jointly influence clustering accuracy. 
Throughout the simulation study, we consider a covariate assisted $K$ block stochastic blockmodel with $K=3$ communities on a bi type heterogeneous network $(T=2)$ with a total $n$ number of nodes. In addition to the network structure, each node is associated with a Bernoulli covariate vector of dimension $R=3$. The proposed clustering method is then applied to networks generated from this heterogeneous stochastic blockmodel framework under a variety of experimental settings.

The simulation model is parameterized by
\[
n, K, T, R, p_1, r_1, p_2, r_2, p_3, r_3,
p^{type1}_{out},
p^{type2}_{out},
d_{cov},
\]
where the network contains an equal number $(n/2)$ of nodes from each type.

The parameters $p_i$ and $r_i$, for $i=1,2,3$, determine the network connectivity structure, while
\[
p^{type1}_{in},
\quad
p^{type1}_{out},
\quad
p^{type2}_{in},
\quad
p^{type2}_{out}
\]
control the covariate expectation matrix $\mathbf{M}$.
The block probability matrix is defined as
\[
\mathbf{P}
=
\begin{pmatrix}
\mathbf{P}_{11} & \mathbf{P}_{12} \\
\mathbf{P}_{21} & \mathbf{P}_{22}
\end{pmatrix},
\]
where each submatrix $\mathbf{P}_{st}$ is of dimension $K\times K$ and represents the connection probabilities between type $s$ and type $t$ nodes, for $s,t\in\{1,2\}$.
Formally, the block matrices are given by
\[
\mathbf{P}_{11}
=
p_1 \mathbf{1}_K \mathbf{1}_K^{T}
+
r_1 \mathbf{I}_K,
\]
\[
\mathbf{P}_{22}
=
p_2 \mathbf{1}_K \mathbf{1}_K^{T}
+
r_2 \mathbf{I}_K,
\]
and
\[
\mathbf{P}_{12}
=
\mathbf{P}_{21}
=
p_3 \mathbf{1}_K \mathbf{1}_K^{T}
+
r_3 \mathbf{I}_K.
\]
Here, $\mathbf{1}_K$ denotes the $K$ dimensional vector of ones and $\mathbf{I}_K$ denotes the identity matrix of order $K$.

The above parameterization provides an interpretable way to control the strength of community structure and heterogeneity in the network. In particular, for the type 1 type 1 homogeneous subnetwork, $p_1$ represents the baseline inter block connection probability, while
$p_1+r_1$
corresponds to the intra block connection probability. Similarly, for the type 2 type 2 homogeneous subnetwork, $p_2$ and $p_2+r_2$ represent the inter block and intra block connection probabilities respectively. Consequently, the parameters $r_1$ and $r_2$ determine the strength of homophily within the homogeneous subnetworks.

For interactions between type 1 and type 2 nodes, the parameter $p_3$ controls the baseline inter type connection probability, whereas $r_3$ determines the strength of inter type homophily. Larger values of $r_3$ therefore correspond to stronger heterogeneous community structure.

Next, we define the covariate expectation matrix as
\[
\mathbf{M}
=
\begin{pmatrix}
M_{type1} \\
M_{type2}
\end{pmatrix},
\]
where
\[
M_{type1}
=
\begin{bmatrix}
p^{type1}_{in} & p^{type1}_{out} & p^{type1}_{out} \\
p^{type1}_{out} & p^{type1}_{in} & p^{type1}_{out} \\
p^{type1}_{out} & p^{type1}_{out} & p^{type1}_{in}
\end{bmatrix},
\]
and
\[
M_{type2}
=
\begin{bmatrix}
p^{type2}_{in} & p^{type2}_{out} & p^{type2}_{out} \\
p^{type2}_{out} & p^{type2}_{in} & p^{type2}_{out} \\
p^{type2}_{out} & p^{type2}_{out} & p^{type2}_{in}
\end{bmatrix}.
\]

The quantities
\[
p^{type1}_{in}-p^{type1}_{out}
\qquad \text{and} \qquad
p^{type2}_{in}-p^{type2}_{out}
\]
are both denoted by $d_{cov}$.
The parameter $d_{cov}$ measures the separation between block specific covariate means. Intuitively, larger values of $d_{cov}$ imply that the covariates carry stronger information about the underlying community memberships, thereby making the clustering task easier.

The primary objective of the simulation study is to investigate how the incorporation of covariate information improves clustering performance in heterogeneous networks. In addition, we compare the proposed method with several benchmark clustering procedures commonly used in the literature.

The clustering performance is evaluated through a Monte Carlo type simulation framework. For each parameter configuration, we independently generate 50 networks according to the model described above. For every realization, we compute the corresponding misclustering error rate, and the reported results correspond to the average error across all replications. The average misclustering error rates are reported separately for the two node types.
The details of the individual simulation experiments are presented in the following subsections.

\subsection{Simulation 1: Effect of the Heterogeneity Parameter $r_3$}

In the first set of experiments, we investigate the effect of the heterogeneity parameter $r_3$ on the clustering performance of the proposed method. Recall that $r_3$ controls the strength of inter type homophily in the heterogeneous network. Consequently, larger values of $r_3$ correspond to stronger alignment between the latent community structures across the two node types.

To isolate the effect of heterogeneity, the covariate separation parameter $d_{cov}$ is fixed at a predetermined level, while $r_3$ is varied systematically over the interval $[0,0.7]$. This range spans settings from weak to strong heterogeneous community structure.
The remaining model parameters are fixed as follows:
\[
p_1 = p_2 = 0.2,
\qquad
r_1 = r_2 = 0.01,
\qquad
p_3 = 0.15,
\]
\[
p^{type1}_{out} = 0.08,
\qquad
p^{type2}_{out} = 0.05,
\qquad
d_{cov} = 0.5.
\]

For each value of $r_3$, we independently generate 50 networks from the heterogeneous stochastic blockmodel and compute the corresponding misclustering error rates. The reported results correspond to the average misclustering error over all replications, computed separately for the two node types.

\begin{figure}[H]
    \centering
    \captionsetup{skip=2pt}
    \includegraphics[width=0.6\linewidth]{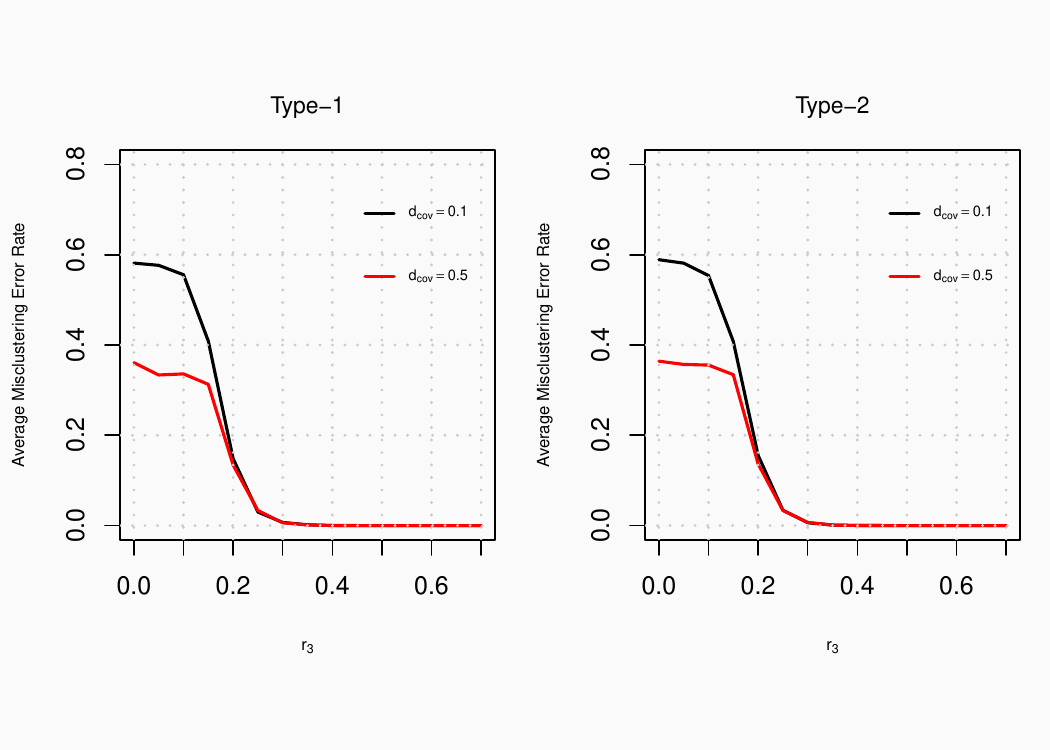}
    \caption{Average misclustering error rate versus the heterogeneity parameter $r_3$. Clustering accuracy improves as the strength of heterogeneous community structure increases, for both node types, when all other parameters are fixed $(n=250)$.}
    \label{d_fixed_r3_vary_n=250}
\end{figure}

\begin{figure}[H]
    \centering
    \includegraphics[width=0.6\linewidth]{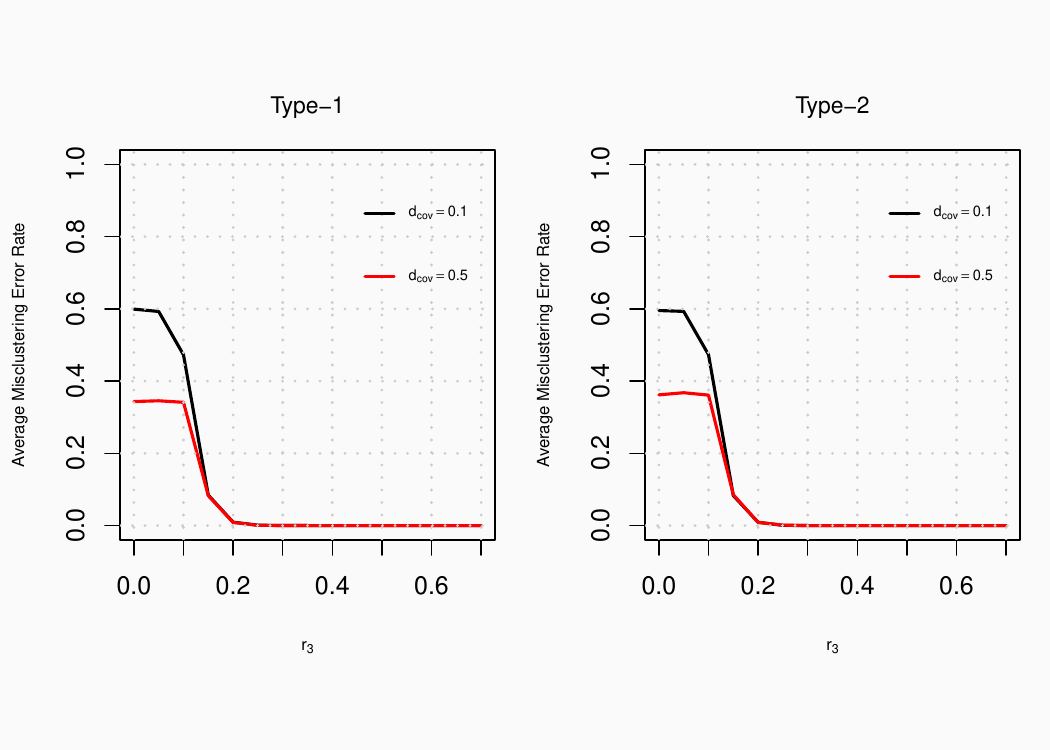}
    \caption{Average misclustering error rate decreases with increasing values of the heterogeneity parameter $r_3$ for both type 1 and type 2 nodes when all other parameters are fixed $(n=500)$.}
    \label{d_fixed_r3_vary_n=500}
\end{figure}

The experiments are carried out for network sizes $n=250$ and $n=500$, respectively, with an equal number of nodes assigned to each type.
Figures~\ref{d_fixed_r3_vary_n=250} and~\ref{d_fixed_r3_vary_n=500} summarize the results. The plots show a clear decreasing trend in the average misclustering error rate as $r_3$ increases. This behaviour is consistent across both node types and for both network sizes.

Intuitively, larger values of $r_3$ strengthen the inter type community signal, thereby making the latent block structure more distinguishable in the spectral embedding. As a result, the clustering algorithm is able to recover the communities more accurately. Furthermore, the improvement becomes more pronounced for larger network sizes, reflecting the consistency behaviour predicted by the theoretical results.

\subsection{Simulation 2: Effect of the Block-Specific Covariate Mean Separation Parameter ($d_{\mathrm{cov}}$) } 

We next investigate the effect of covariate separation on clustering performance. Recall that 
$d_{cov}$ measures the difference between within block and between block covariate expectations. Larger values of $d_{cov}$ therefore correspond to stronger covariate signals and better separation among latent communities through the nodal covariates.

In this experiment, all network parameters are kept fixed at the values specified earlier, while $d_{cov}$ is varied over a suitable range. The heterogeneity parameter $r_3$ is fixed at selected levels in order to isolate the effect of covariate information on clustering accuracy.

The experiments are conducted for network sizes $n=250$ and $n=500$, with an equal number of nodes assigned to each type. For every parameter configuration, we independently generate 50 networks and compute the average misclustering error rate separately for the two node types.
The objective of this experiment is to examine whether stronger covariate separation improves the recovery of latent communities in heterogeneous networks.

\begin{figure}[H]
    \centering
    \captionsetup{skip=2pt}
    \includegraphics[width=0.6\linewidth]{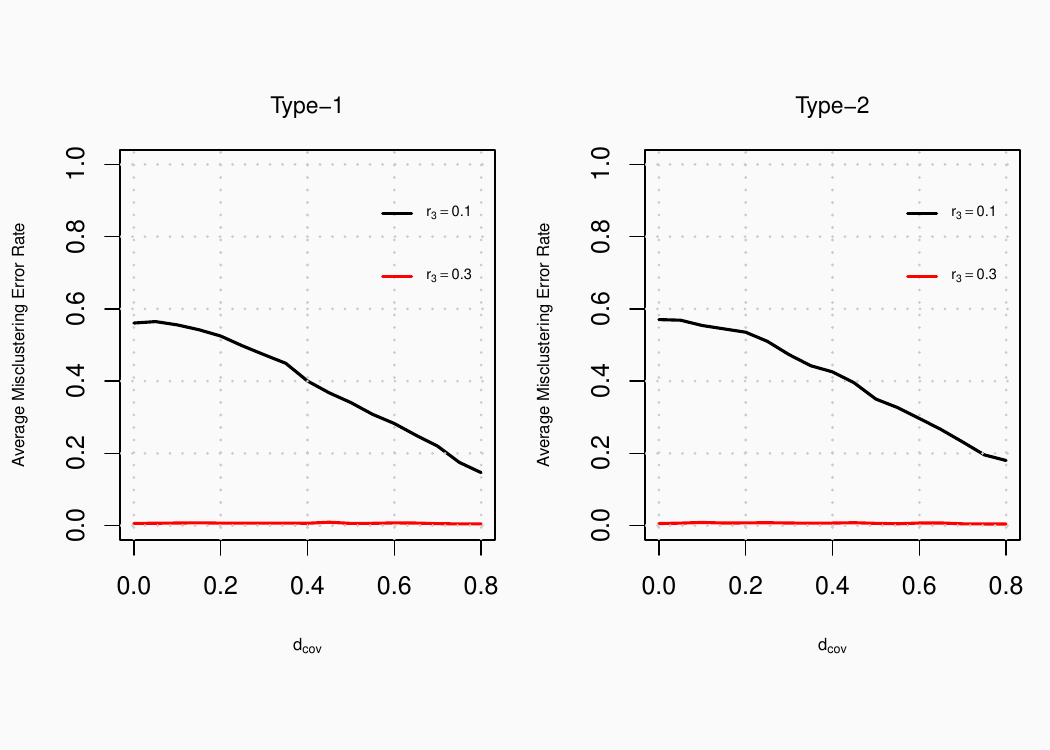}
    \caption{Average misclustering error rate decreases as the covariate separation parameter $d_{cov}$ increases, while keeping all other parameters fixed $(n=250)$.}
    \label{r3_fixed_d_vary_n=250}
\end{figure}

\vspace{-1em}

\begin{figure}[H]
    \centering
    \captionsetup{skip=2pt}
    \includegraphics[width=0.6\linewidth]{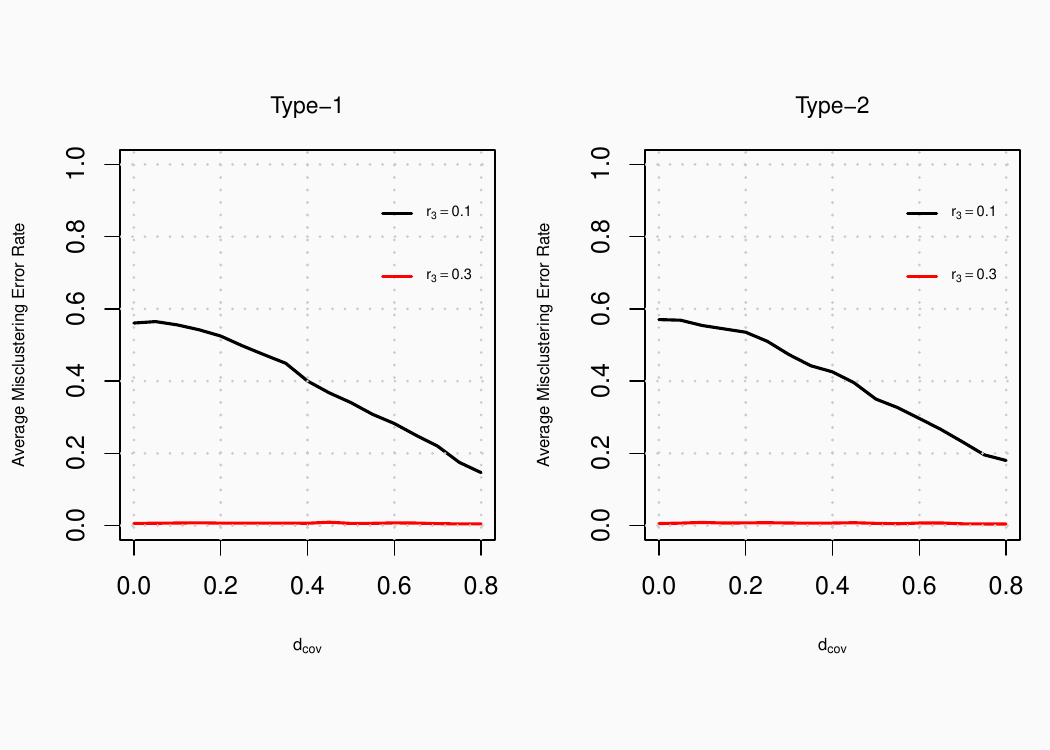}
    \caption{Average misclustering error rate decreases with increasing $d_{cov}$ for both node types when $n=500$.}
    \label{r3_fixed_d_vary_n=500}
\end{figure}

Figures~\ref{r3_fixed_d_vary_n=250} and~\ref{r3_fixed_d_vary_n=500} show a clear decreasing trend in the average misclustering error rate as $d_{cov}$ increases. This behaviour is consistent with the intuition that stronger separation in covariate distributions provides additional information for identifying latent communities. The improvement becomes more pronounced for larger network sizes, where the combined effect of network structure and covariate information leads to more accurate recovery of the underlying blocks.

\subsection{Simulation 3: Joint Effect of $r_3$ and $d_{cov}$}

The previous experiments separately examined the effects of the heterogeneity parameter $r_3$ and the covariate separation parameter $d_{cov}$. We now study their joint effect on clustering performance.

In this experiment, the pair $(r_3,d_{cov})$ is varied simultaneously over the range $(0,0)$ to $(0.55,0.55)$. The purpose is to understand how the interaction between heterogeneity and covariate information influences the recovery of latent communities.

Intuitively, larger values of $r_3$ strengthen the alignment between node types across blocks, while larger values of $d_{cov}$ increase the separation among block specific covariate distributions. Therefore, increasing both parameters simultaneously should improve the quality of clustering.

\begin{figure}[H]
    \centering
    \includegraphics[width=0.7\textwidth]{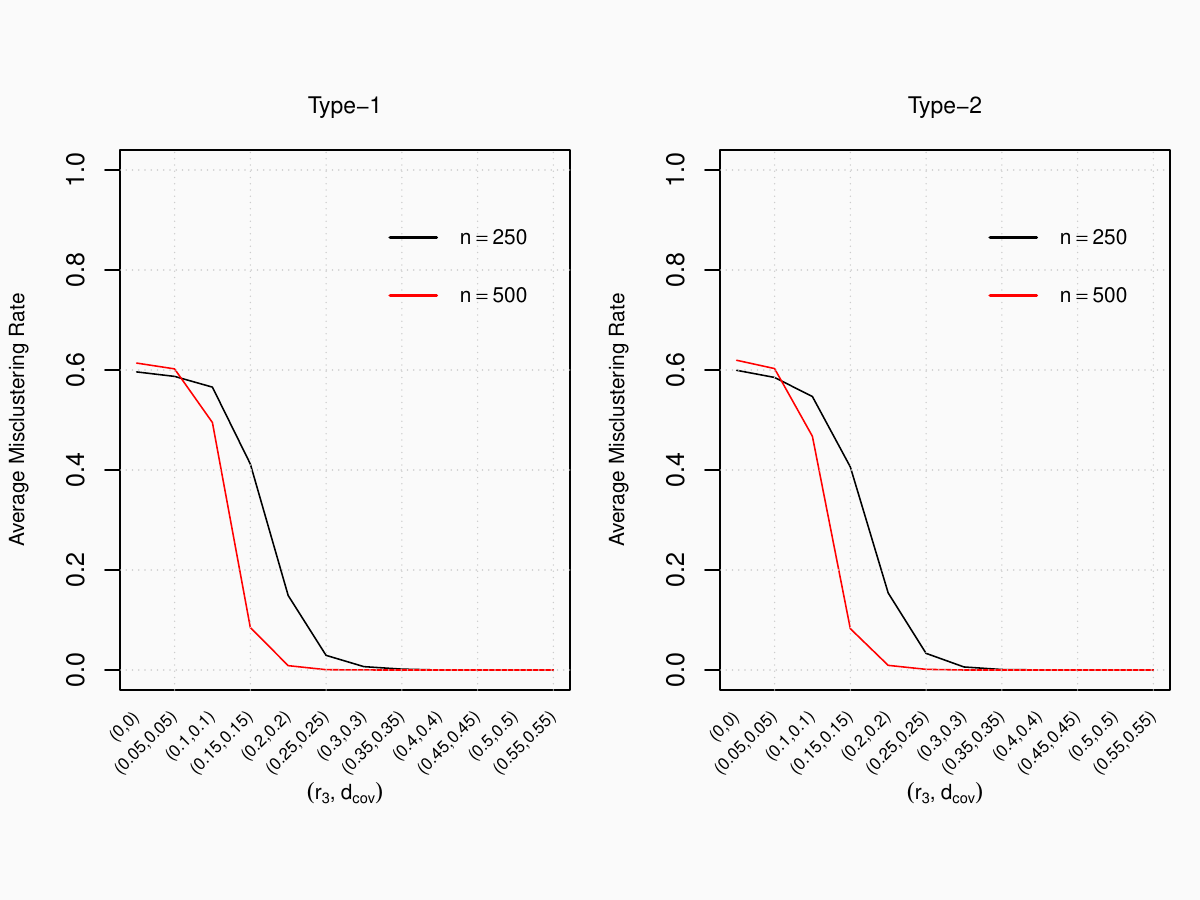}
    \caption{Average misclustering error rate decreases as the pair $(r_3,d_{cov})$ increases jointly from $(0,0)$ to $(0.55,0.55)$.}
    \label{varyboth}
\end{figure}

Figure~\ref{varyboth} shows that the average misclustering error rate decreases steadily as both $r_3$ and $d_{cov}$ increase. In particular, the error rate approaches zero for sufficiently large values of the two parameters. This demonstrates that heterogeneous connectivity information and nodal covariates complement each other in recovering the latent block structure.

\subsection{Simulation 4: Comparison with Benchmark Methods}

We now compare the proposed methodology with several benchmark clustering procedures commonly used in the literature.

The competing methods considered in this experiment are:
\begin{enumerate}
    \item Heterogeneous spectral clustering proposed in \cite{het_sengupta}, denoted by ``Het'',
    \item Covariate assisted spectral clustering for homogeneous networks proposed in \cite{cov_sc_bink}, denoted by ``Hom cov'',
    \item Standard spectral clustering for homogeneous networks, denoted by ``Hom''.
\end{enumerate}

These methods are natural benchmark procedures since each captures only a part of the information available in the present setting. The ``Het'' method utilizes heterogeneous network structure but ignores covariates, while ``Hom cov'' incorporates covariates but does not explicitly model heterogeneity. The standard ``Hom'' procedure ignores both heterogeneity and auxiliary covariate information.
The objective of this experiment is to examine whether simultaneously incorporating heterogeneous connectivity patterns and nodal covariates leads to improved clustering performance.

\paragraph{Varying $r_3$}

We first vary the heterogeneity parameter $r_3$ while keeping all remaining parameters fixed. As before, the reported values correspond to average misclustering error rates computed over 50 independently generated networks.

The network size is fixed at $n=250$, with equal representation of type 1 and type 2 nodes. The model parameters are chosen as follows:
\[
(p_1,p_2,p_3)=(0.2,0.2,0.15), \qquad
(r_1,r_2)=(0.01,0.01),
\]
\[
p^{type1}_{out}=0.08, \qquad
p^{type2}_{out}=0.05,
\]
and
\[
d_{cov}=0.5.
\]

\begin{figure}[H]
\centering
\begin{subfigure}{0.48\textwidth}
\centering
\includegraphics[width=\linewidth]{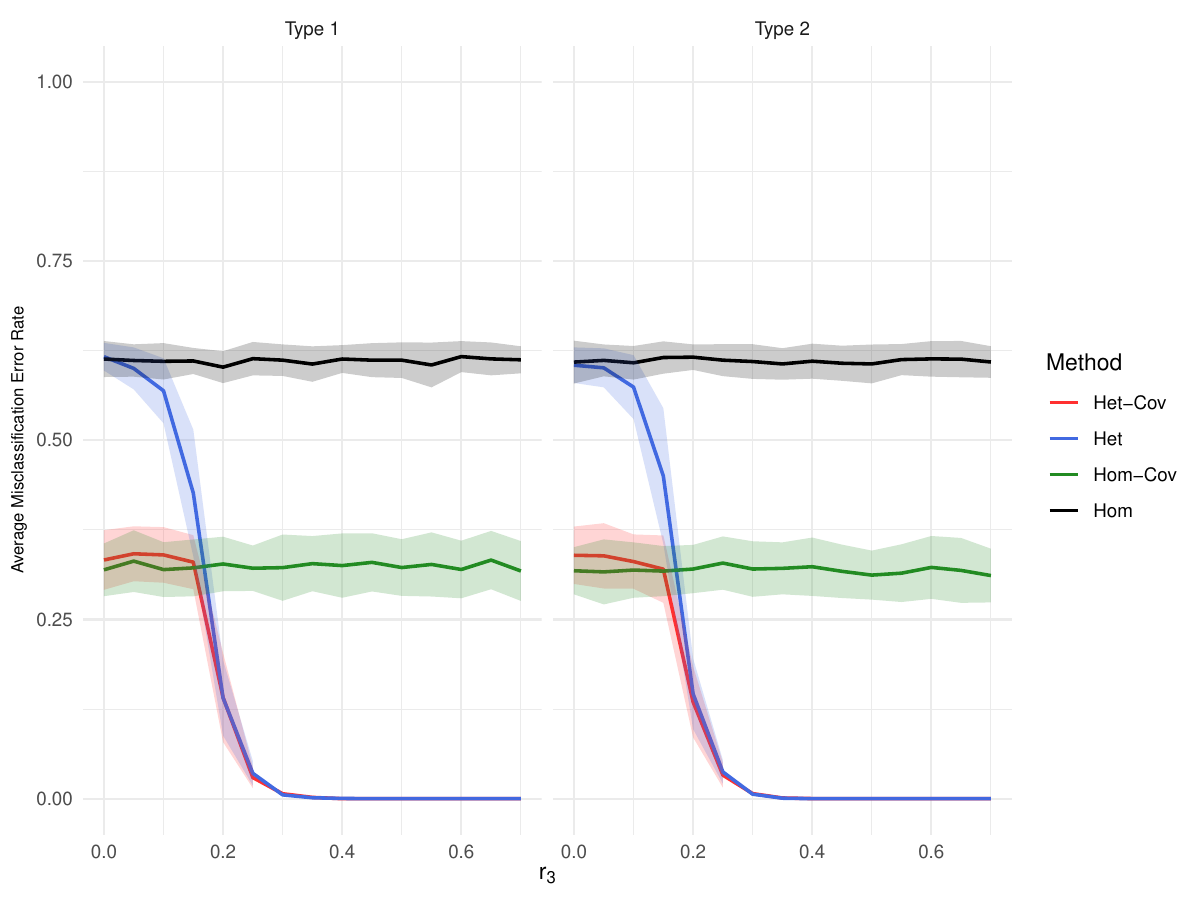}
\caption{}
\label{vary_r3_with_ci}
\end{subfigure}
\hfill
\begin{subfigure}{0.48\textwidth}
\centering
\includegraphics[width=\linewidth]{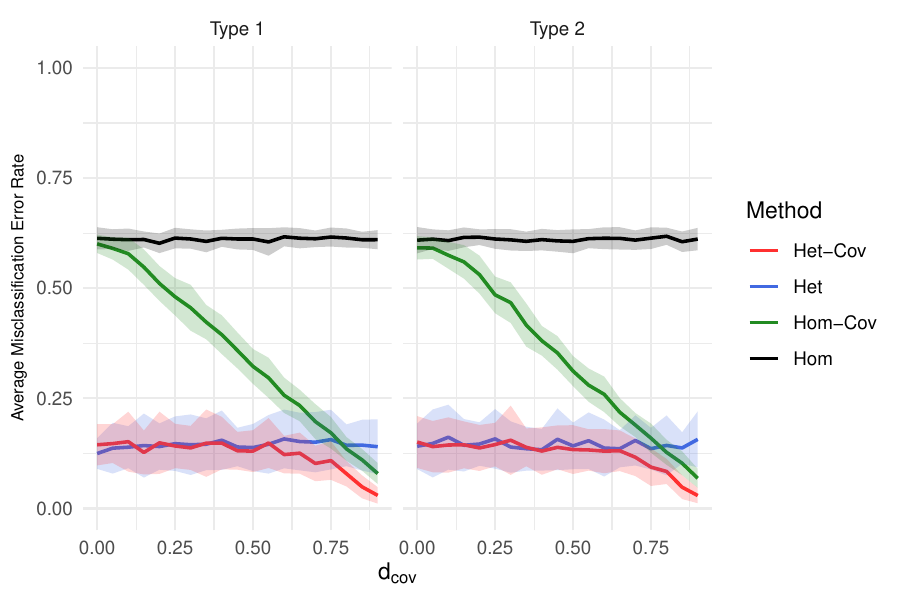}
\caption{}
\label{vary_d_with_ci}
\end{subfigure}

\caption{Comparison of average misclustering error rates for different clustering methods. Panel (a) corresponds to varying $r_3$, while panel (b) corresponds to varying $d_{\mathrm{cov}}$. }

\end{figure}

\paragraph{Varying $d_{cov}$}

Next, we vary the covariate separation parameter $d_{cov}$ while keeping the remaining parameters fixed. The average misclustering error rate is again computed using 50 independently generated networks of size $n=250$.
The parameters are fixed at the following values:
\[
p_1=p_2=0.2, \qquad
p_3=0.15,
\]
\[
r_1=r_2=0.01, \qquad
r_3=0.2,
\]
and
\[
p^{type1}_{out}=0.08, \qquad
p^{type2}_{out}=0.05.
\]

Figures~\ref{vary_r3_with_ci} and~\ref{vary_d_with_ci} present a comparison between the proposed method and the benchmark procedures for different values of $r_3$ and $d_{\mathrm{cov}}$, respectively.
To assess the variability of the estimated misclustering rates, confidence bands are superimposed on the error curves in both figures. As $r_3$ and $d_{\mathrm{cov}}$ increase, the confidence bands become progressively narrower, reflecting reduced variability across simulation replicates. This suggests that stronger community separation and covariate signal strength lead to more stable clustering performance and greater robustness to stochastic variations in the observed network.

These results show that the proposed covariate assisted heterogeneous spectral clustering framework consistently achieves lower misclustering error rates compared to the competing methods. In particular, the advantage becomes more substantial when both heterogeneity and covariate signals are strong. This indicates that jointly utilizing heterogeneous network structure and nodal covariates provides significant improvements in recovering latent communities.

Motivated by these encouraging simulation results, we next apply the proposed methodology to the United Nations General Assembly voting dataset. The empirical analysis demonstrates that incorporating covariate information within a heterogeneous network framework can reveal important geopolitical structures and voting alignments that are not captured effectively by existing benchmark clustering procedures.

\section{Real Data Application: United Nations General Assembly Voting Data}

In this section, we illustrate the effectiveness of the proposed heterogeneous covariate-assisted spectral clustering method using voting records from the United Nations General Assembly (UNGA). The UNGA provides a rich source of relational data in which countries interact through their voting behavior on resolutions spanning a wide range of geopolitical and economic issues. This setting is particularly suitable for our framework, as it naturally gives rise to a heterogeneous network consisting of countries and resolutions, while also allowing the incorporation of node-specific covariate information.

We focus on the post-Cold War period from 1990 to 2000 and investigate whether the proposed method can uncover meaningful geopolitical structures and evolving patterns of international alignment. Beyond evaluating clustering stability and comparing against competing approaches, we examine the extent to which the identified clusters reflect the emergence of major international groupings and provide insights beyond those obtainable from voting behavior alone.

\subsection{Description of the United Nations General Assembly (UNGA) Data}

The United Nations General Assembly (UNGA), established in 1945 under the Charter of the United Nations, is the principal deliberative and representative body of the United Nations. It comprises all 193 member states and serves as a forum where countries discuss and vote on a broad range of international issues. During each annual session, member states cast votes on resolutions, with voting outcomes recorded as \textit{Yes}, \textit{No}, \textit{Abstain}, or \textit{Absent}.

For this study, we use the \texttt{unvotes} R package \cite{unvotes_rpkg}, which provides historical roll call voting records from UNGA sessions across multiple years. The dataset contains the voting decision of each member state on every resolution considered during a given session, making it a widely used source for studying patterns of international alignment and diplomatic behaviour.

\subsection{Data Preparation}
\label{data_prep}

For each year, the voting records are transformed into a network representation and augmented with node level covariate information. Since the proposed method operates on heterogeneous networks, we construct a bipartite network consisting of countries and resolutions. For comparison with baseline methods, a homogeneous country network is also constructed. The data preparation steps are described below.

\paragraph{Heterogeneous Network Construction.}
For a given year, a bipartite network is formed with countries and resolutions as two distinct node types. An edge is placed between a country and a resolution if the country casts a \textit{Yes} vote on that resolution. Votes recorded as \textit{No}, \textit{Abstain}, or \textit{Absent} are treated as the absence of an edge. 

\paragraph{Homogeneous Network Construction.}
To construct a country only network, each country is represented by a voting vector with entries coded as $1$, $-1$, and $0$ for \textit{Yes}, \textit{No}, and \textit{Abstain/Absent}, respectively. Pairwise cosine similarities between voting vectors are then computed to obtain a weighted similarity network. Following thresholding, a binary adjacency matrix is obtained, where an edge is present if the similarity exceeds a chosen cutoff. Throughout the analysis, the median similarity is used as the threshold. Negative similarities are set to zero.

\paragraph{Node Level Covariates.}
The proposed method incorporates covariate information for both countries and resolutions.
For country nodes, GDP per capita is used as the covariate. The data are obtained from the \texttt{WDI} package \cite{wdi_rpkg}. Missing values are imputed using regional mean values to preserve the underlying economic context.
For resolution nodes, the \texttt{unvotes} package classifies resolutions into eight issue categories: Arms Control and Disarmament, Colonialism, Economic Development, Environment, Human Rights, Nuclear Weapons and Nuclear Material, Palestinian Conflict, and Other. Missing categories are assigned to the final group.

Each node is represented by a common feature vector. The first component contains the country level covariate (GDP per capita), while the remaining eight components correspond to the resolution categories. For country nodes, the category components are set to zero. For resolution nodes, the GDP component is zero and the relevant category indicators take value one. Since a resolution may belong to multiple categories, multiple category indicators can be active simultaneously.

This construction allows country and resolution specific information to be embedded within a unified covariate space while preserving the heterogeneous structure of the network. In contrast, the Hom Cov spectral clustering baseline operates only on the homogeneous country network and therefore uses GDP per capita as the sole node level covariate.

\subsection{Clustering Results and Geopolitical Insights}

We now apply the proposed heterogeneous covariate assisted spectral clustering method to the UNGA voting data. The analysis proceeds in several stages. We first examine the choice of the number of clusters and assess the temporal stability of the resulting partitions. We then investigate the geopolitical structures revealed by the clustering and study their relationship with the formation of major international organizations. Finally, we compare the insights obtained from our clustering framework with those arising from the ideal point approach commonly used in the political science literature.

\subsubsection{Selecting the Number of Clusters $K$}

\begin{figure}
    \centering
    
\begin{subfigure}{0.3\textwidth}
        \includegraphics[width=0.6\linewidth]{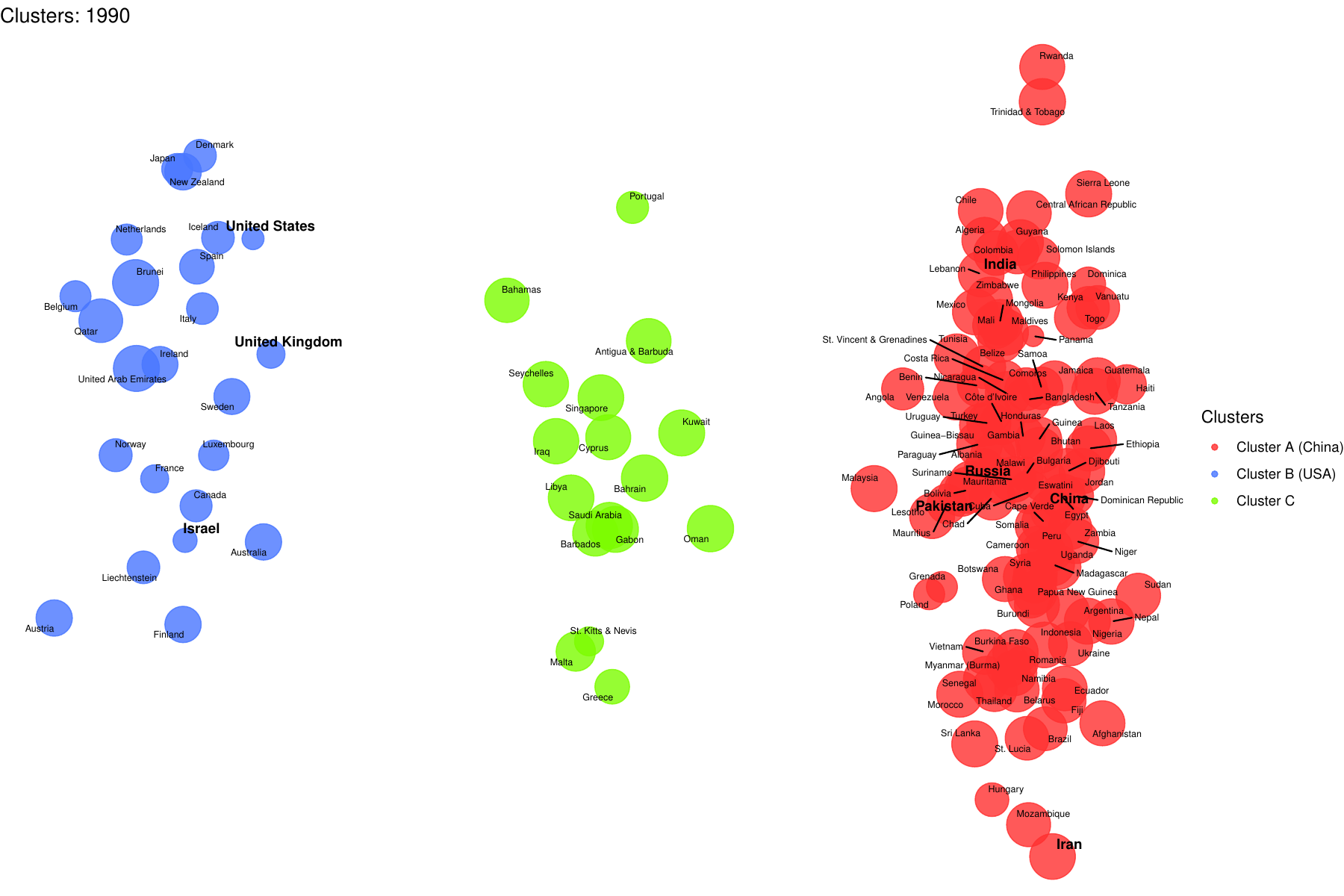}
        \caption{1990, $K=3$}
    \end{subfigure}
    \hfill
   \begin{subfigure}{0.3\textwidth}
        \includegraphics[width=0.6\linewidth]{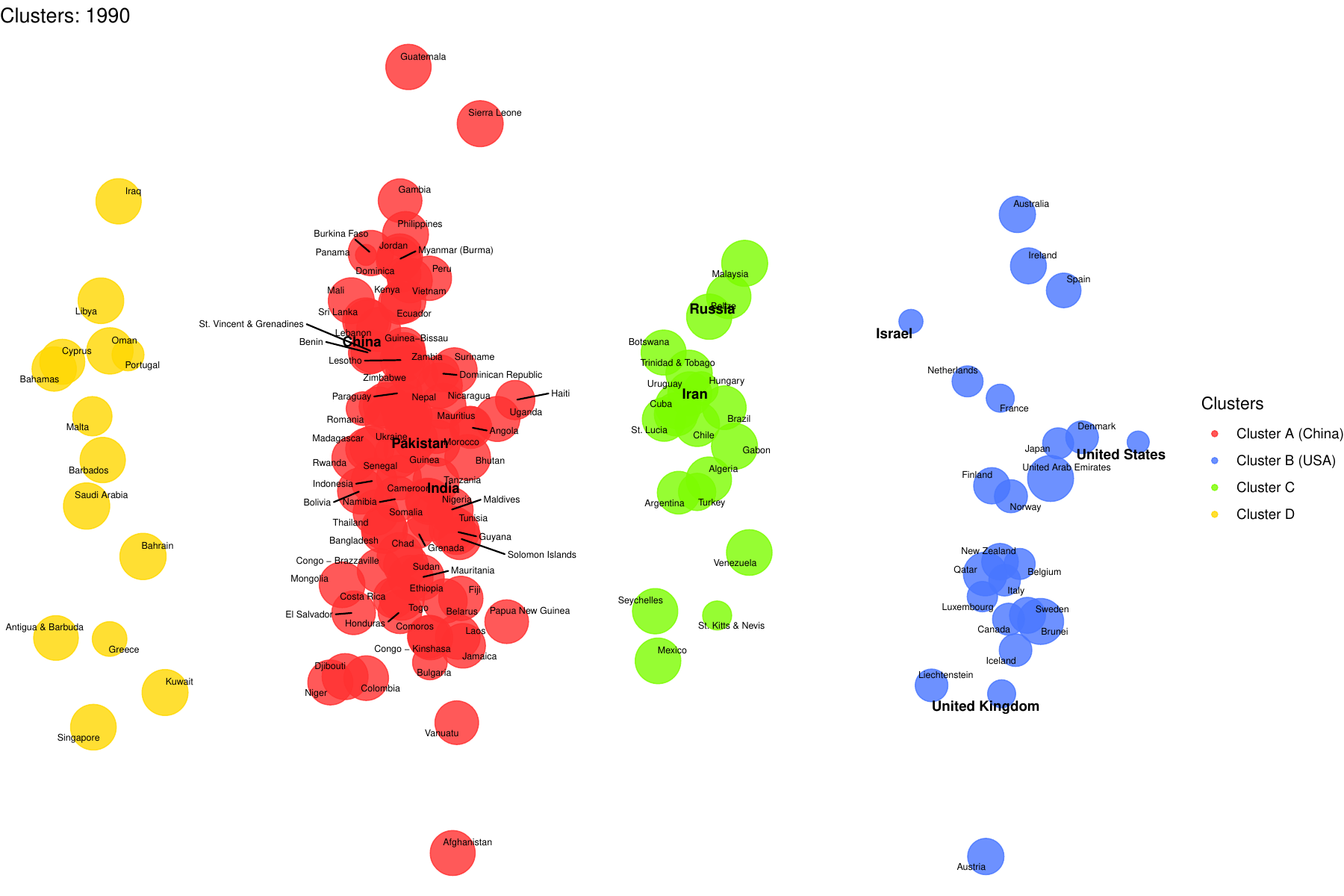}
        \caption{1990, $K=4$}
    \end{subfigure}
    \hfill
\begin{subfigure}{0.3\textwidth}
        \includegraphics[width=0.6\linewidth]{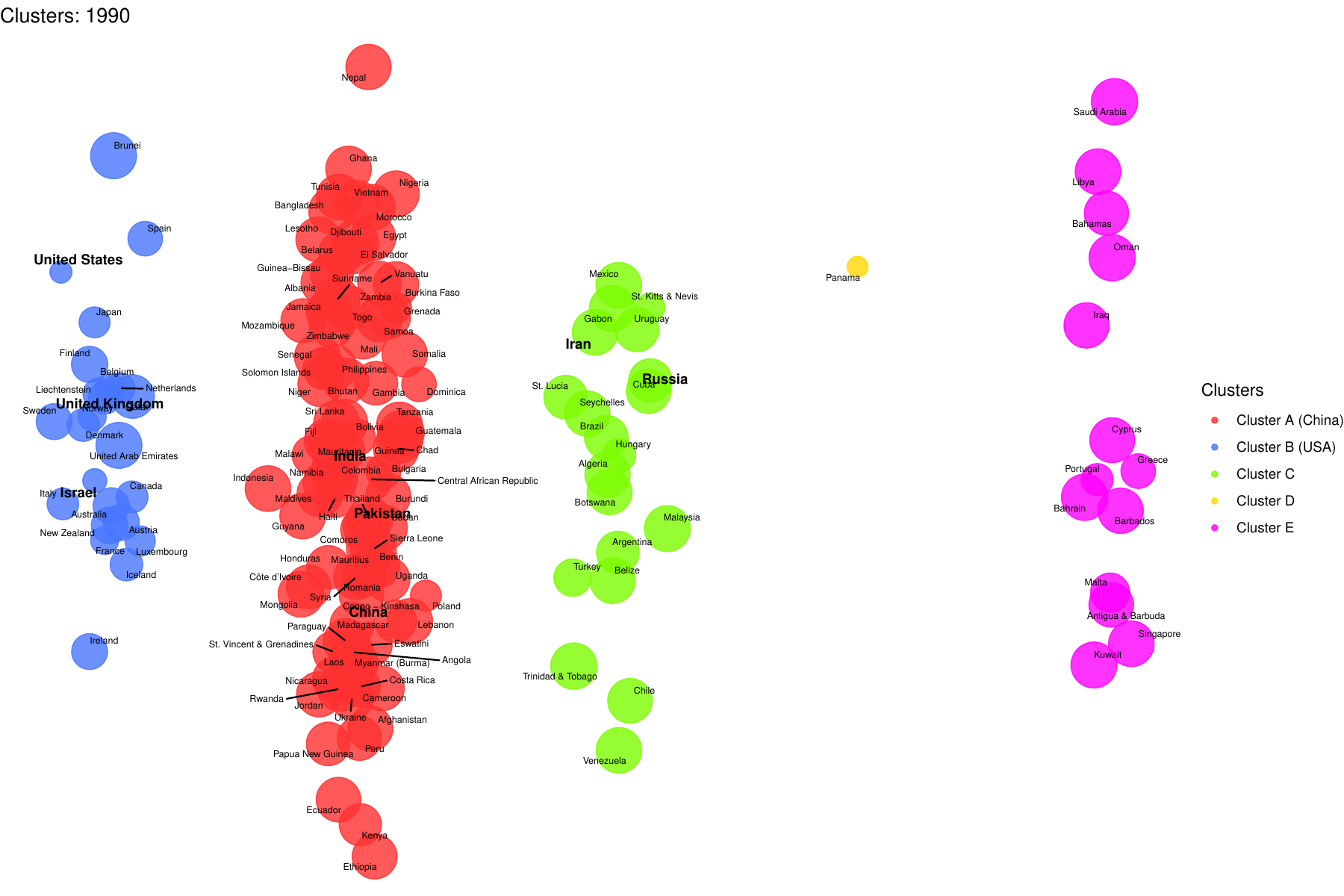}
        \caption{1990, $K=5$}
    \end{subfigure}
    \hfill
\begin{subfigure}{0.3\textwidth}
        \includegraphics[width=0.6\linewidth]{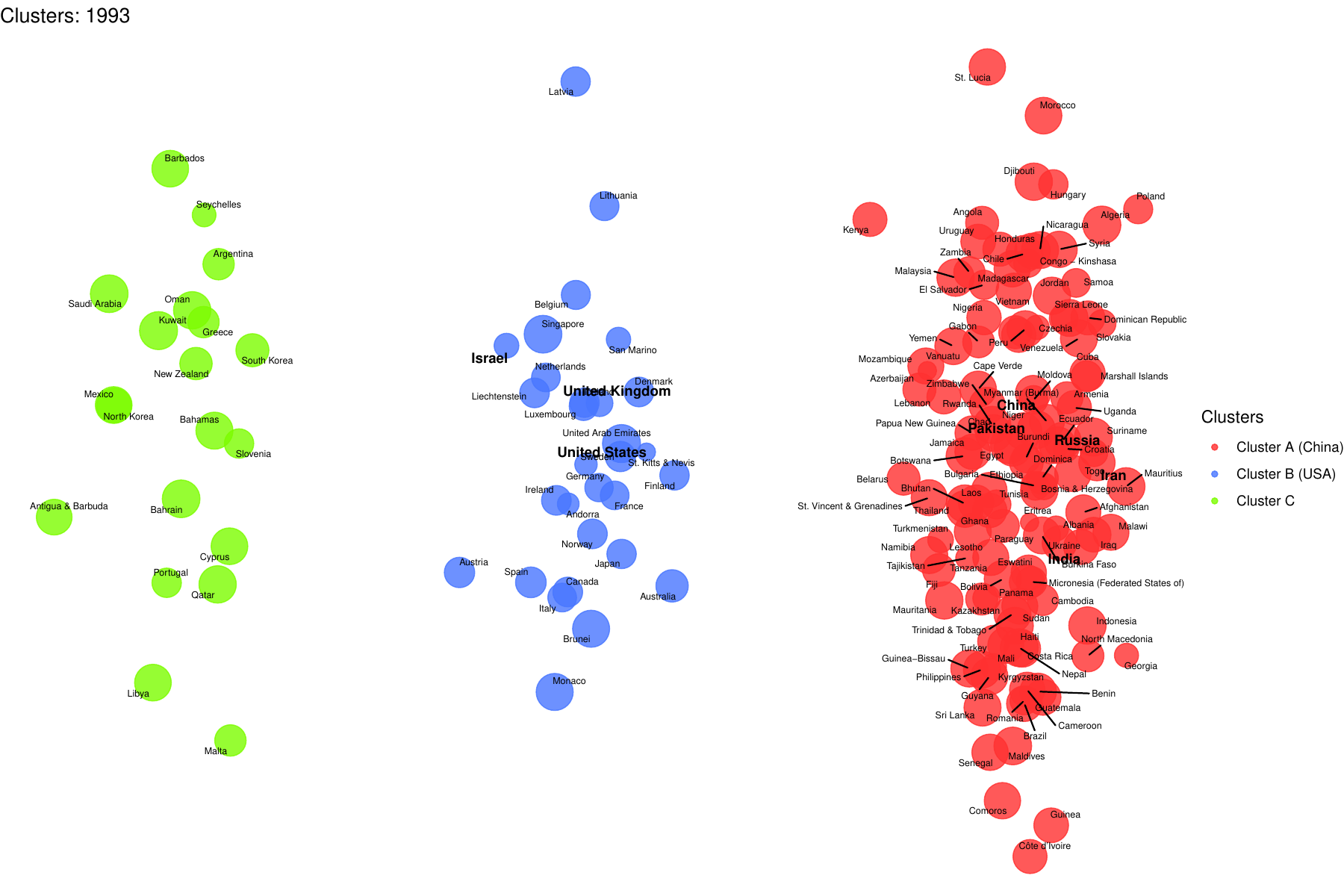}
        \caption{1993, $K=3$}
    \end{subfigure}
    \hfill
   \begin{subfigure}{0.3\textwidth}
        \includegraphics[width=0.6\linewidth]{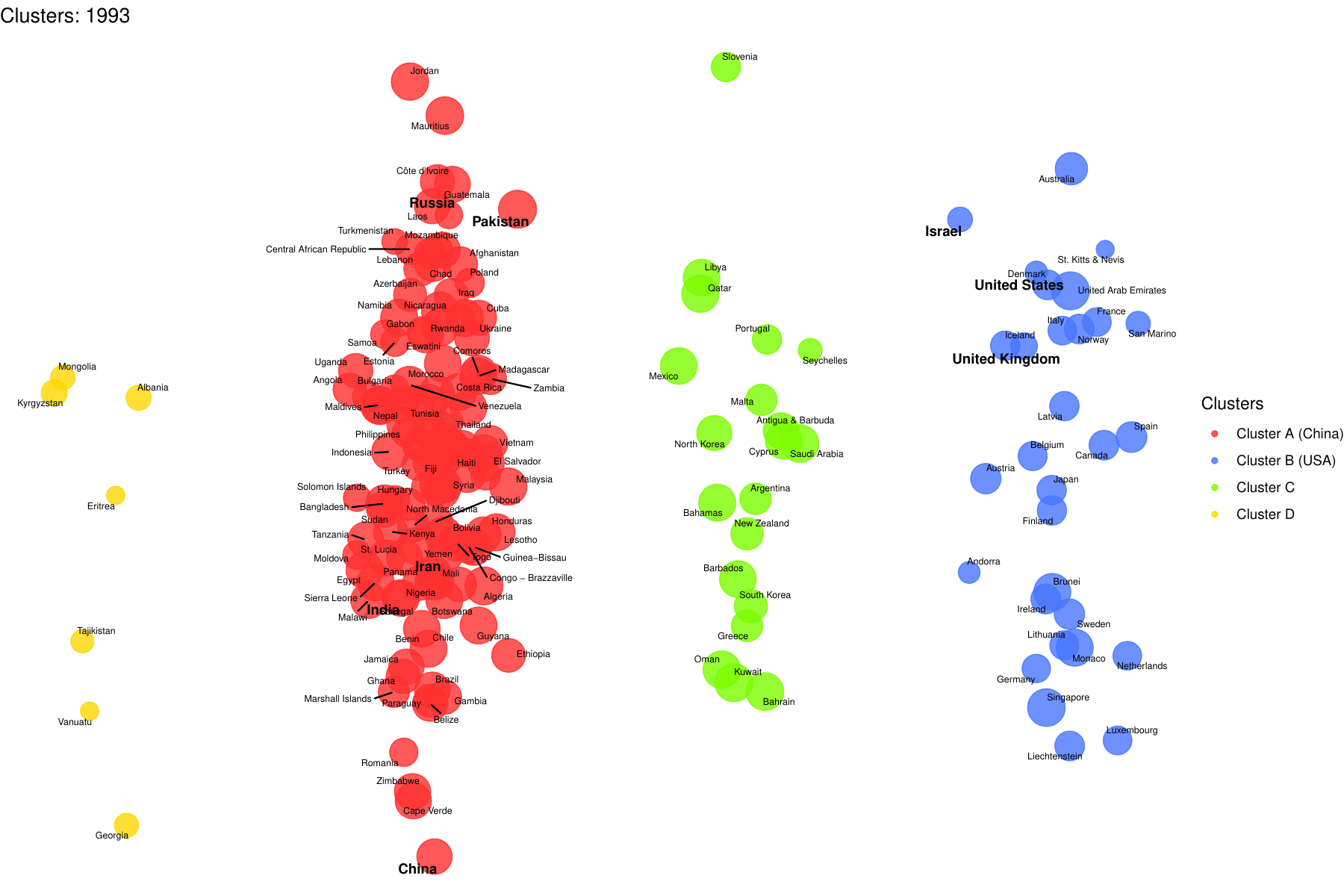}
        \caption{1993, $K=4$}
    \end{subfigure}
    \hfill
\begin{subfigure}{0.3\textwidth}
        \includegraphics[width=0.6\linewidth]{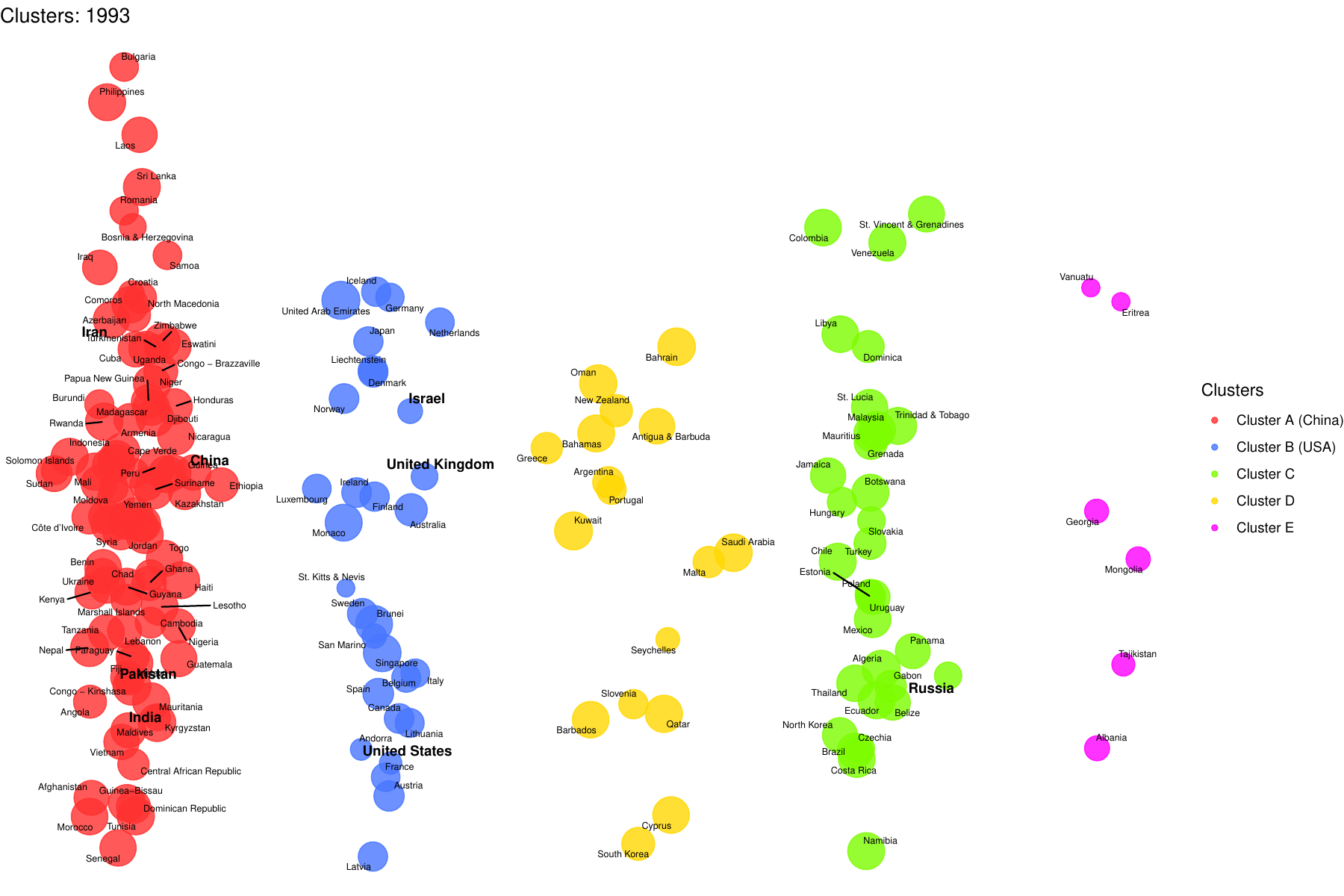}
        \caption{1993, $K=5$}
    \end{subfigure}
    \hfill
\begin{subfigure}{0.3\textwidth}
        \includegraphics[width=0.6\linewidth]{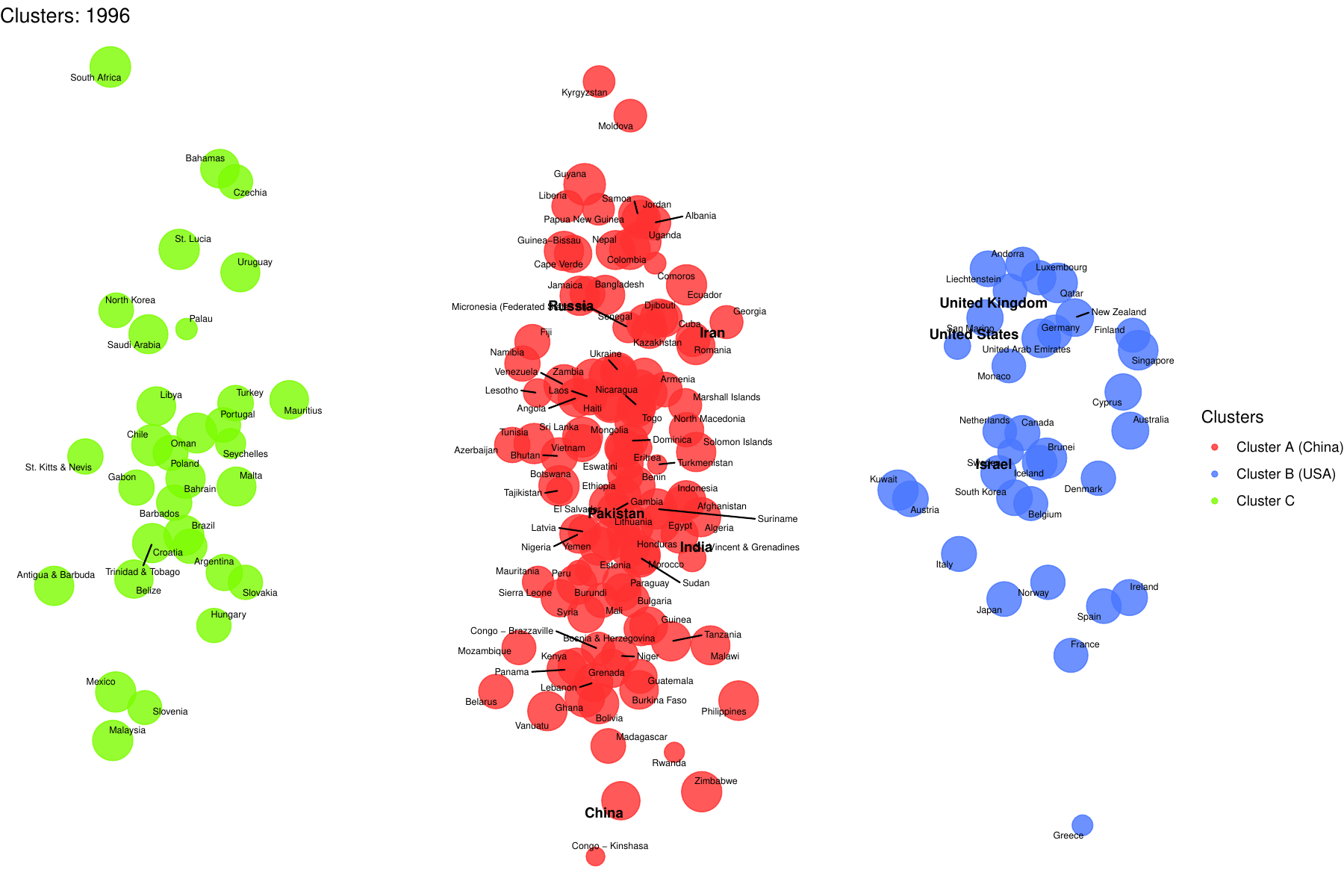}
        \caption{1996, $K=3$}
    \end{subfigure}
    \hfill
   \begin{subfigure}{0.3\textwidth}
        \includegraphics[width=0.6\linewidth]{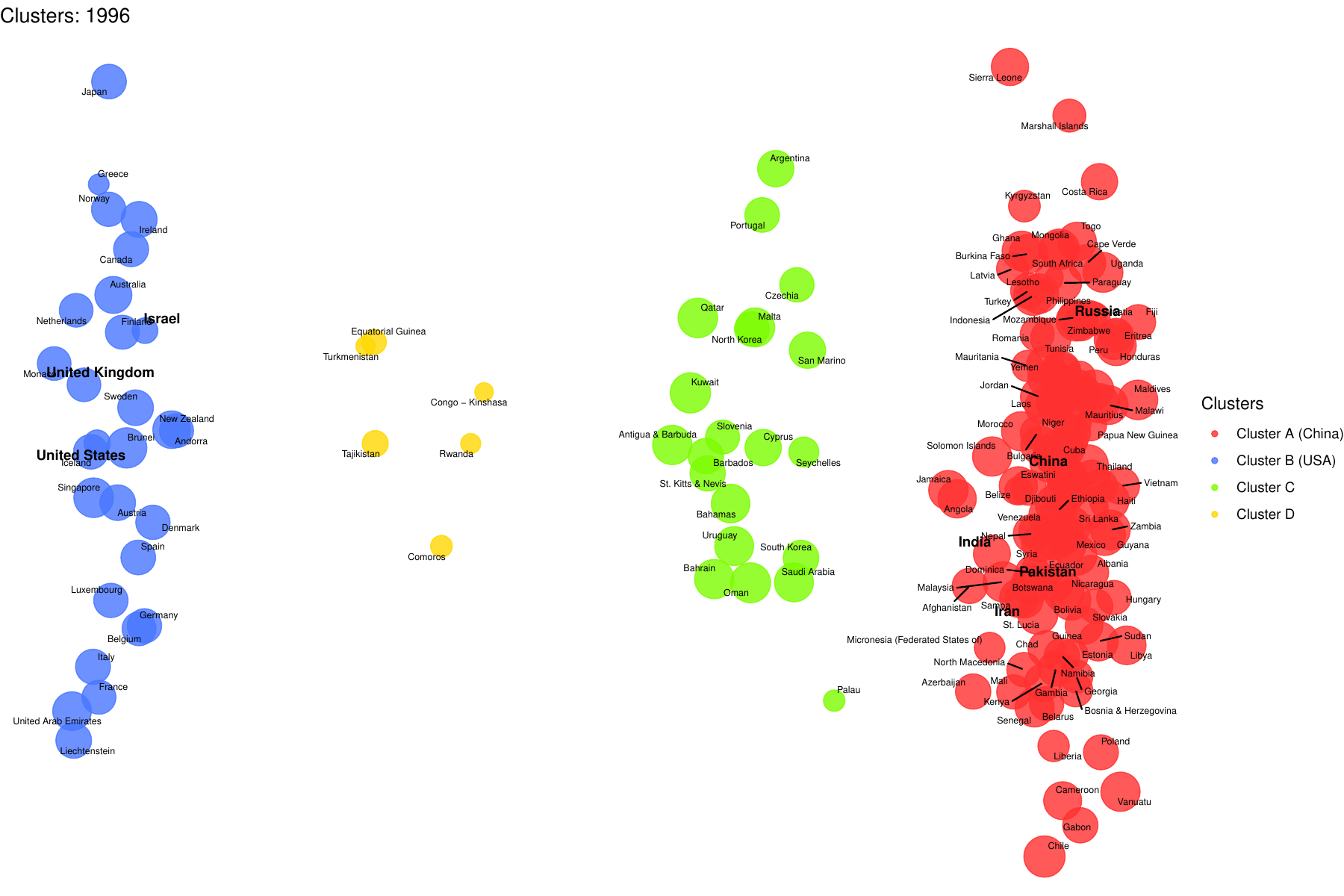}
        \caption{1996, $K=4$}
    \end{subfigure}
    \hfill
\begin{subfigure}{0.3\textwidth}
        \includegraphics[width=0.6\linewidth]{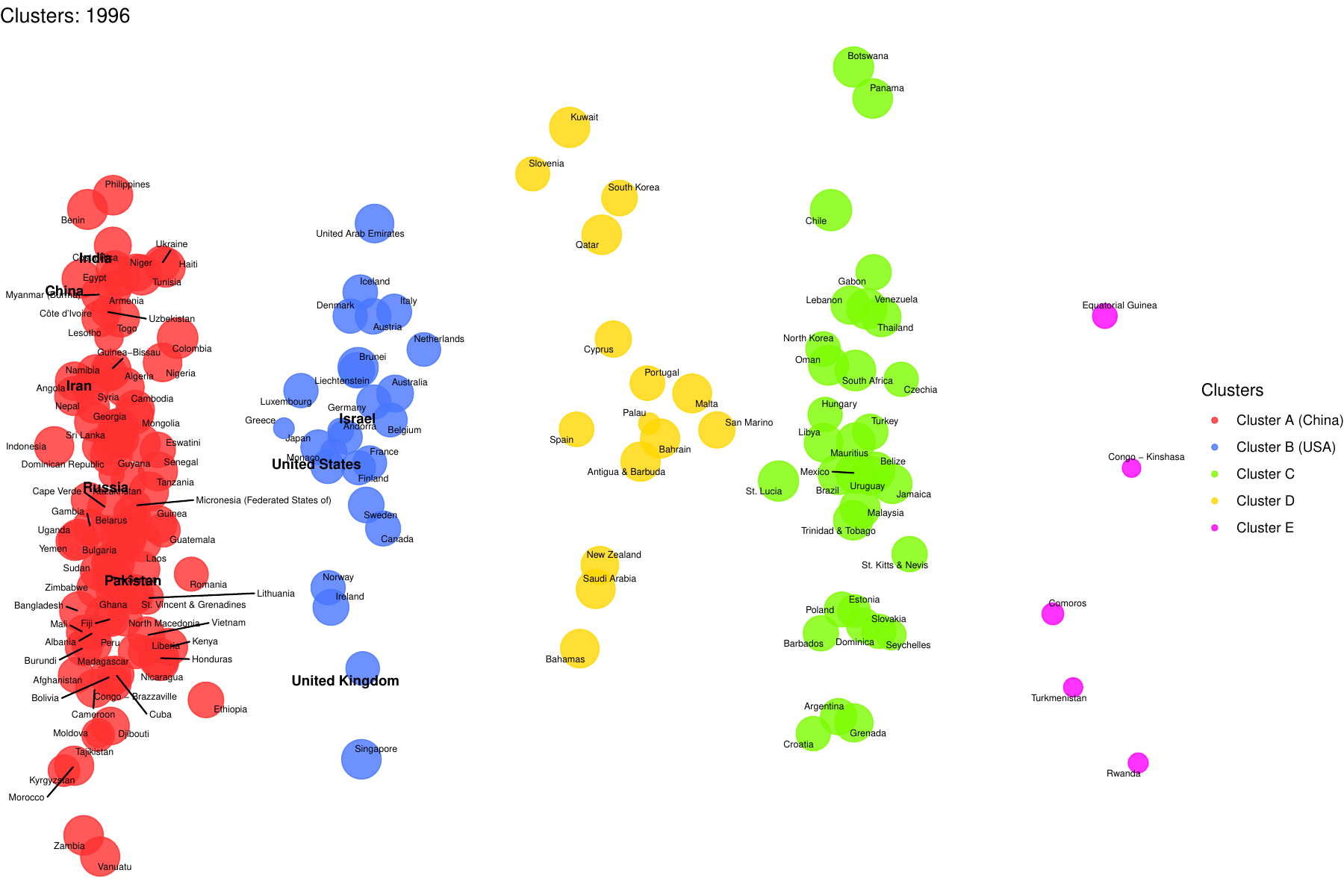}
        \caption{1996, $K=5$}
    \end{subfigure}
    \hfill
\begin{subfigure}{0.3\textwidth}
        \includegraphics[width=0.6\linewidth]{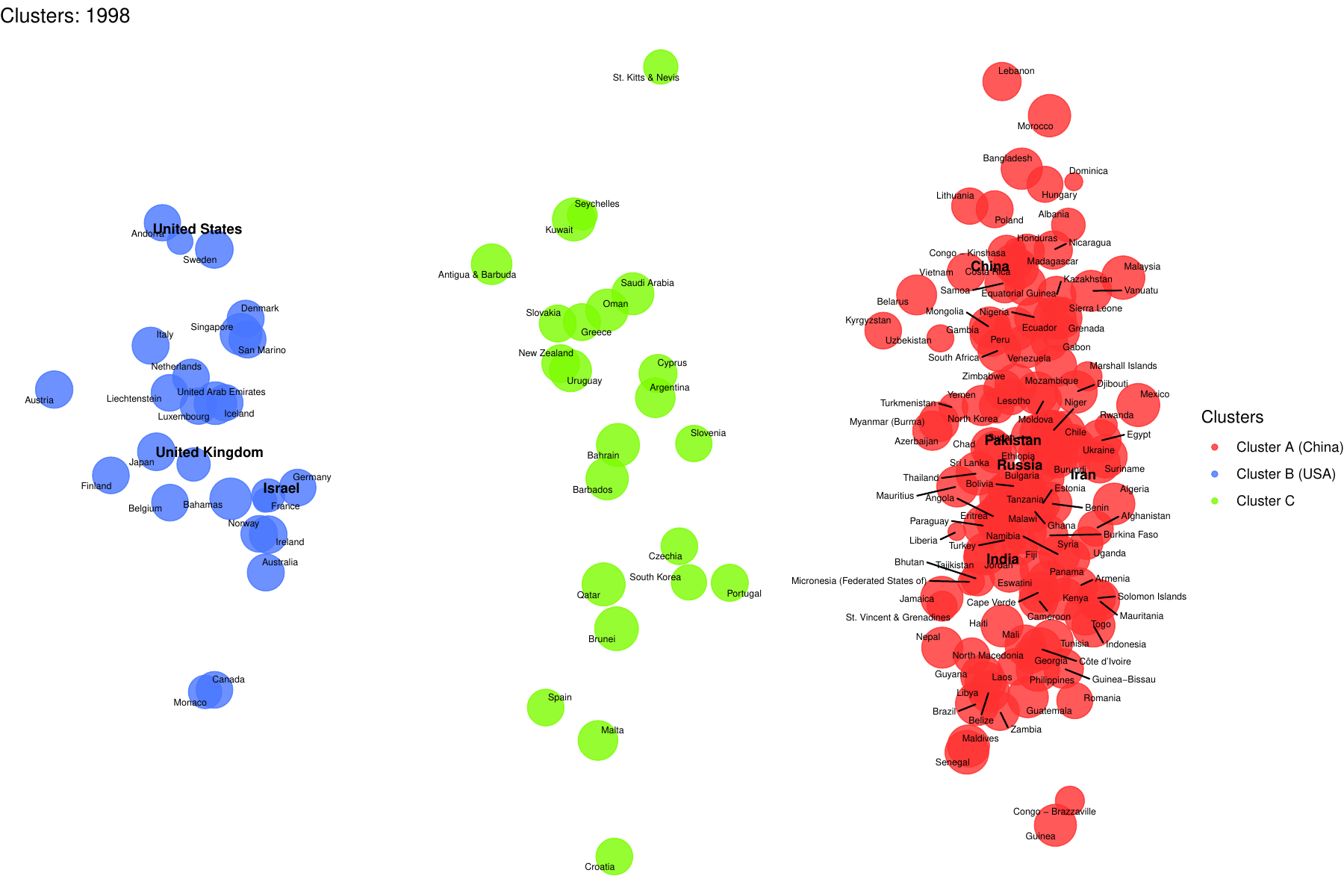}
        \caption{1998, $K=3$}
    \end{subfigure}
    \hfill
   \begin{subfigure}{0.3\textwidth}
        \includegraphics[width=0.6\linewidth]{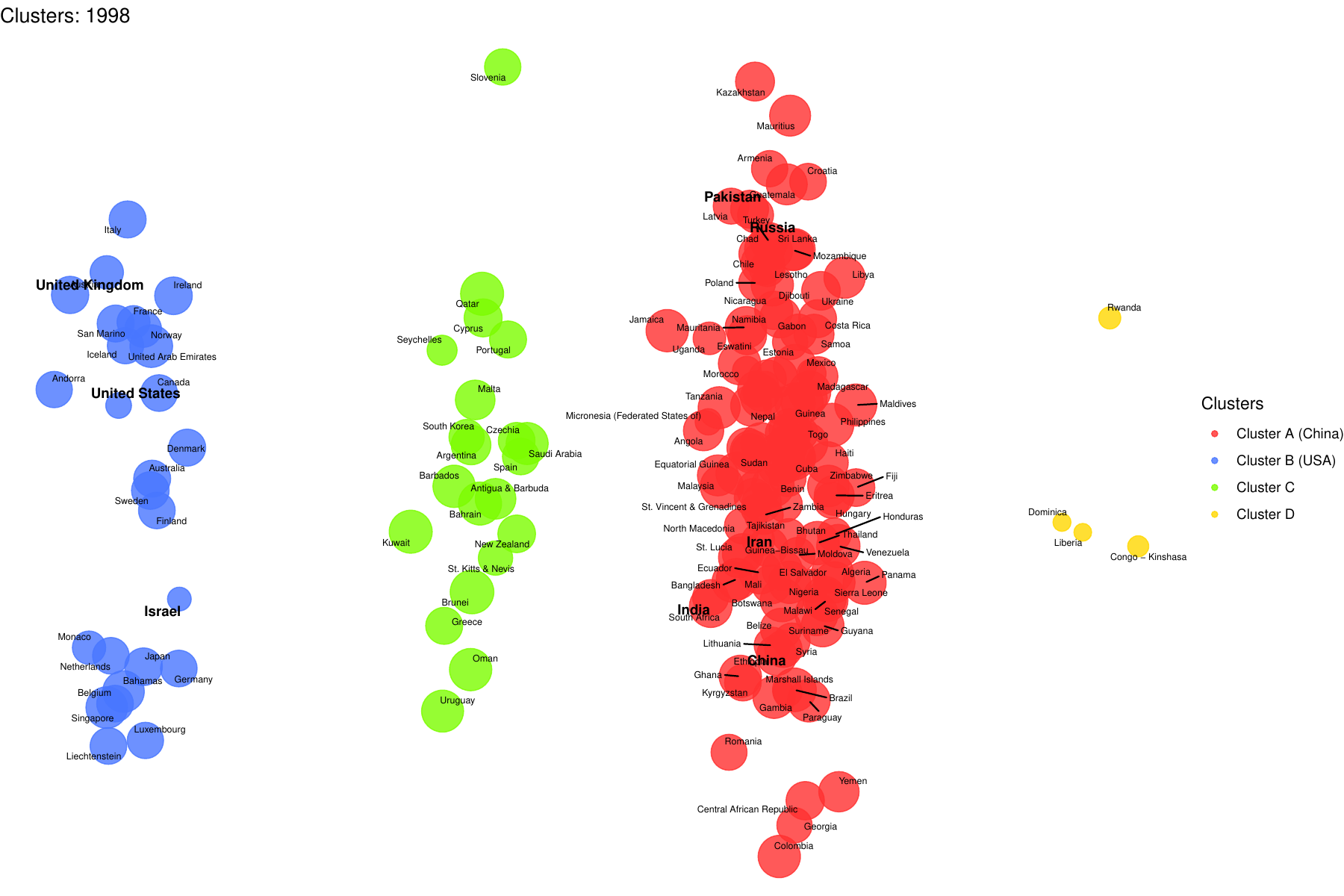}
        \caption{1998, $K=4$}
    \end{subfigure}
    \hfill
\begin{subfigure}{0.3\textwidth}
        \includegraphics[width=0.6\linewidth]{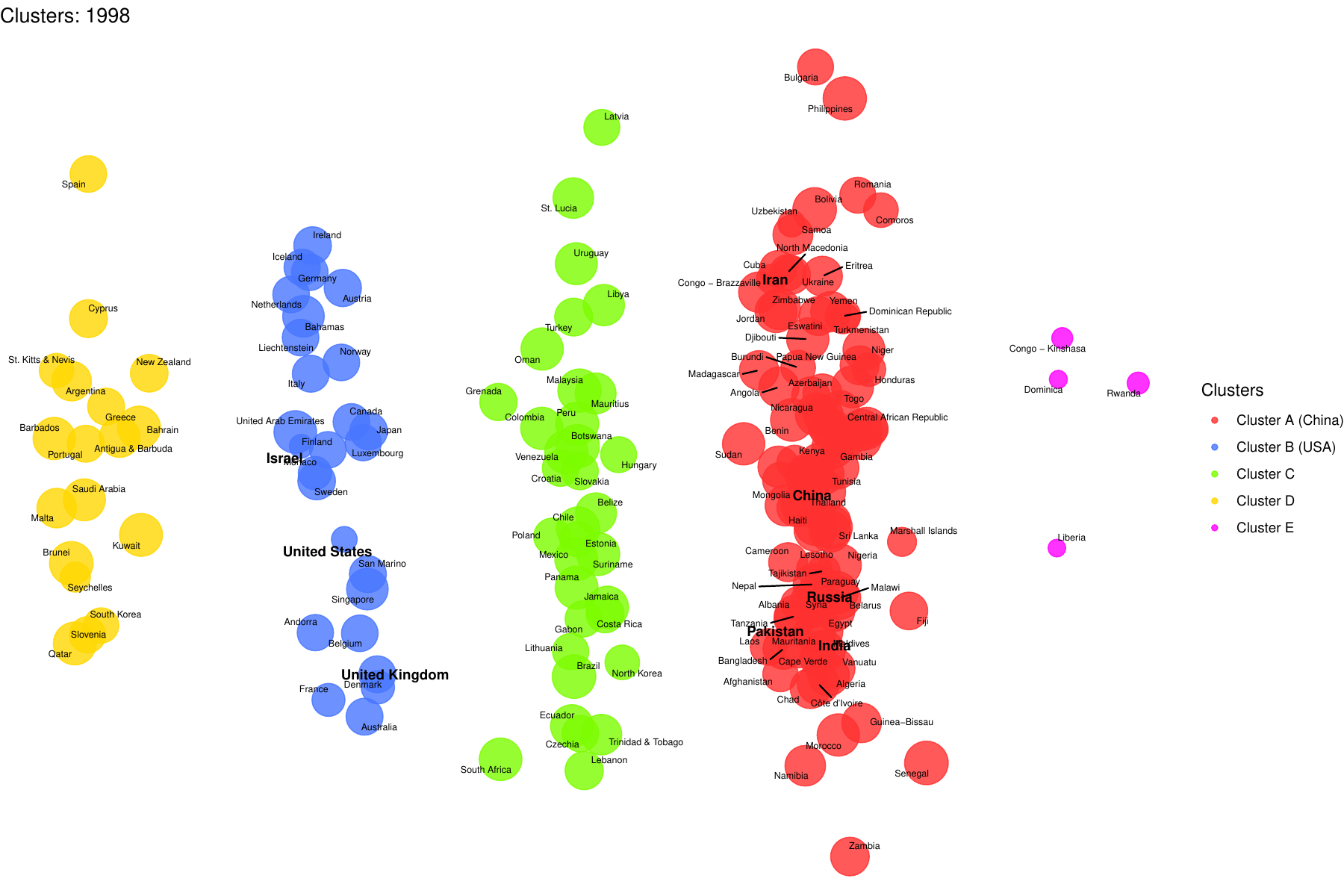}
        \caption{1998, $K=5$}
    \end{subfigure}
    \hfill
    \begin{subfigure}{0.3\textwidth}
        \includegraphics[width=0.6\linewidth]{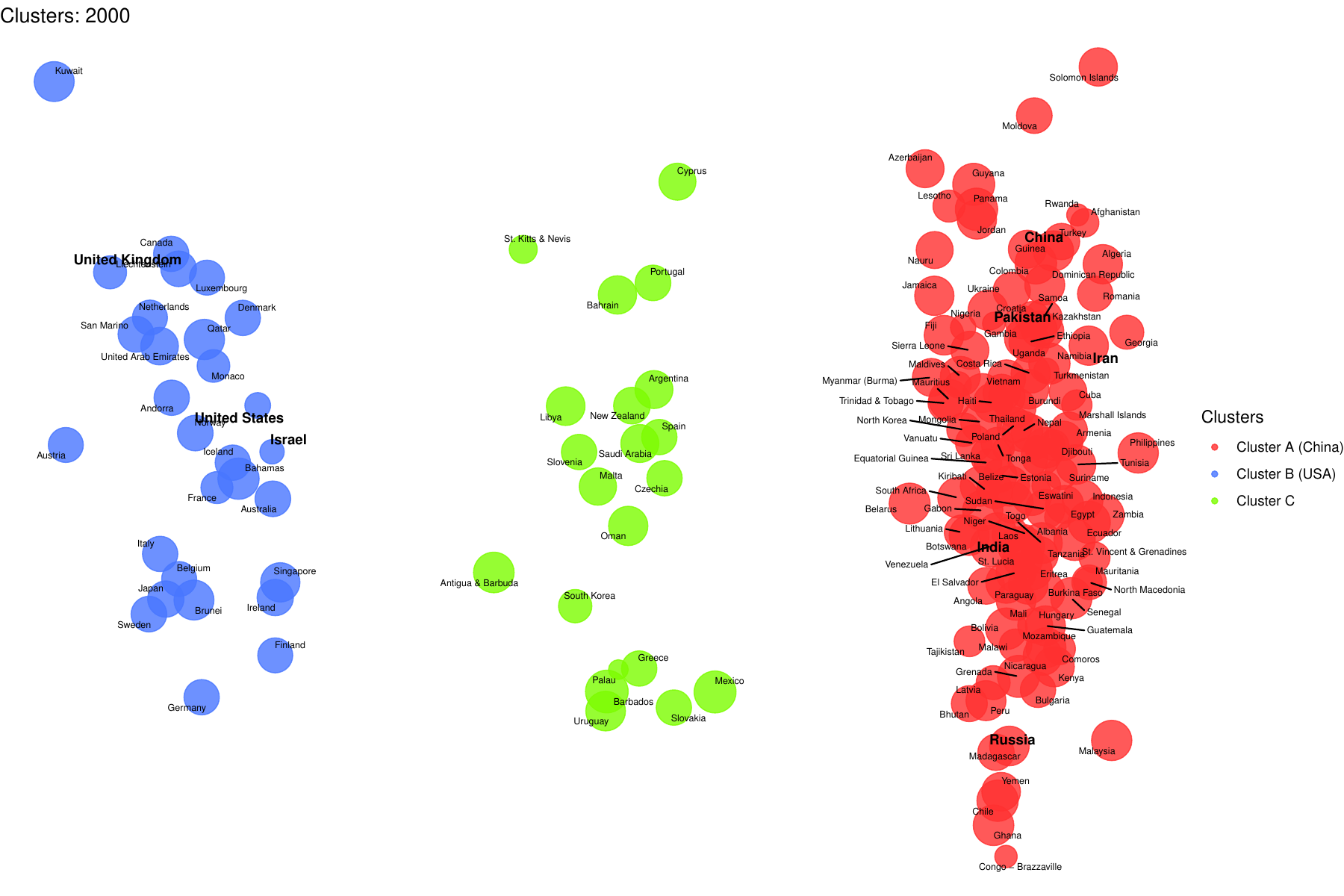}
        \caption{2000, $K=3$}
    \end{subfigure}
    \hfill
   \begin{subfigure}{0.3\textwidth}
        \includegraphics[width=0.6\linewidth]{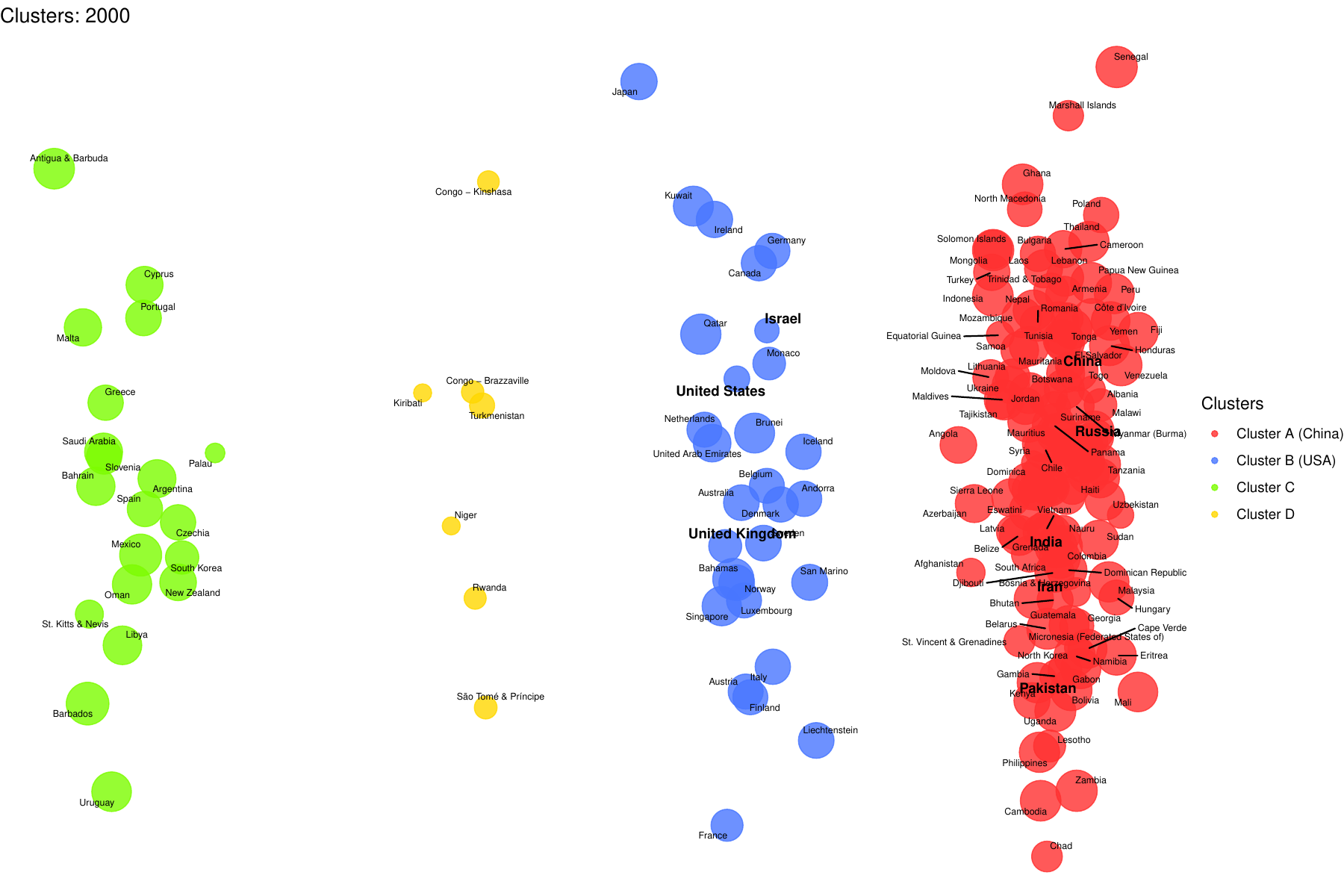}
        \caption{2000, $K=4$}
    \end{subfigure}
    \hfill
\begin{subfigure}{0.3\textwidth}
        \includegraphics[width=0.6\linewidth]{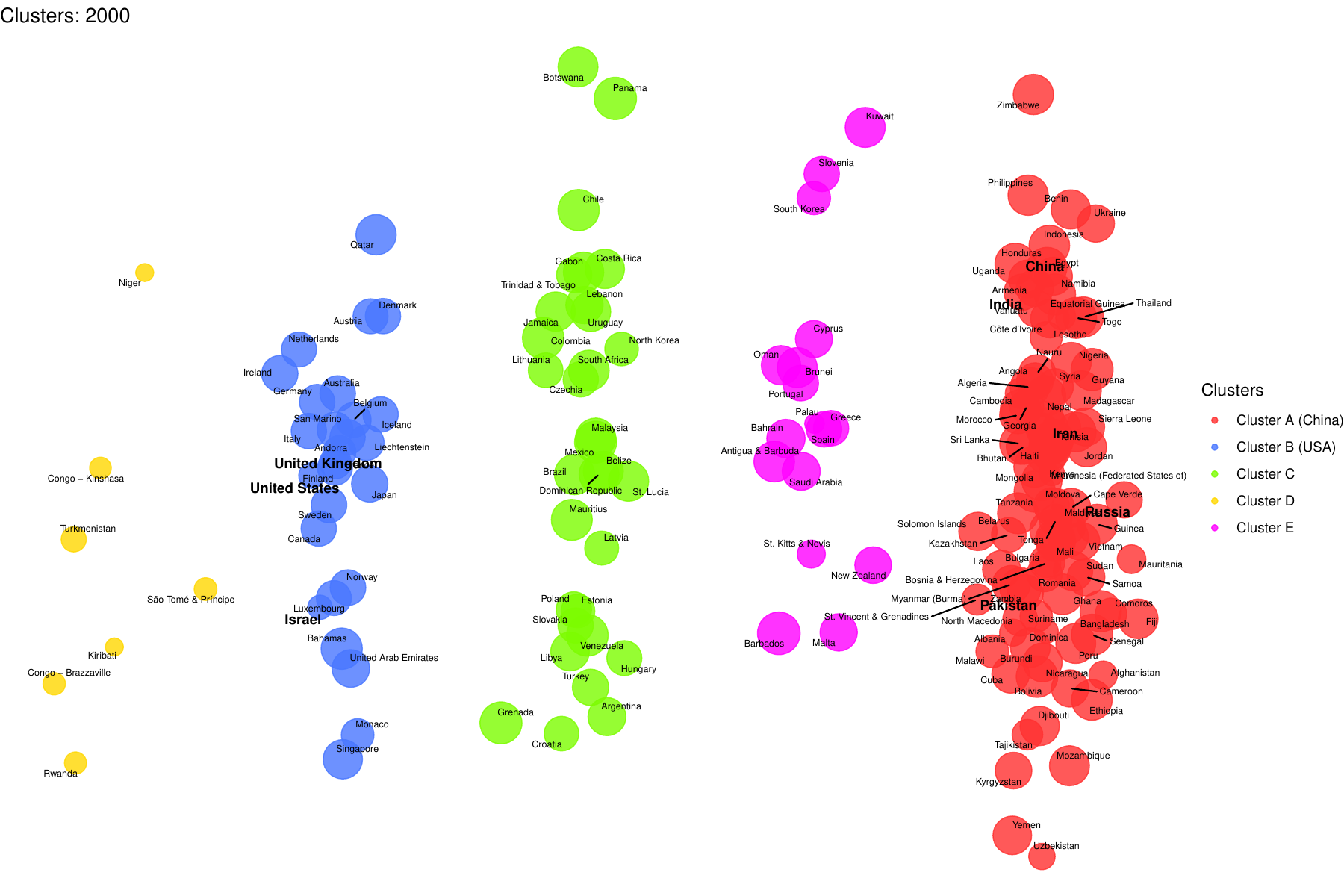}
        \caption{2000, $K=5$}
    \end{subfigure}
    \hfill
 \vspace{0.2cm}

    \caption{Cluster Compositions (Het-Cov) for selected years 1990, 1993, 1996, 1998 and 2000.}
    \label{cluster_plots}
\end{figure}

Spectral clustering requires the number of clusters to be specified in advance. From a geopolitical perspective, the post Cold War world is often viewed as consisting of a small number of broad political and economic blocs, although the exact classification remains debated \cite{country_classification}. Motivated by this observation, we investigate the clustering structure of UNGA voting data during the period 1990--2000 using several candidate values of $K$.
More precisely, we perform a grid search over $K \in {3,4,5}$. Figure \ref{cluster_plots} displays the resulting cluster compositions for selected years. The overall clustering structure remains remarkably stable across different choices of $K$. In particular, the major geopolitical groupings, including the United States aligned and China aligned blocs, persist throughout the period.

Increasing the number of clusters primarily results in further subdivision of existing groups rather than the emergence of qualitatively new structures. Since the results for $K=3$ capture the dominant patterns while providing the most parsimonious representation, we adopt $K=3$ for all subsequent analyses.

\subsubsection{Stability of Het Cov Clustering}

To assess the temporal stability of the identified partitions, we compare clustering results across consecutive years using the Adjusted Rand Index (ARI), a widely used measure of similarity between two clusterings that corrects for chance agreement. Higher ARI values indicate greater consistency in cluster membership over time.

The results reported in Table \ref{tab:ari_years} show that the proposed Het Cov spectral clustering method produces highly stable partitions across most consecutive years. In particular, Het Cov attains the highest ARI values in the majority of year pairs, suggesting that the joint use of network heterogeneity and covariate information yields robust and persistent clustering structures. A noticeable decline in ARI is observed during the late 1990s, which may reflect shifts in geopolitical alignments and voting behavior during that period. As discussed later in section \ref{ip_discussion}, such variations could be associated with major political and economic developments that influenced international relations. Overall, the results indicate that the proposed method captures stable country groupings while remaining sensitive to meaningful changes in the global political landscape.

\begin{table}[H]
\centering
\small
\setlength{\tabcolsep}{4pt} 
\renewcommand{\arraystretch}{1.2}
\caption{Pairwise ARI comparison across consecutive years for proposed method and baseline algorithms}
\label{tab:ari_years}

\begin{tabular}{lcccccccccc}
\toprule
\textbf{Method} & \textbf{(90,91)} & \textbf{(91,92)} & \textbf{(92,93)} & \textbf{(93,94)} & \textbf{(94,95)} & \textbf{(95,96)} & \textbf{(96,97)} & \textbf{(97,98)} & \textbf{(98,99)} & \textbf{(99,00)} \\
\midrule

Het-Cov & \textbf{0.9476} & \textbf{0.9794} & \textbf{0.9273} & \textbf{0.9371} & \textbf{0.9077} & 0.6888 & \textbf{0.8959} & \textbf{0.7889} & \textbf{0.8245} & \textbf{0.8470} \\

Het & 0.5001 & 0.4758 & 0.4891 & 0.5437 & 0.6038 & 0.6431 & 0.6414 & 0.7033 & 0.6857 & 0.6282 \\

Hom-Cov &  0.3224 & 0.7437 & 0.7644 & 0.7471 & 0.6818 & \textbf{0.7398} & 0.7960 & 0.7358 & 0.7025 & 0.5637 \\

Hom & 0.5485 & 0.3435 & 0.3299 & 0.3748 & 0.2467 & 0.3749 & 0.3207 & 0.3042 & 0.4206 & 0.4855 \\

\bottomrule
\end{tabular}
\end{table}


\subsubsection{Cluster Structure and International Alignments of Intergovernmental Organizations}
To interpret the geopolitical significance of the identified clusters, we examine several major regional and international organizations whose memberships are shaped by common economic, political, or strategic interests. These include BRICS, the Shanghai Cooperation Organization (SCO), and the Latin American Integration Association (ALADI). Although some of these organizations were formally established after the study period, their origins can often be traced to patterns of cooperation and shared interests that existed earlier. Consequently, if countries that later became members of the same organization are grouped together by the proposed method, it would provide evidence that the clustering captures meaningful underlying international alignments rather than merely similarities in voting behavior. We therefore briefly review the historical background of these organizations and compare their memberships with the clusters obtained from the UNGA voting network.

\paragraph{BRICS:}
BRICS is a grouping of major emerging economies comprising Brazil, Russia, India, China, and South Africa. The term ``BRIC'' was coined by economist Jim O'Neill in 2001, although the idea of cooperation among these countries had been discussed during the late 1990s. Formal collaboration began with meetings of foreign ministers in 2006, followed by the first BRIC summit in 2009. South Africa joined in 2010, leading to the formation of BRICS. The bloc has since expanded economic cooperation through initiatives such as the New Development Bank and has emerged as an important platform for promoting a more multipolar global economic order.

\paragraph{Shanghai Cooperation Organization (SCO):}
The SCO originated from the ``Shanghai Five,'' established in 1996 by China, Kazakhstan, Kyrgyzstan, Russia, and Tajikistan to strengthen regional cooperation and security. Following the inclusion of Uzbekistan in 2000, the organization was formally established as the SCO in 2001. Over time, it expanded to include countries such as India and Pakistan in 2017 and Iran in 2023. The SCO serves as a major Eurasian forum for economic, political, and security cooperation and is often viewed as an important instrument in promoting a multipolar world order.

\paragraph{Latin American Integration Association (ALADI):}
ALADI was established under the Montevideo Treaty of 1980 as the successor to the Latin American Free Trade Association. It is the largest regional integration organization in Latin America, bringing together countries such as Argentina, Bolivia, Brazil, Chile, Paraguay, Peru, Uruguay, and others. The organization aims to promote regional economic integration through preferential trade agreements and gradual market convergence while accommodating differences in the economic development levels of its member states.

\paragraph{Remark.}
The European Union (EU), despite being one of the most prominent examples of regional cooperation, is intentionally excluded from the scope of our analysis. This exclusion is motivated by conceptual considerations rather than empirical findings.

The organizations considered in this study, namely BRICS, SCO, and ALADI, are primarily intergovernmental arrangements that seek to facilitate cooperation among sovereign states while largely preserving the autonomy of their members. Their objectives focus on coordination in areas such as trade, economic development, security, and regional cooperation. In contrast, the EU represents a substantially deeper and more institutionalized form of integration. It possesses supranational legislative, judicial, and monetary institutions, including the European Parliament and the Court of Justice of the European Union (CJEU), and supports policies such as the single market, the free movement of people, goods, services, and capital, and, for many member states, a common currency. In fact, the European Union Customs Union (EUCU) establishes a single customs area for all EU member states, allowing goods to move freely without internal tariffs while applying a common external tariff to imports from non-EU countries. These features place the EU in a category that is qualitatively different from the organizations examined in this paper.

A second distinction concerns historical development. The idea of European political integration predates not only the formation of BRICS, SCO, and ALADI, but also the establishment of the United Nations itself. Intellectual and political movements advocating a united Europe emerged long before the post-war era. For example, the Pan-European Movement may be viewed as an early effort toward a politically integrated Europe. Following World War II, decades of political, economic, and institutional developments transformed these ideas into a concrete political project and laid the foundations of modern European integration. By comparison, the organizations studied here are relatively recent creations whose origins lie primarily in late twentieth-century and early twenty-first-century efforts at intergovernmental cooperation.

Because our objective is to study the formation and evolution of cooperative structures among sovereign states within relatively loose institutional frameworks, including the EU would introduce a fundamentally different mechanism of integration into the analysis. The EU is therefore best viewed not as another observation within the same class of organizations, but rather as a distinct political and economic project whose historical trajectory and institutional architecture demand a separate investigation.

\subsubsection{Metrics for Evaluating and Interpreting Clustering Results}

Assessing the quality of clustering results in geopolitical settings requires measures that are both interpretable and relevant to the underlying political structure. To examine whether the identified clusters reflect meaningful international alignments, we focus on the emergence and cohesion of several intergovernmental organizations, including BRICS, SCO and ALADI. Our analysis indicates that the proposed method captures patterns of alignment among member countries that are less apparent under competing approaches, including early signs of cohesion preceding the formal establishment of some organizations.

To quantify these patterns, we introduce two measures: \textit{Largest Cluster Proportion (LCP)} and \textit{Affiliation}. Together, these metrics assess the extent to which member countries of a given organization are grouped within the same cluster and the role of key countries in shaping these alignments over time. 

\paragraph{Largest Cluster Proportion (LCP):}
Let $\mathcal{N}=\{1,\ldots,N\}$ denote the set of countries and let $\mathcal{C}=\{C_1,\ldots,C_K\}$ be the clustering partition. For a group of countries $G \subseteq \mathcal{N}$, define
\[
C^*=\arg\max_{C_k\in\mathcal{C}} |C_k|.
\]
The largest cluster proportion is given by
\[
\mathrm{LCP}(G)=\frac{|G\cap C^*|}{|G|}.
\]

LCP measures the proportion of group members belonging to the largest cluster. Higher values indicate stronger collective alignment among member countries. Persistent high LCP values over time suggest the gradual formation and consolidation of a common geopolitical or economic bloc.

\paragraph{Affiliation:}
For a reference country $j\in\mathcal{N}$, let $C(j)$ denote the cluster containing $j$. The affiliation of group $G$ with respect to country $j$ is defined as
\[
A_G(j)=\frac{|G\cap C(j)|}{|G|}.
\]

This metric measures the proportion of group members that share a cluster with the reference country. For example, the affiliation of China with respect to SCO quantifies how closely other SCO member states are clustered with China in a given year. High affiliation values indicate a strong association of the country with the group, whereas persistently low values may suggest weaker alignment.

\begin{figure}[H]
    \centering
    
    \begin{subfigure}{0.32\textwidth}
        \centering
        \includegraphics[width=\linewidth]{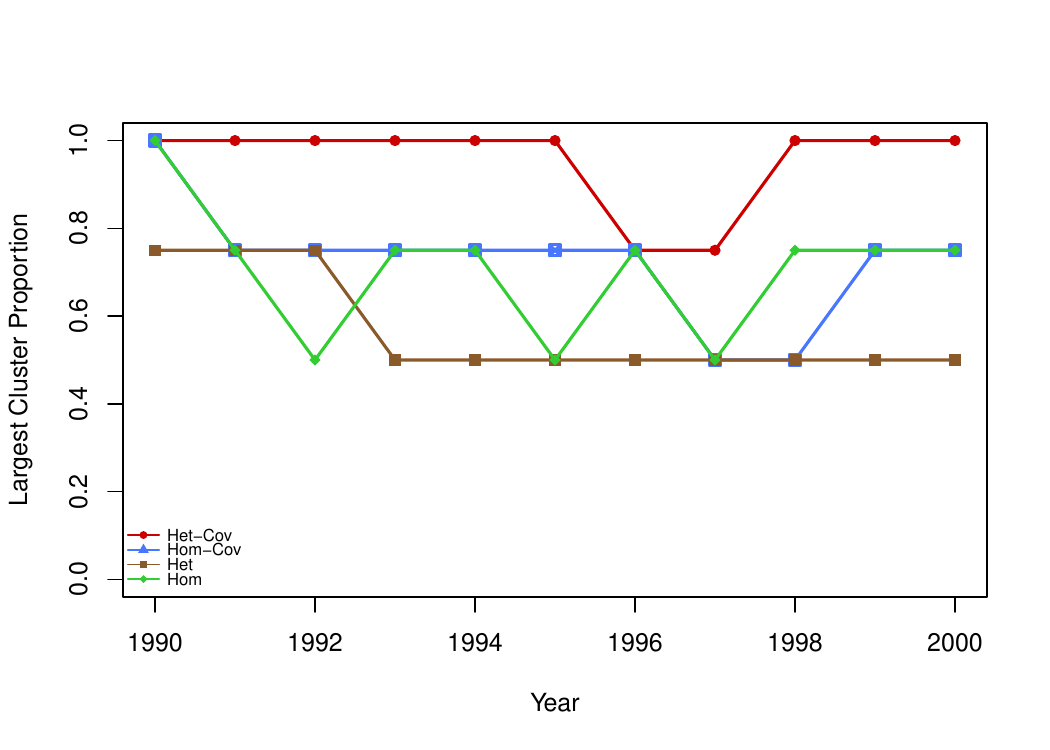}
        \caption{LCP for BRICS countries over the years.}
        \label{lc_brics}
    \end{subfigure}
    \hfill
    \begin{subfigure}{0.32\textwidth}
        \centering
    \includegraphics[width=\linewidth]{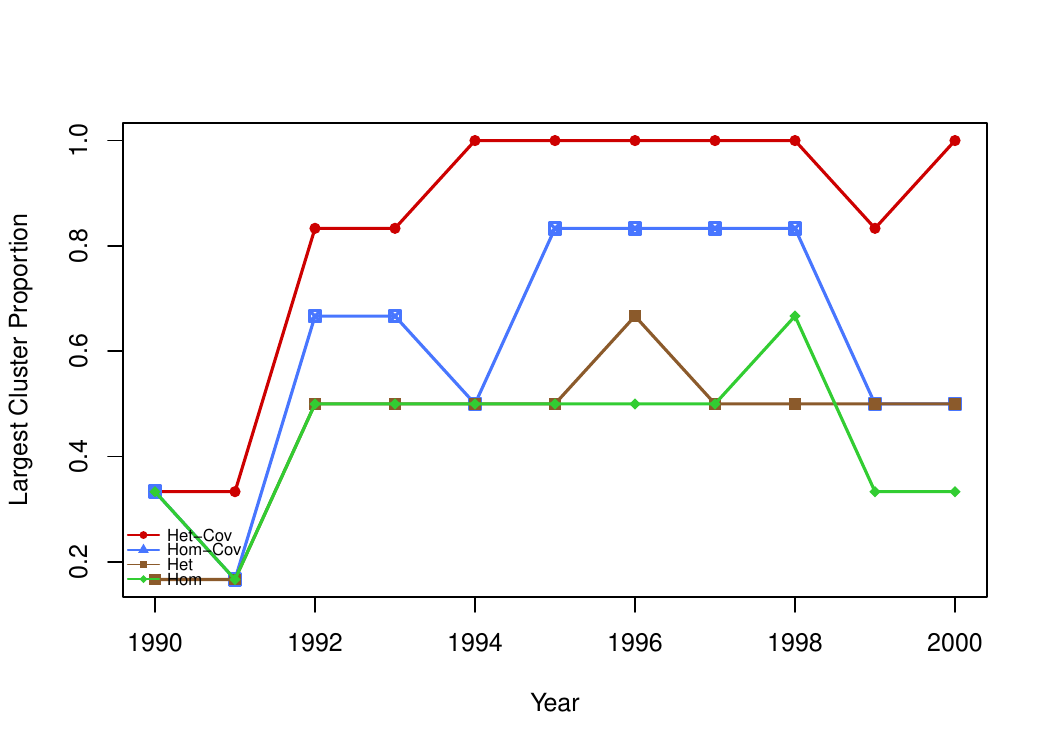}
        \caption{LCP for SCO countries over the years.}
        \label{lc_sco}
    \end{subfigure}
    \hfill
    \begin{subfigure}{0.32\textwidth}
        \centering
        \includegraphics[width=\linewidth]{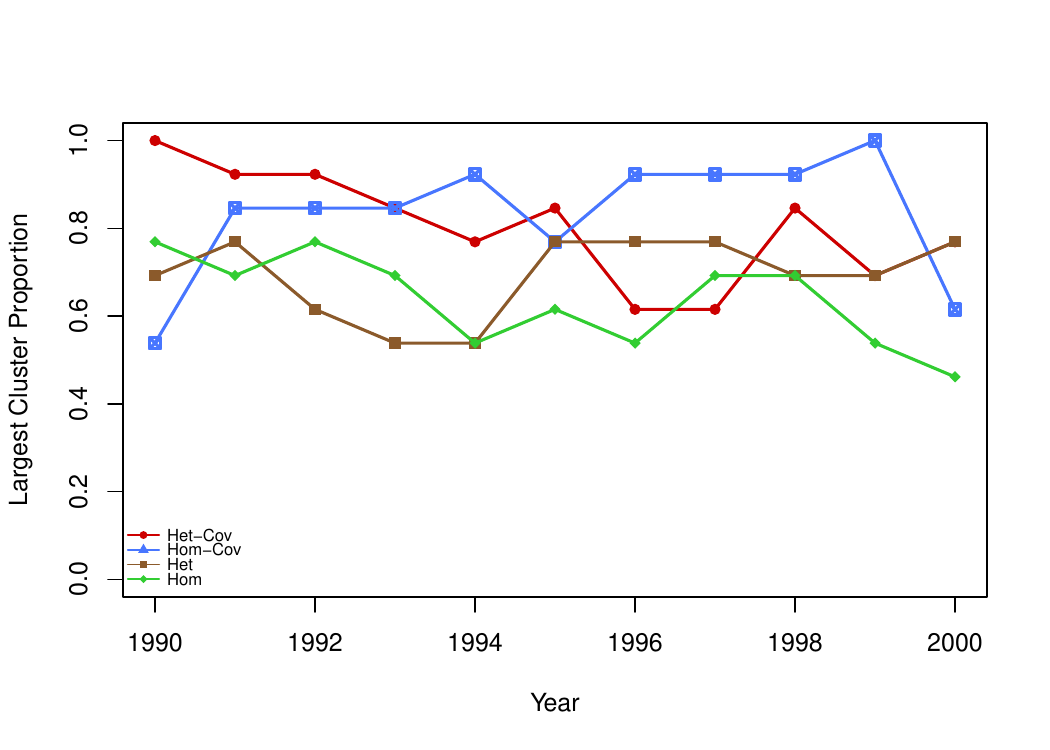}
        \caption{LCP for ALADI countries over the years.}
        \label{lc_la}
    \end{subfigure}
    \caption{Largest Cluster Proportion (LCP) for selected international organizations across the study period. The red, blue, brown, and green curves correspond to methods Het-Cov, Hom-Cov, Het, and Hom respectively.}
    \label{lcp}
\end{figure}

\begin{figure}[H]
    \centering
    
    \begin{subfigure}{0.45\textwidth}
        \centering
        \includegraphics[width=\linewidth]{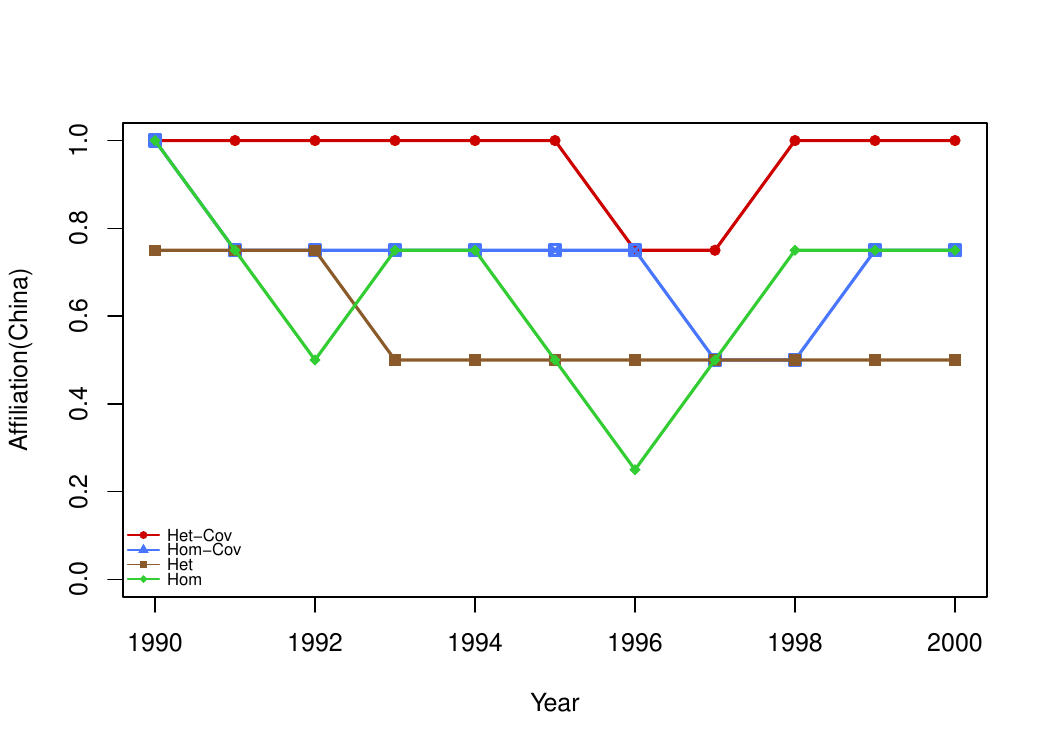}
        \caption{Affiliation for China in BRICS over the years.}
        \label{ch_aff_brics}
    \end{subfigure}
    \hfill
    \begin{subfigure}{0.45\textwidth}
        \centering
        \includegraphics[width=\linewidth]{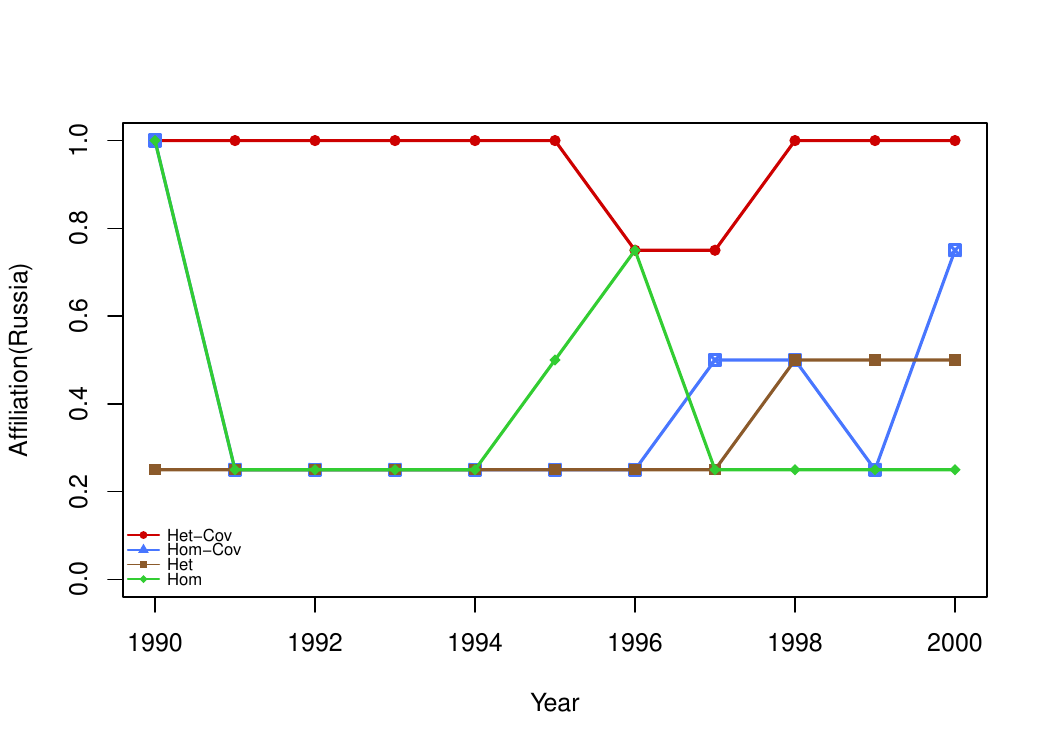}
        \caption{Affiliation for Russia in BRICS over the years.}
        \label{rus_aff_brics}
    \end{subfigure}

    \begin{subfigure}{0.45\textwidth}
        \centering
        \includegraphics[width=\linewidth]{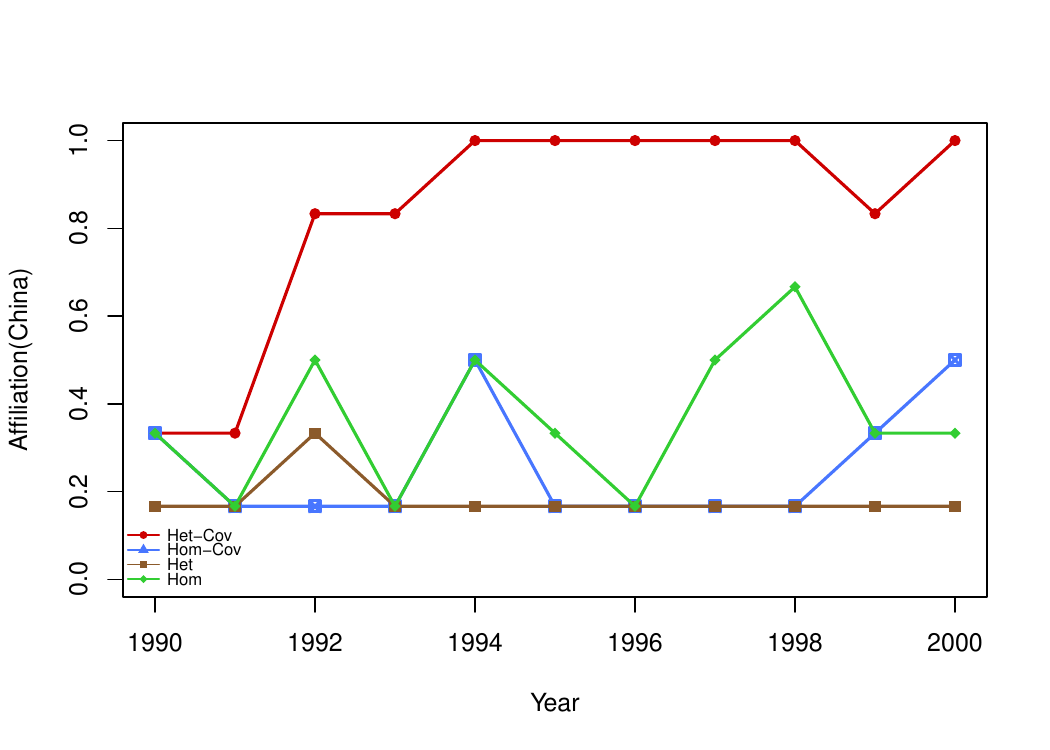}
        \caption{Affiliation for China in SCO over the years.}
        \label{aff_ch_sco}
    \end{subfigure}
    \hfill
    \begin{subfigure}{0.45\textwidth}
        \centering
        \includegraphics[width=\linewidth]{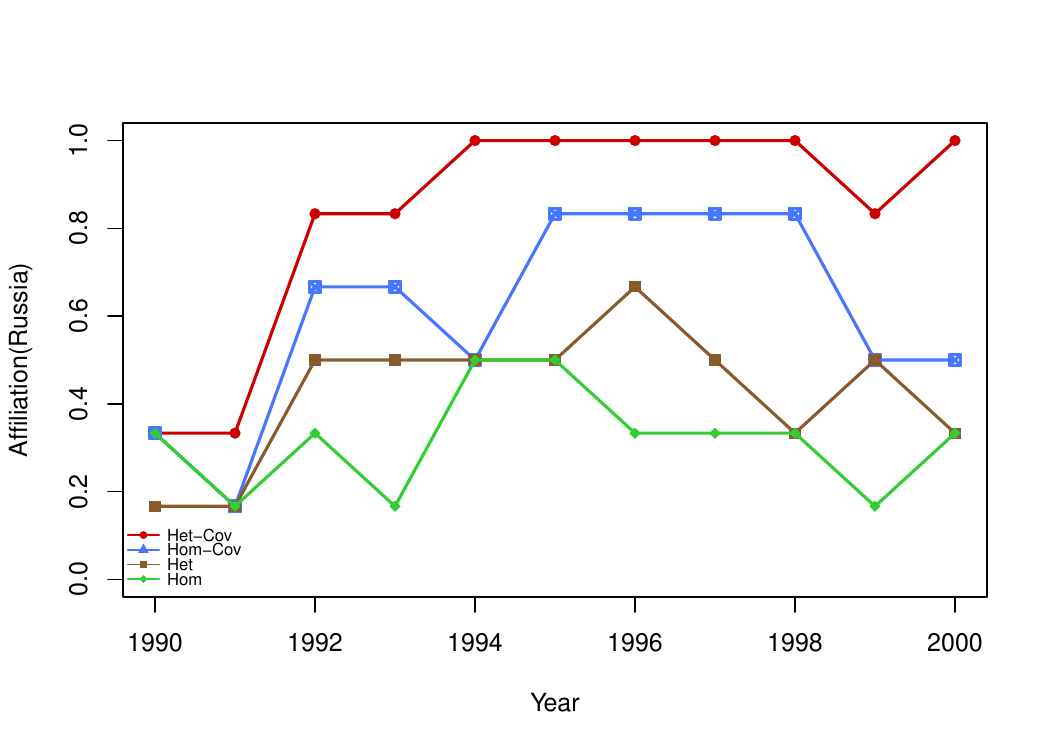}
        \caption{Affiliation for Russia in SCO over the years.}
        \label{aff_rus_sco}
    \end{subfigure}
    
\caption{Affiliation trends of key countries (China and Russia) within BRICS and SCO. Panel (a) shows China's affiliation with BRICS countries, Panel (b) presents Russia's affiliation with BRICS countries, Panels (c) and (d) show the affiliation of China and Russia with SCO countries, respectively. } \label{affiliation plots}
\end{figure}

\paragraph{Affiliation and LCP plots corresponding to BRICS, SCO and ALADI}
Figures \ref{lcp} and \ref{affiliation plots}  present the temporal evolution of LCP and affiliation of key countries for  organizations SCO, BRICS and ALADI, during the 1990s. The results show that the proposed Het-Cov spectral clustering method effectively captures the emergence and consolidation of major international groupings, often revealing patterns that are less evident under the baseline methods.

The proposed Het Cov spectral clustering method consistently attains higher LCP values than the competing approaches, indicating stronger clustering of member countries within the same geopolitical bloc. The only notable exception occurs for ALADI during the final years of the study period. 

 We can clearly see in the affiliation plots \ref{affiliation plots}, that the trend identified by our proposed method consistently dominates those obtained from the competing methods. The results suggest that the proposed Het Cov spectral clustering method more effectively captures the evolving geopolitical alignment of major countries within the organizations SCO and BRICS.

\paragraph{Interpretation of Affiliation and Largest Cluster Proportion Trends}

Figures \ref{lcp} and \ref{affiliation plots} summarize the temporal evolution of the largest cluster proportion (LCP) and affiliation measures for BRICS, SCO and ALADI countries during the 1990s. Across most years, the proposed Het Cov spectral clustering method attains higher LCP and affiliation values than the competing approaches, indicating a stronger ability to identify cohesive geopolitical groupings and evolving international alignments. The only notable exception occurs for ALADI during the final years of the study period.

The observed trends are also consistent with several important historical developments. For BRICS and SCO, both of which were formally established only after 2000, the consistently high affiliation of key countries, particularly China and Russia, suggests that the underlying geopolitical partnerships were already emerging during the 1990s. Similarly, strong clustering among ALADI member countries reflects the growing regional integration within Latin America, including the period leading up to Cuba's accession in 1999. Overall, the proposed method recovers several historically meaningful patterns and provides a coherent picture of the evolution of international alignments during the 1990s.

\subsection{Complementing the Ideal Point Framework with Het-Cov Clustering} 
\label{ip_discussion}

The dynamic ideal point framework has been widely used to study political alignments in the United Nations General Assembly through country-specific ideological positions derived from voting records. Since our proposed method aims to identify clusters of countries by jointly incorporating voting behaviour and economic characteristics, it is natural to compare its findings with those obtained from ideal point models. In this subsection, we examine areas of agreement and divergence between the two approaches and discuss how the inclusion of economic covariates can provide additional insights into the geopolitical and economic alignments of countries during the 1990s.

\subsubsection{Conceptual Differences Between Ideal Point and Cluster-Based Analyses}
Estimating dynamic ideal points using an ordinal spatial model and making them comparable across periods has long been useful in political science and related disciplines. Using roll-call voting data, \cite{voeten_2017} estimated annual country-specific ideal points along a single ideological dimension that broadly captures a country's placement with respect to the US-led liberal order. Subsequently, \cite{voeten_2018} demonstrated the emergence of an additional dimension associated with North--South tensions, reflecting the influence of groups such as the Non-Aligned Movement and the Group of 77, which was particularly prominent during the period 1960--1980.

While ideal point models focus on estimating latent political preferences of individual countries in relation to the US-led world order, the objective of our analysis is fundamentally different. Rather than placing countries on a continuous ideological scale, we study the formation of country clusters by jointly incorporating voting behaviour and economic characteristics, particularly GDP per capita. Consequently, the two approaches address complementary questions. Ideal points describe where countries are positioned politically, whereas our method identifies groups of countries that exhibit similar geopolitical and economic profiles. Moreover, we are primarily interested in interpreting the resulting cluster structures in light of major historical developments during the 1990s.

\subsubsection{Empirical Relationship Between Ideal Point Estimates and Cluster Structure}
To facilitate comparison, we first examine voting trajectories across major UNGA issue categories. Figure~\ref{fig:traj} shows that China voted ``Yes'' on an overwhelming majority of resolutions throughout the decade across most issue areas, with human rights resolutions constituting the primary exception. The United States displays an almost opposite pattern, voting ``No'' on a large proportion of resolutions in several categories. Russia exhibits a more mixed voting profile, particularly during the middle of the decade. These patterns closely mirror the ideological ordering reflected in the ideal point estimates.

The relationship between the ideal point framework and our clustering results is illustrated further through the boxplots in Figure~\ref{fig:boxplots}. Countries belonging to the USA-aligned cluster occupy the upper end of the ideal point scale and are concentrated within a relatively narrow range. Countries in the China-aligned cluster occupy the lower end of the scale. The remaining cluster spans a much wider range of ideal point values and overlaps substantially with both extremes throughout the decade. This observation suggests that the third cluster captures factors beyond voting ideology alone and highlights the possibility that economic characteristics contribute meaningfully to cluster formation. In the following discussion, we examine several country groups and historical episodes to better understand these differences.

Several of the most interesting distinctions between the ideal point framework and our clustering approach emerge while examining specific countries and regional groupings. The examples below illustrate how the incorporation of economic information can lead to interpretations that differ from those based solely on voting behaviour.

\paragraph{Gulf States and Libya}

The Gulf states provide an instructive example of the additional information captured by our framework. Figures~\ref{fig:saudi_traj}, \ref{fig:qa_traj}, \ref{fig:uae_traj}, and \ref{fig:om_traj} show voting patterns that closely resemble those of China. Correspondingly, ideal point estimates place these countries near the anti-US end of the ideological spectrum. Nevertheless, our method frequently assigns them to the intermediate cluster because of their high-income status.

The cases of the United Arab Emirates, Qatar, and Kuwait are particularly informative. During parts of the late 1990s, these countries move closer to the USA-aligned cluster, reflecting their increasing strategic and security ties with the United States. UAE, despite having a voting trajectory similar to that of other Gulf states, consistently belongs to the USA-specific blue cluster, while Kuwait and Qatar, especially during the late 1990s, transition from the intermediate cluster to the USA cluster. Such transitions demonstrate the sensitivity of the proposed method to geopolitical changes that are difficult to infer from voting behaviour alone.

Libya provides a contrasting example. Its voting trajectory and ideal point estimates consistently indicate a strongly anti-US position (Figure~\ref{fig:libya_traj}). Yet Libya is frequently assigned to the intermediate cluster because of its relatively high per capita income derived from oil revenues. This finding illustrates how countries with similar voting behaviour may occupy different positions in the clustering structure once economic characteristics are incorporated.

\paragraph{Eastern European Transition Economies}

\begin{figure}[H]
\centering
    \includegraphics[width = 15cm]{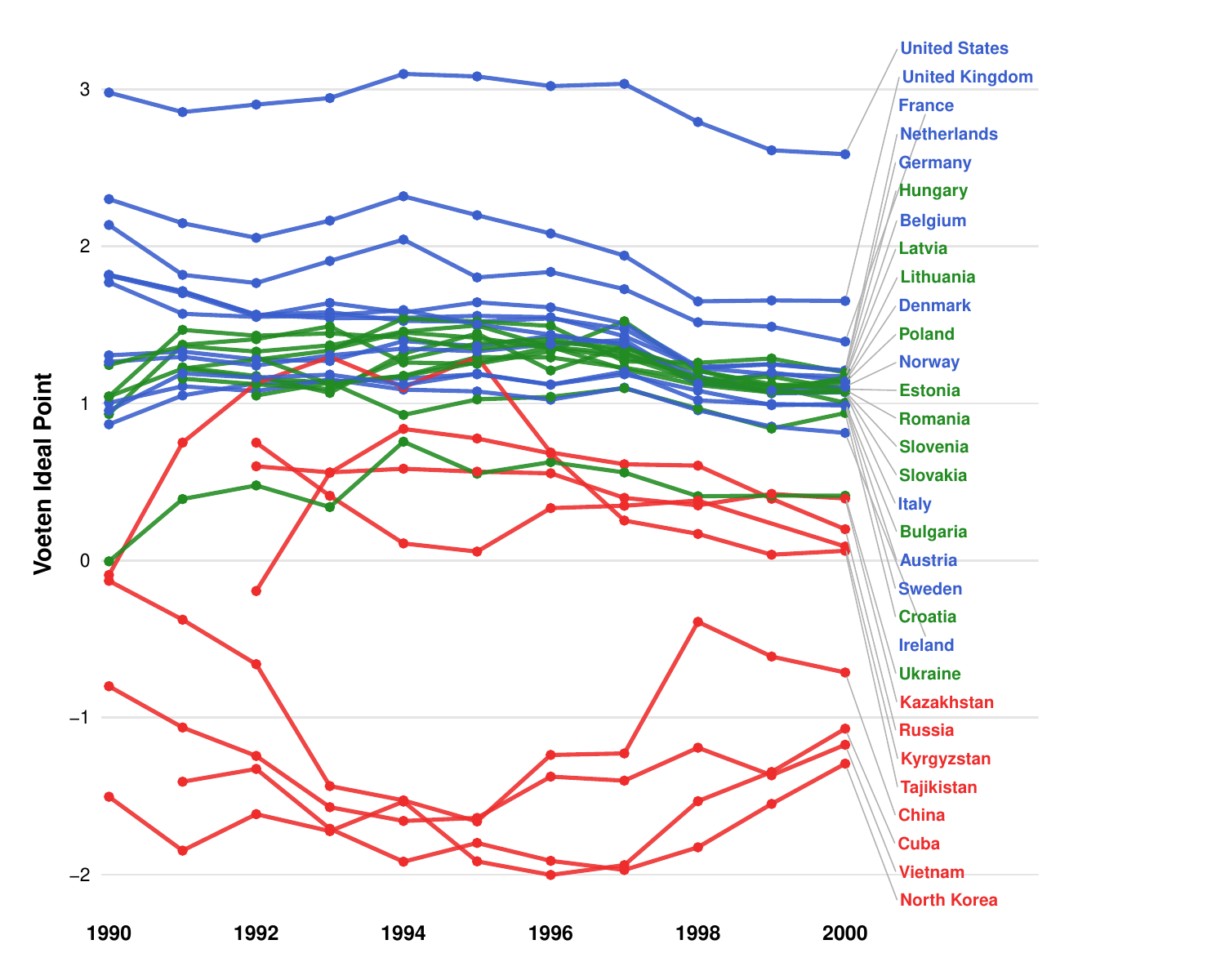}
    \caption{The movement of ideal points of the countries in West Europe bloc, East Europe bloc and China bloc during 1990 -2000.}
    \label{fig:ip_trend}
\end{figure}

The most substantial differences between the two approaches emerge in Eastern Europe. Following the collapse of the Soviet bloc, ideal point trajectories suggest that many Eastern European countries gradually converged toward Western Europe and the broader pro-US camp (Figure~\ref{fig:ip_trend}). Based solely on voting behaviour, these countries appear increasingly similar to Western European states.

\begin{figure}[H]
\centering

\begin{subfigure}[b]{0.9\textwidth}
    \centering
    \includegraphics[height=6cm]
    {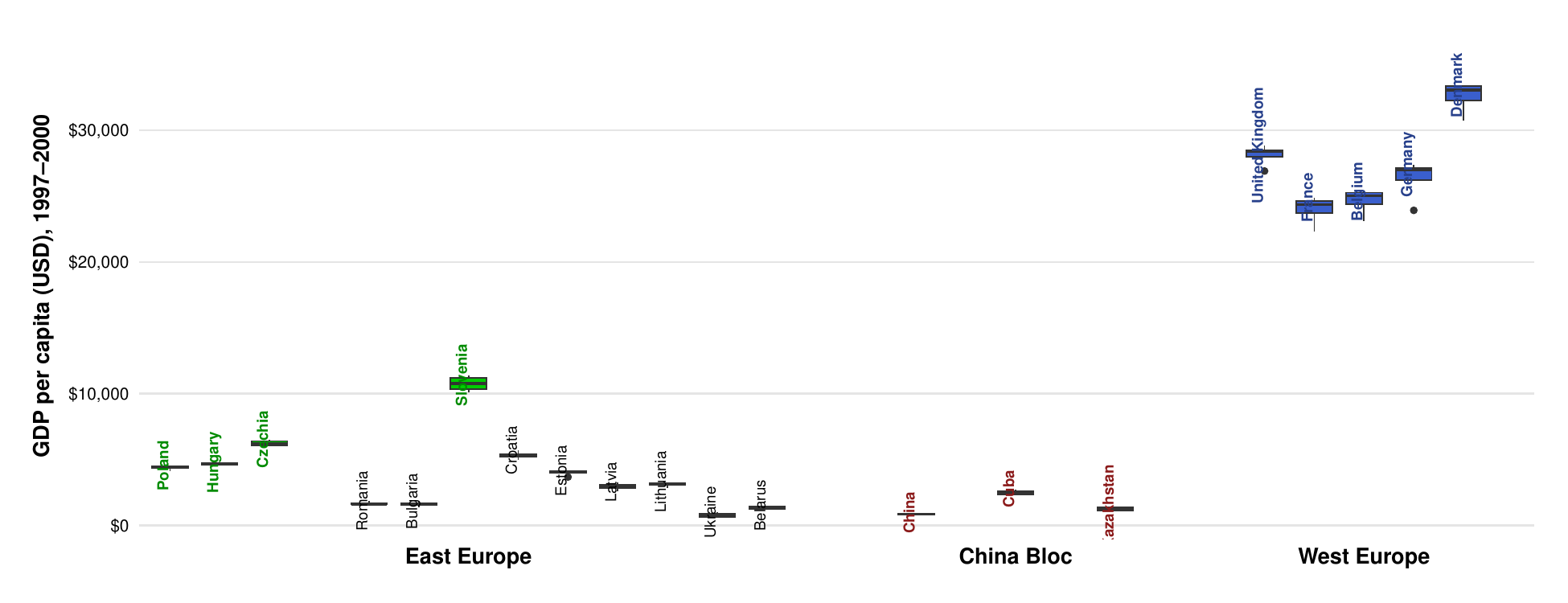}
    \caption{GDP per capita values for countries in Eastern Europe, the China-aligned bloc, and Western Europe during the closing years of the twentieth century.}
    \label{fig:gdp_pc_ee_ch_we}
\end{subfigure}
\vspace{0.3cm}
\begin{subfigure}[b] {0.9\textwidth}
    \centering
    \includegraphics[height=6cm]
    {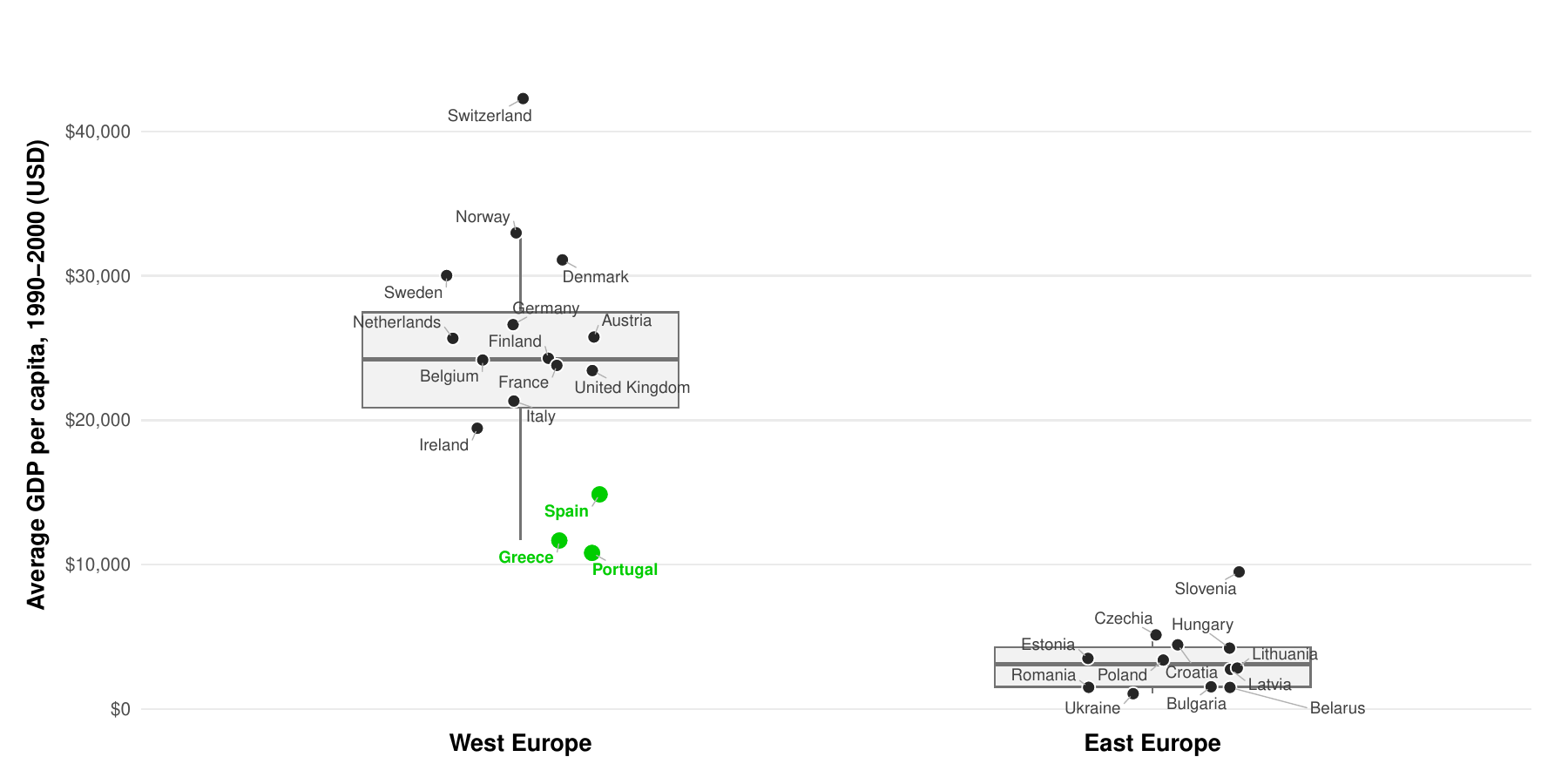}
    \caption{GDP-per-capita averaged over the years 1990-2000 for Western as well as Eastern European Countries, respectively. }
    \label{fig:east_vs_west_europe_gdp}
\end{subfigure}

\caption
{\footnotesize Boxplots depicting the disparity between GDP per capita income distribution among different demographic and poltical blocs. The boundary countries, Portugal, Spain and Greece, have been featured in green colour.}
\label{fig:bp}
\end{figure}

However, our method identifies a distinct intermediate grouping during approximately 1993--1998 that includes countries such as Slovenia, Czechia, Poland, and Hungary. Figure~\ref{fig:gdp_pc_ee_ch_we} provides a possible explanation. Although these countries increasingly aligned with Western Europe politically, their economic conditions remained substantially below those of core Western European economies. Their GDP per capita distributions were often much closer to those of the developing world than to those of Western Europe.

Historical studies of post-socialist transition economies (\cite{russian_crisis_2,russian_crisis_1}) indicate that countries such as Slovenia, Czechia, Poland, and Hungary implemented market reforms more successfully and attracted foreign investment earlier than many of their regional neighbours. Consistent with this historical account, these countries remain in the intermediate cluster longer than other Eastern European states. In contrast, much of South-Eastern Europe (for example, Albania, Bulgaria, and Romania) remains grouped with Russia throughout the decade, reflecting slower economic transition due to uneven progress in liberalisation, large debt burdens, weaker institutional development, and greater distance from Western European markets.

\paragraph{Western European Heterogeneity}

The proposed method also reveals meaningful differences within Western Europe. Countries such as Spain, Portugal, and Greece possess strongly positive ideal point estimates and therefore appear firmly embedded within the pro-US camp. However, our clustering results distinguish them from the wealthiest Western European economies.

Portugal remains predominantly in the intermediate cluster, while Spain alternates between the intermediate and USA-aligned clusters. Figure~\ref{fig:east_vs_west_europe_gdp} suggests that these assignments reflect economic differences within Western Europe itself. Although politically aligned with the West, these countries occupied a lower income tier relative to the core Western European economies. Their location near the boundary between clusters naturally leads to occasional oscillations over time.

\paragraph{Mexico and Latin America}

Mexico represents another informative case. Throughout the decade, Mexico maintains negative ideal point estimates, placing it on the anti-US side of the ideological spectrum. At the same time, its economic profile differs substantially from that of many developing countries. Despite joining NAFTA in 1994 and deepening economic integration with the United States, Mexico continued to vote similarly to many developing nations (Figure~\ref{fig:mexico_traj}). Our method therefore places Mexico primarily in the intermediate cluster, reflecting the tension between its voting behaviour and economic position.

\begin{figure}[H]
\centering
    \includegraphics[width = 10cm]{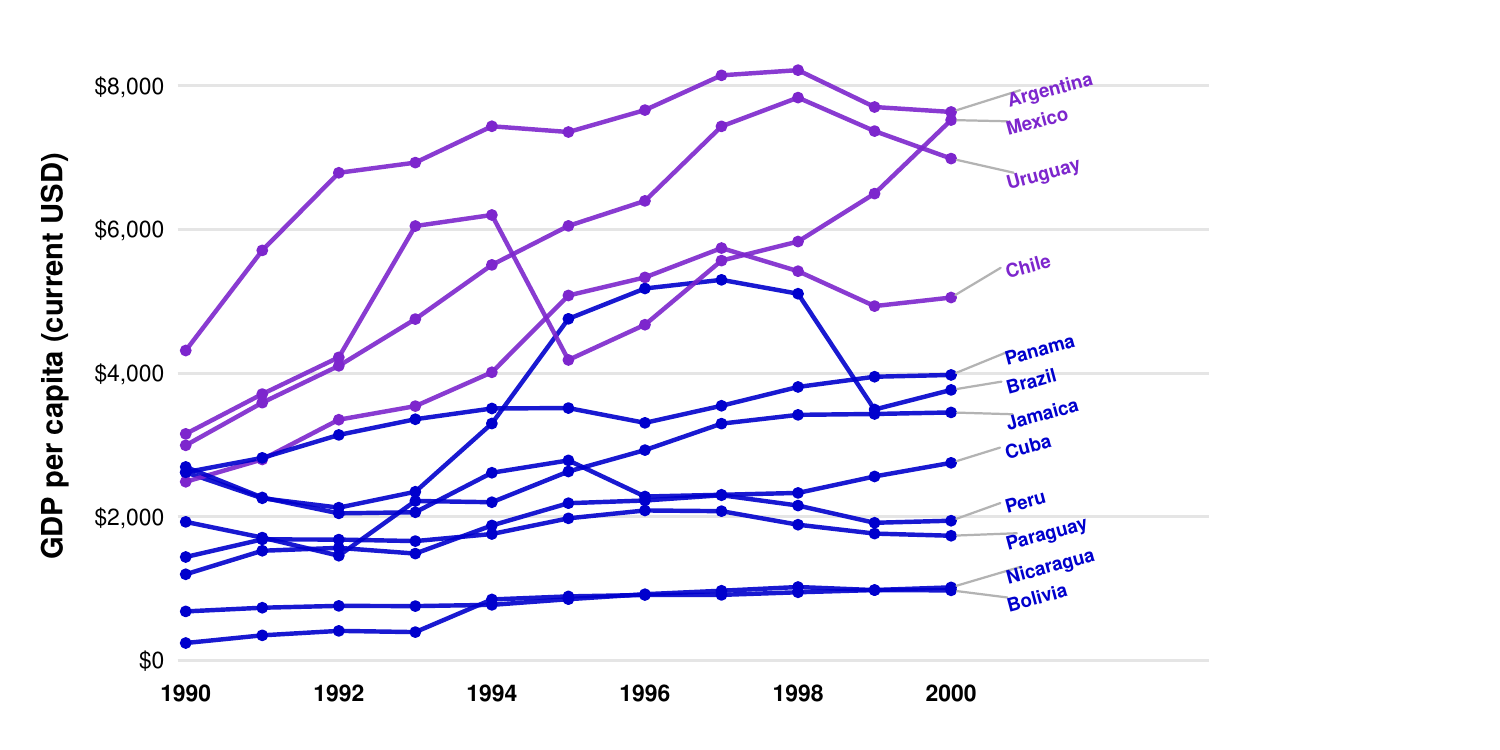}
    \caption{Trajectory of GDP per capita income for major Latin American countries during 1990 -2000. The top countries are marked in purple: Argentina, Mexico, Uruguay, and Chile.}
    \label{fig:latin_america_gdp}
\end{figure}

A related pattern emerges across Latin America more broadly. Although many Latin American countries move together between clusters over time, the movement is not uniform. Argentina, Chile, Mexico, and Uruguay tend to remain in the intermediate cluster even when several neighbouring countries move toward the China-aligned cluster. Figure~\ref{fig:latin_america_gdp} indicates that these countries consistently occupied the upper end of the regional income distribution during the 1990s, suggesting that economic differences play an important role in explaining the observed clustering patterns.

One notable exception is North Korea. Throughout the study period, North Korea is frequently assigned to the same cluster as South Korea despite possessing strongly negative ideal point estimates and maintaining a consistently anti-US position. This result is primarily driven by the imputation procedure used to address missing economic data. Since GDP-related information for North Korea was unavailable in the WDI database, region-wise mean imputation assigned economic characteristics similar to neighbouring countries. Consequently, this particular clustering outcome should be interpreted with caution.

\subsubsection{Discussion: A Joint Perspective on Ideal Points and Clustering}

Overall, the comparison highlights the complementary nature of the two approaches. Ideal point models provide a powerful framework for measuring ideological positioning based on voting behaviour, whereas the proposed Het-Cov clustering framework combines voting patterns with economic characteristics to uncover broader geopolitical and economic alignments. As a result, the proposed method reveals several historically meaningful distinctions that are not readily apparent from ideal point estimates alone. In particular, it captures the role of economic heterogeneity within politically similar groups and provides a richer interpretation of the evolving international alignments of the 1990s.

\begin{center}
\captionsetup{
type=figure,
font=small,
skip=2pt
}
\centering

\begin{subfigure}[b]{0.3\textwidth}
    \centering
    \includegraphics[width=\linewidth]{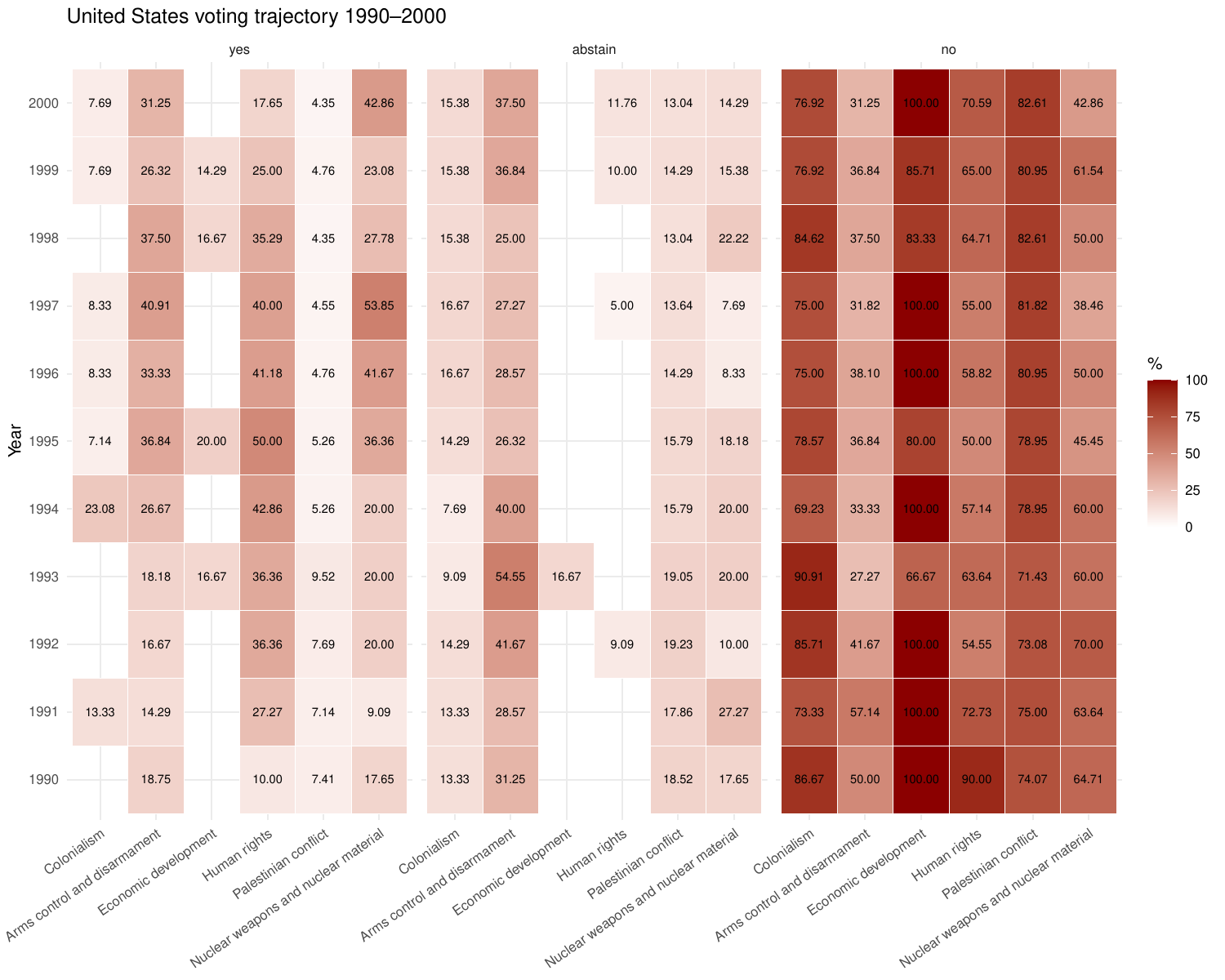}
    \caption{United States}
    \label{fig:us_traj}
\end{subfigure}
\hfill
\begin{subfigure}[b]{0.3\textwidth}
    \centering
    \includegraphics[width=\linewidth]{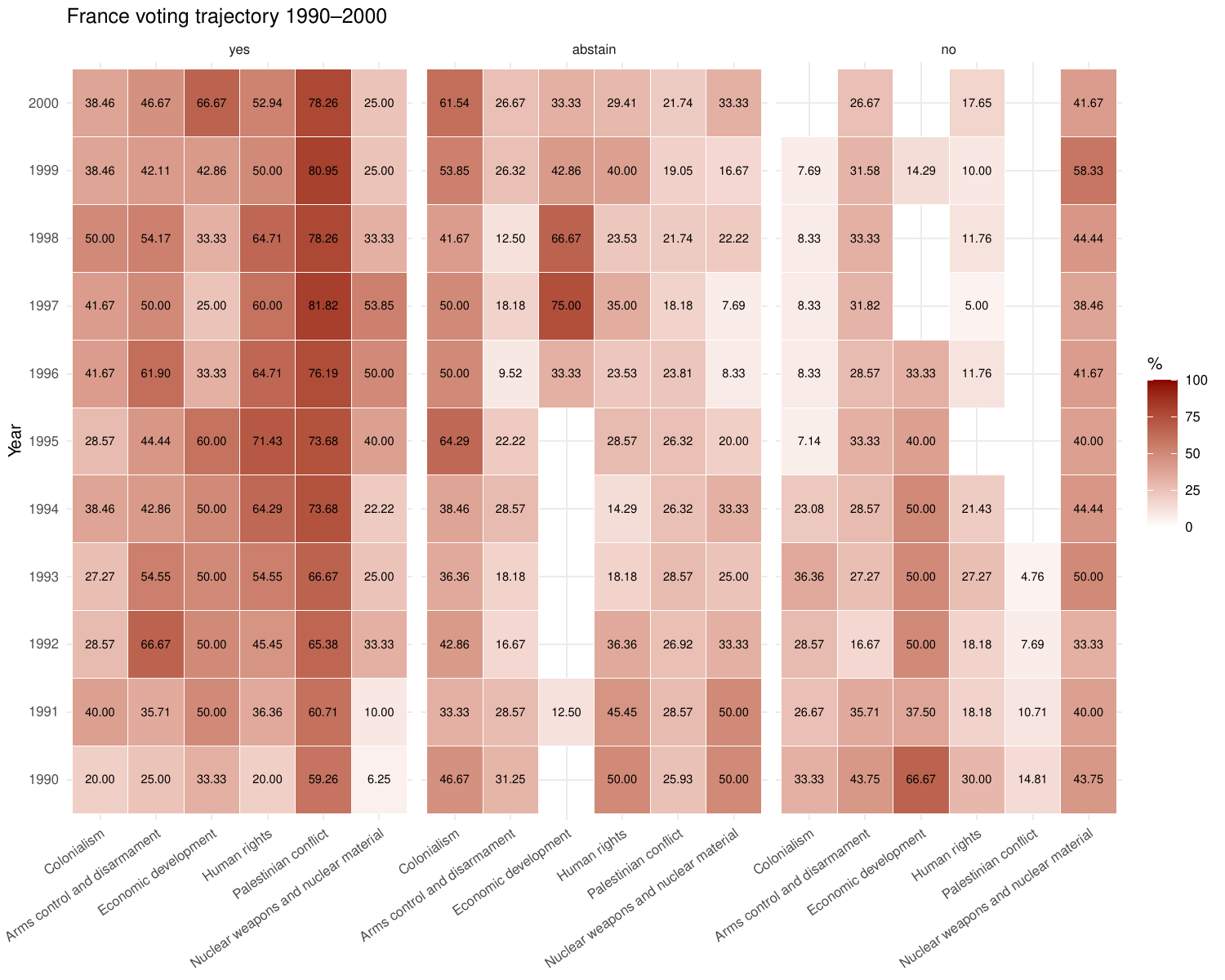}
    \caption{France}
    \label{fig:fr_traj}
\end{subfigure}
\hfill
\begin{subfigure}[b]{0.3\textwidth}
    \centering
    \includegraphics[width=\linewidth]{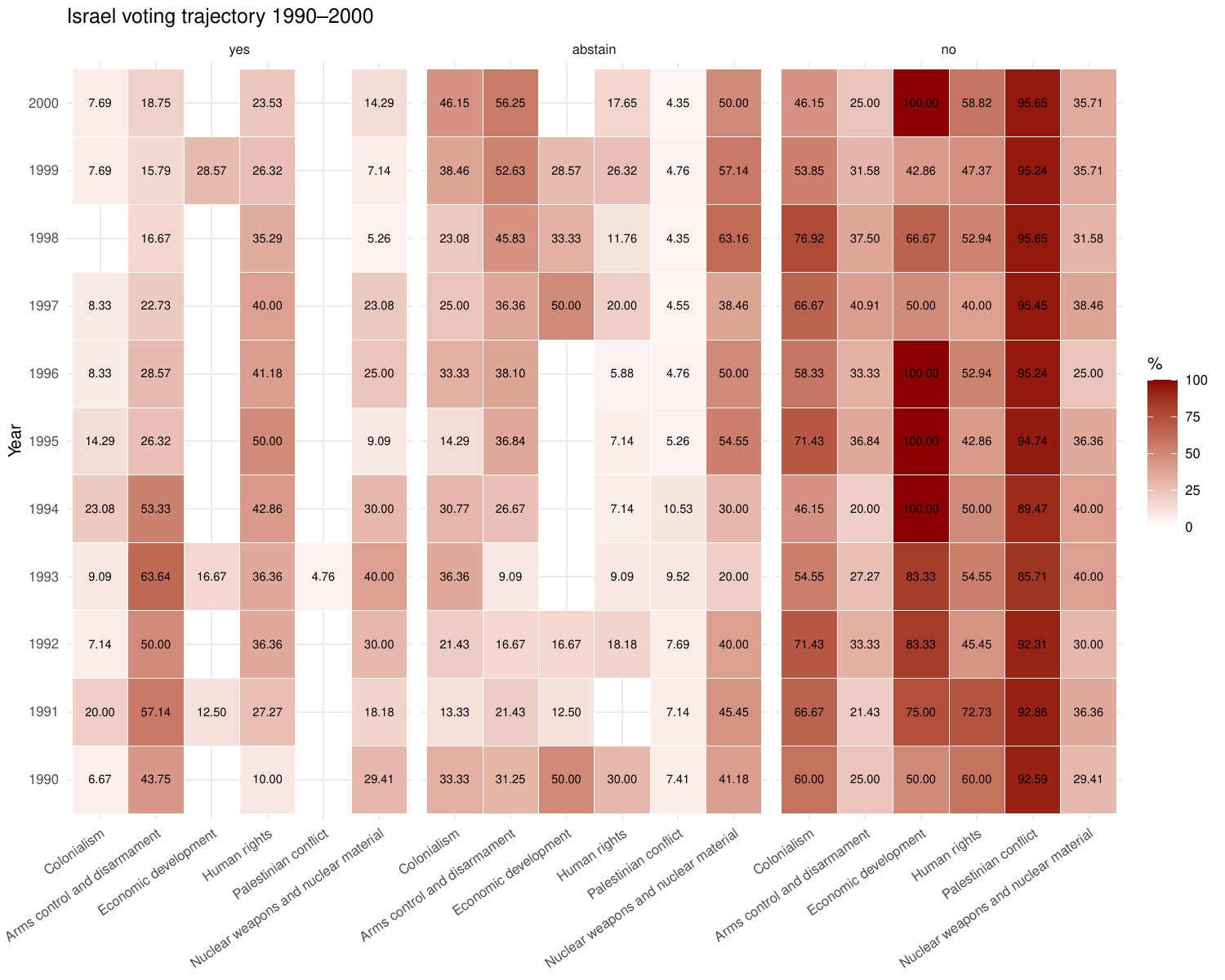}
    \caption{Israel}
    \label{fig:isr_traj}
\end{subfigure}

\vspace{0.5cm}

\begin{subfigure}[b]{0.3\textwidth}
    \centering
    \includegraphics[width=\linewidth]{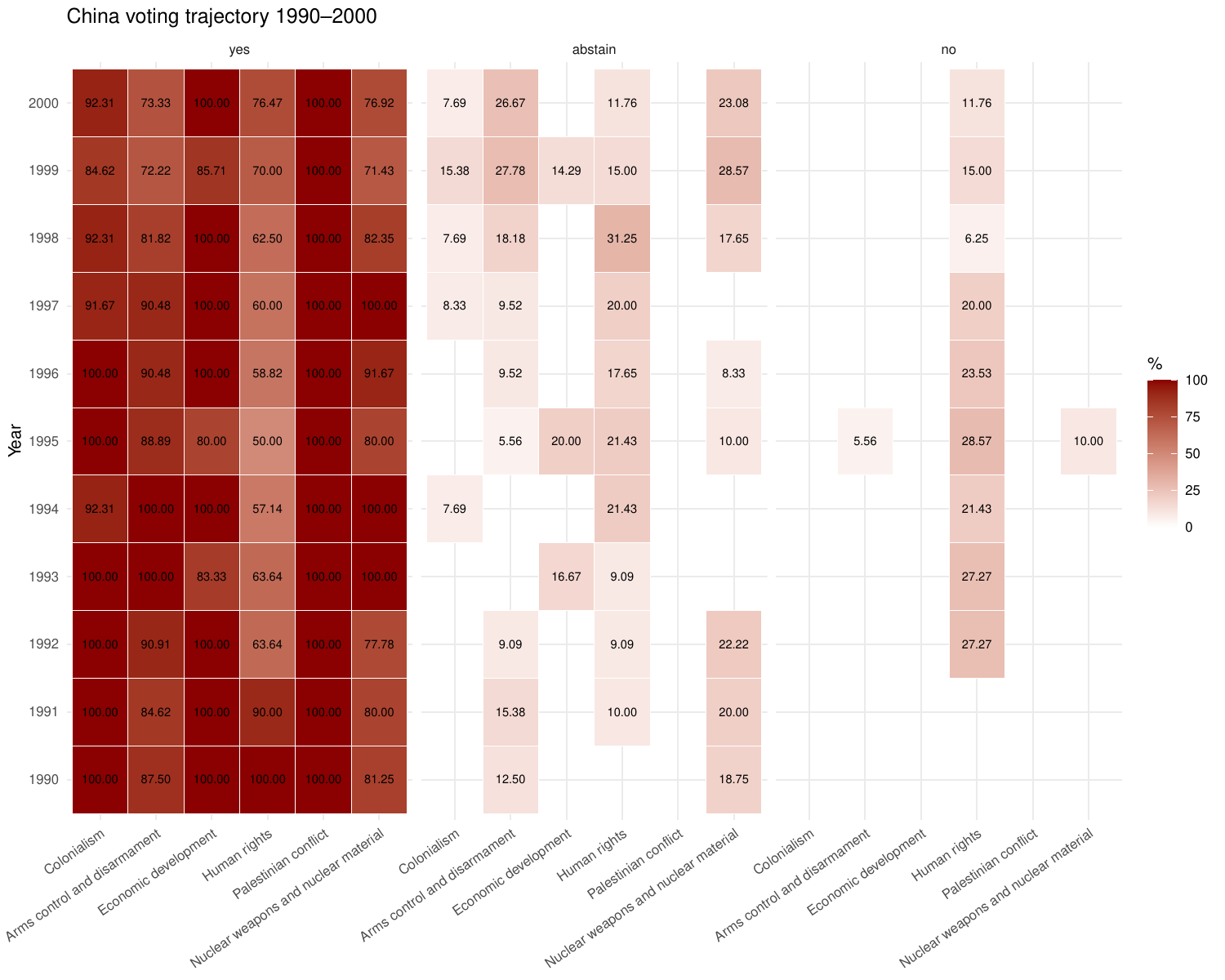}
    \caption{China}
    \label{fig:ch_traj}
\end{subfigure}
\hfill
\begin{subfigure}[b]{0.3\textwidth}
    \centering
    \includegraphics[width=\linewidth]{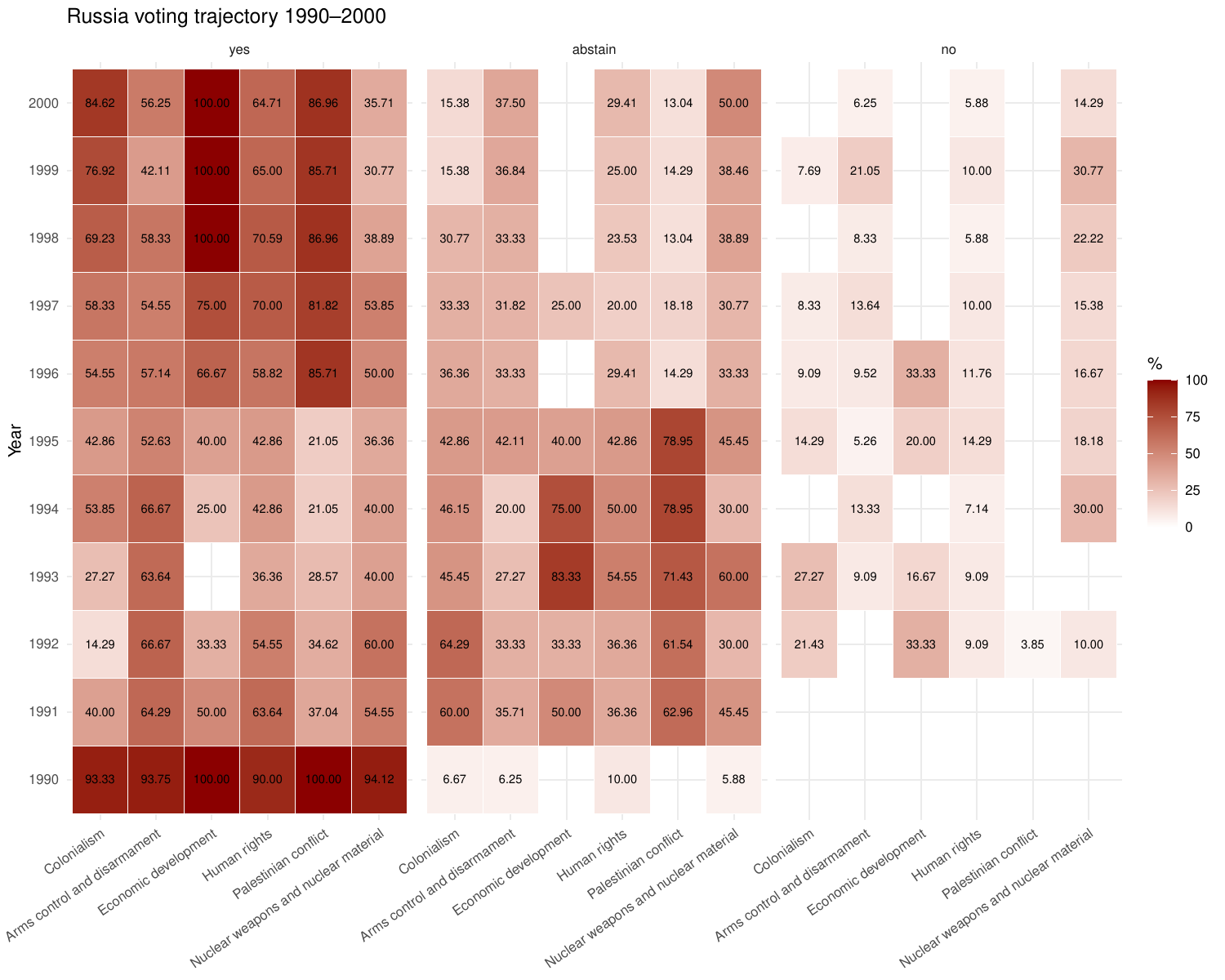}
    \caption{Russia}
    \label{fig:rus_traj}
\end{subfigure}
\hfill
\begin{subfigure}[b]{0.3\textwidth}
    \centering
    \includegraphics[width=\linewidth]{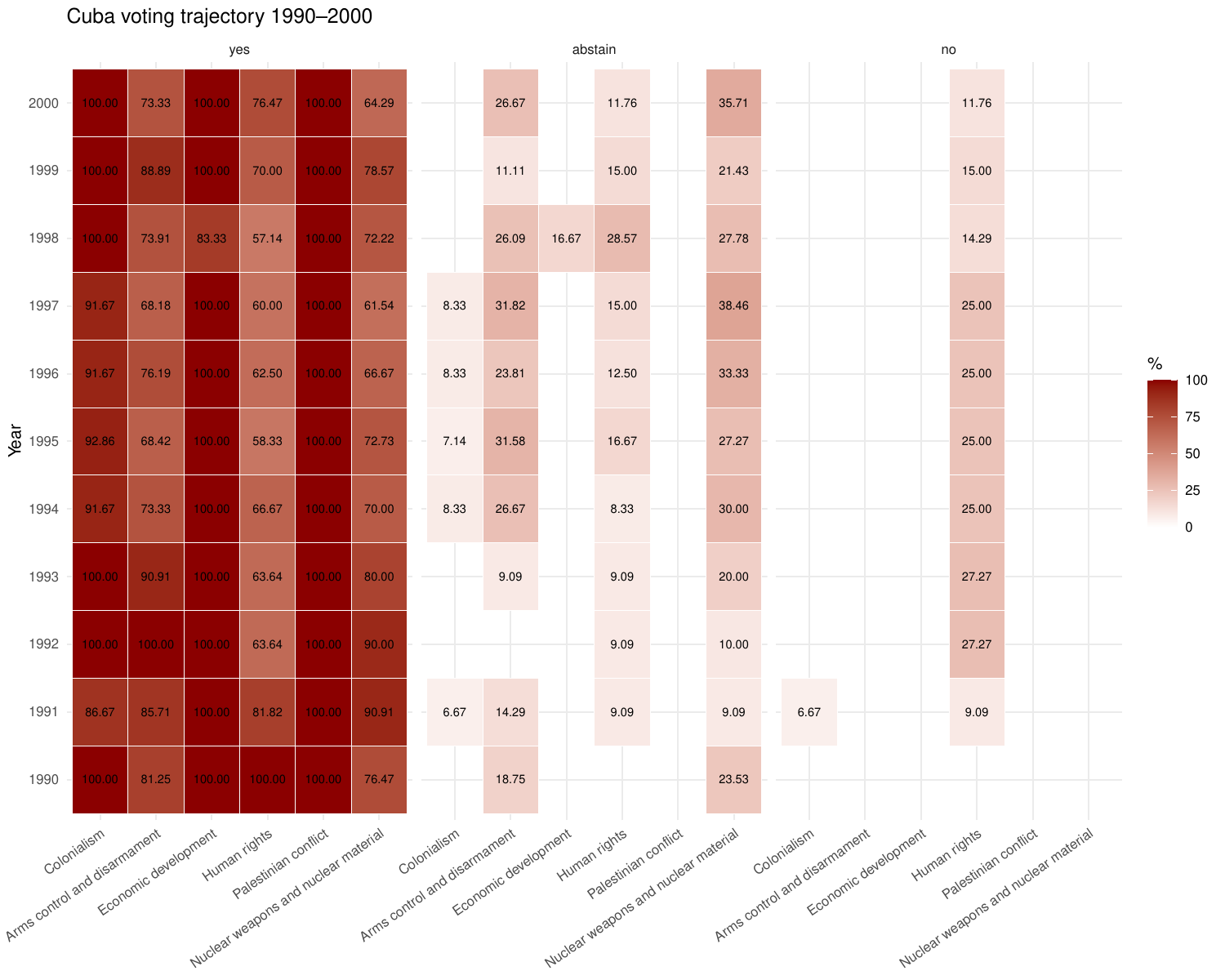}
    \caption{Cuba}
    \label{fig:cuba_traj}
\end{subfigure}

\vspace{0.5cm}
\begin{subfigure}[b]{0.3\textwidth}
    \centering
    \includegraphics[width=\linewidth]{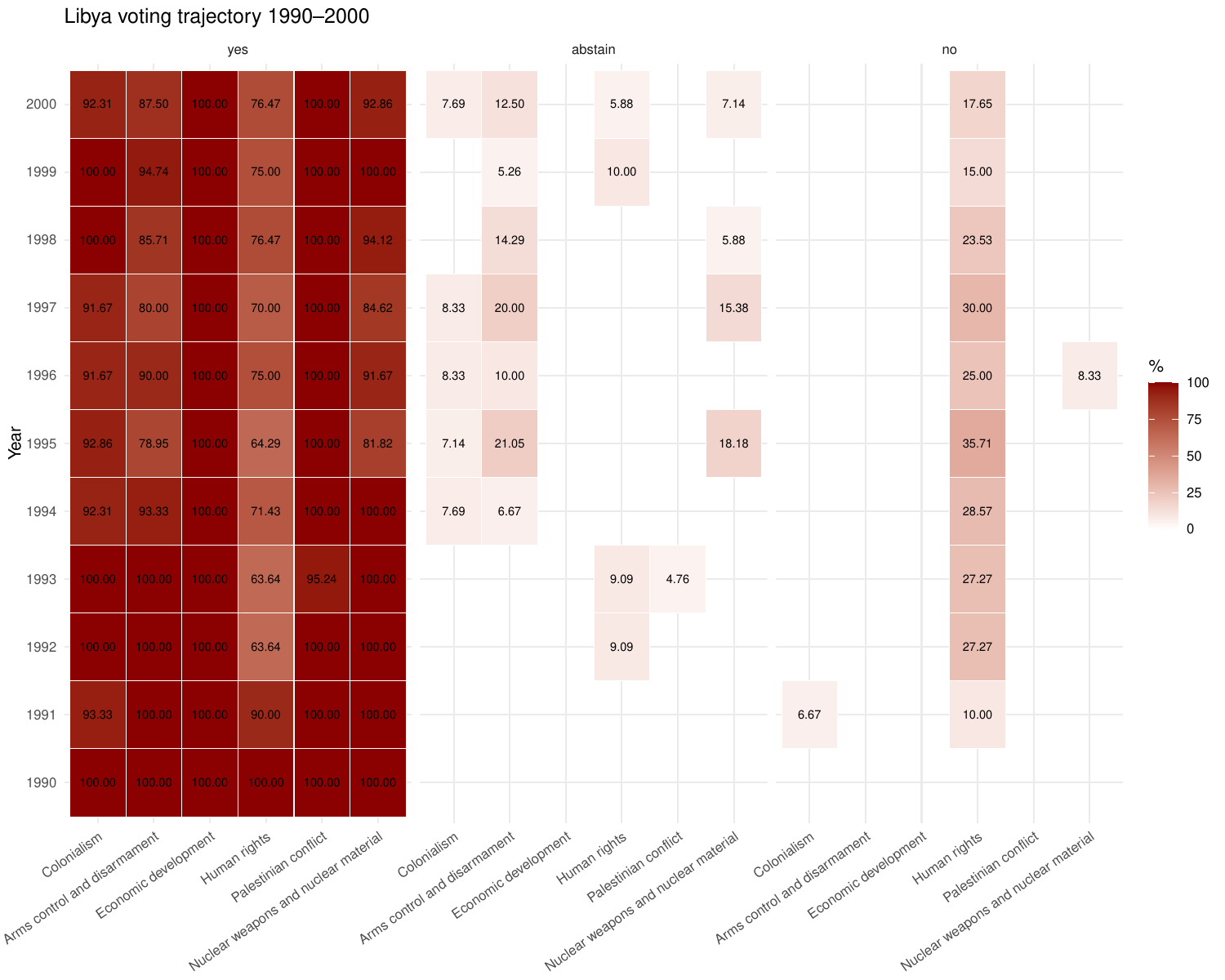}
    \caption{Libya}
    \label{fig:libya_traj}
\end{subfigure}
\begin{subfigure}[b]{0.3\textwidth}
    \centering
    \includegraphics[width=\linewidth]{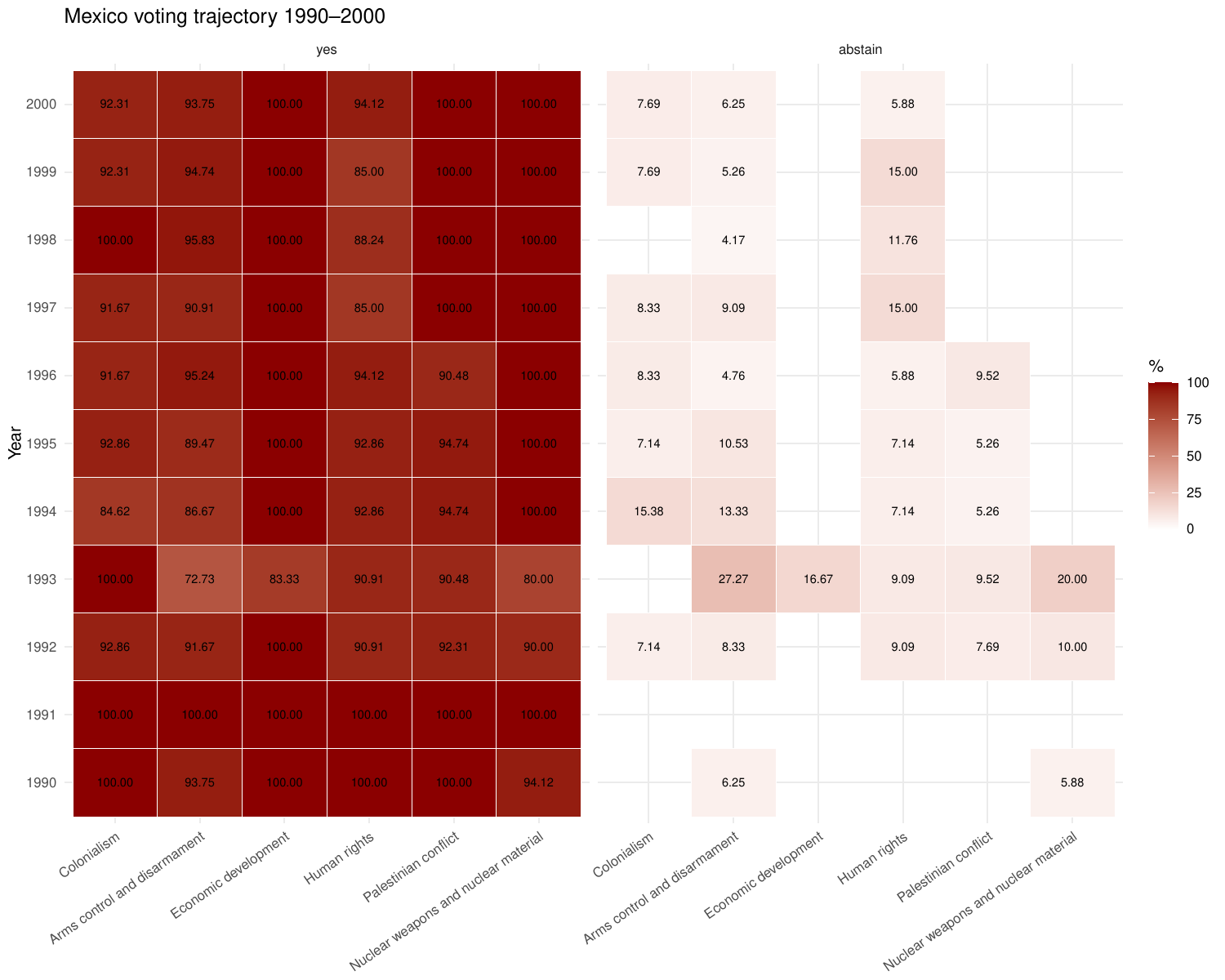}
    \caption{Mexico}
    \label{fig:mexico_traj}
\end{subfigure}

\vspace{0.5cm}
\begin{subfigure}[b]{0.3\textwidth}
    \centering
    \includegraphics[width=\linewidth]{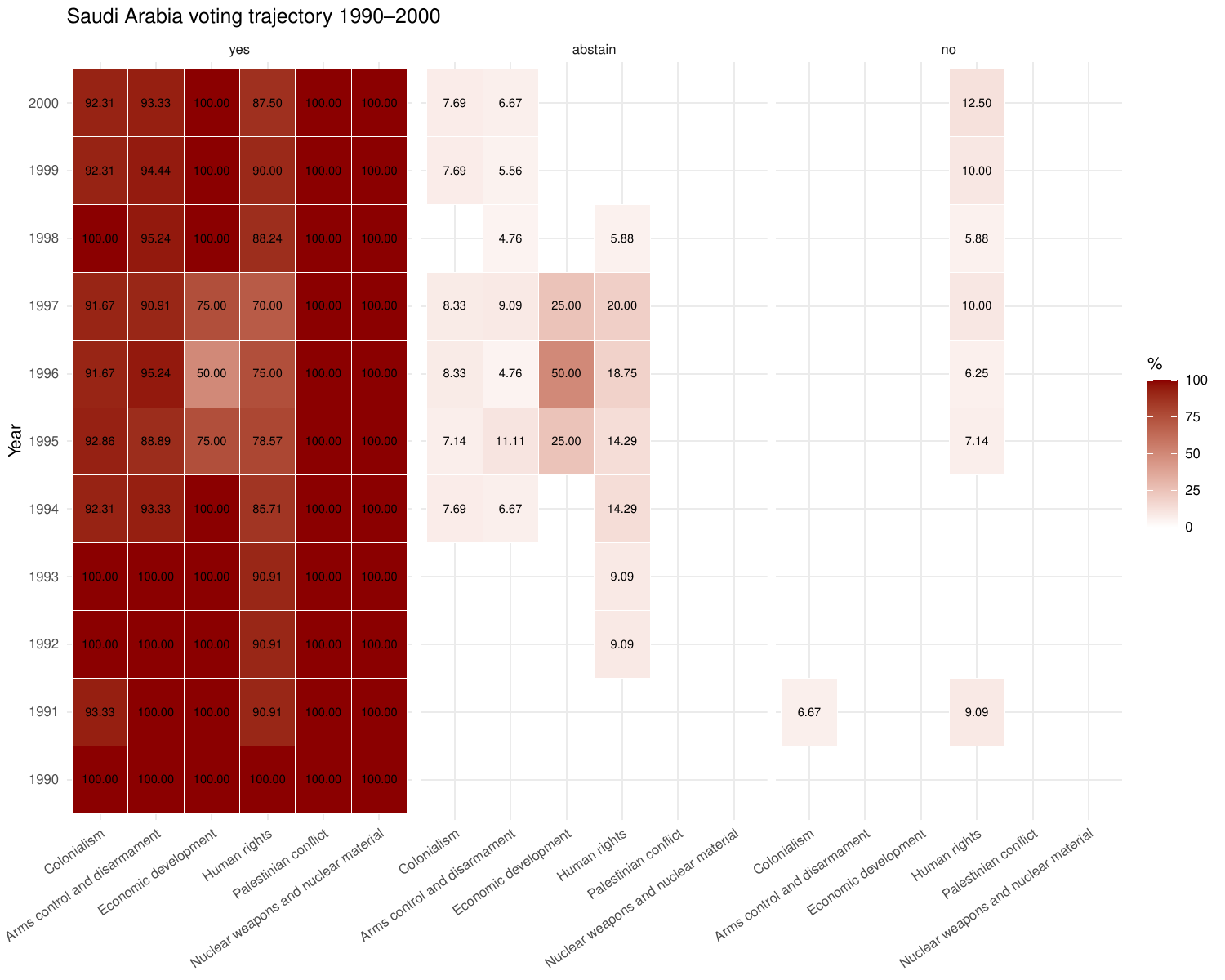}
    \caption{Saudi Arabia}
    \label{fig:saudi_traj}
\end{subfigure}
\hfill
\begin{subfigure}[b]{0.3\textwidth}
    \centering
    \includegraphics[width=\linewidth]{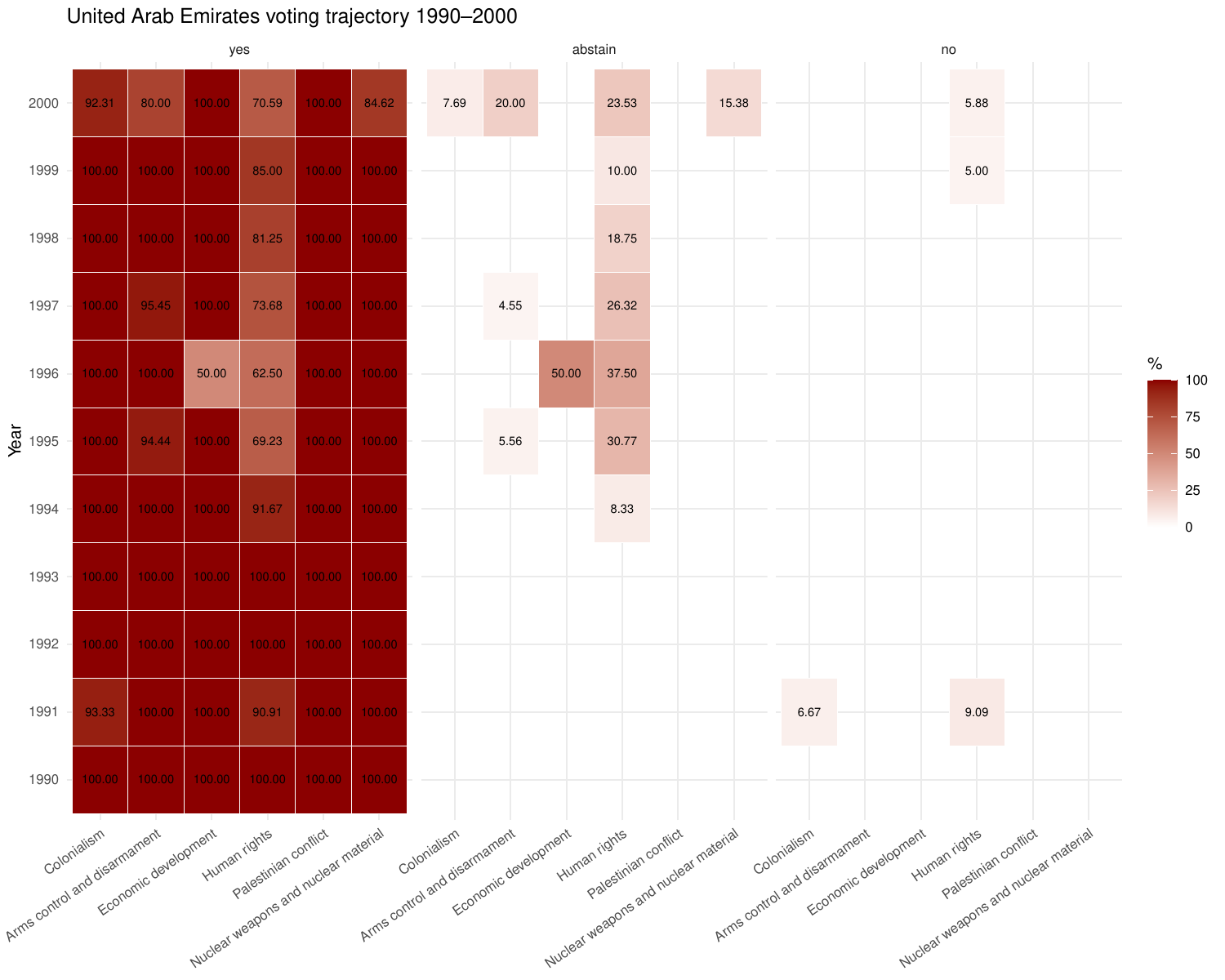}
    \caption{United Arab Emirates}
    \label{fig:uae_traj}
\end{subfigure}
\hfill
\begin{subfigure}[b]{0.3\textwidth}
    \centering
    \includegraphics[width=\linewidth]{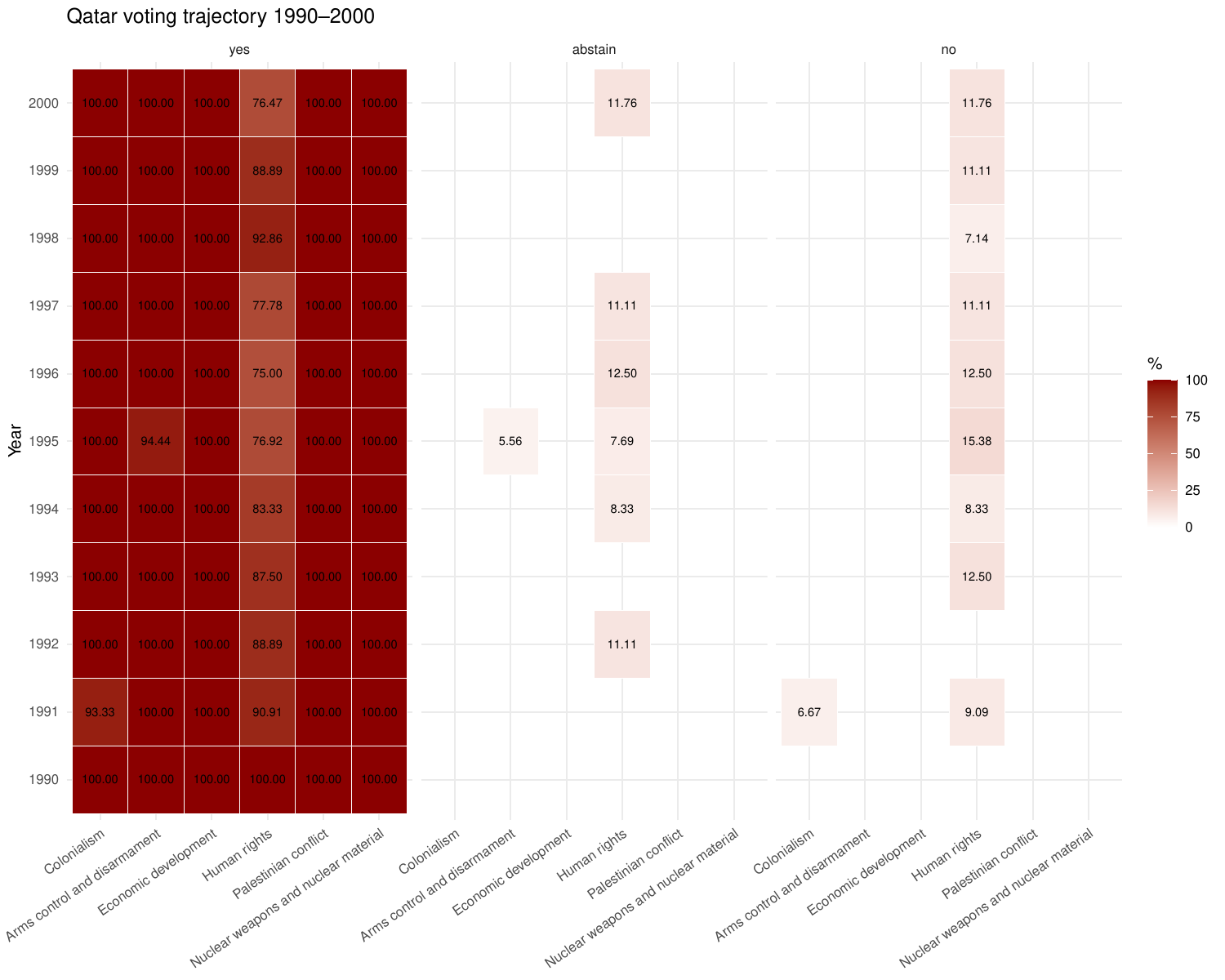}
    \caption{Qatar}
    \label{fig:qa_traj}
\end{subfigure}

\vspace{0.5cm}

\begin{subfigure}[b]{0.3\textwidth}
    \centering
    \includegraphics[width=\linewidth]{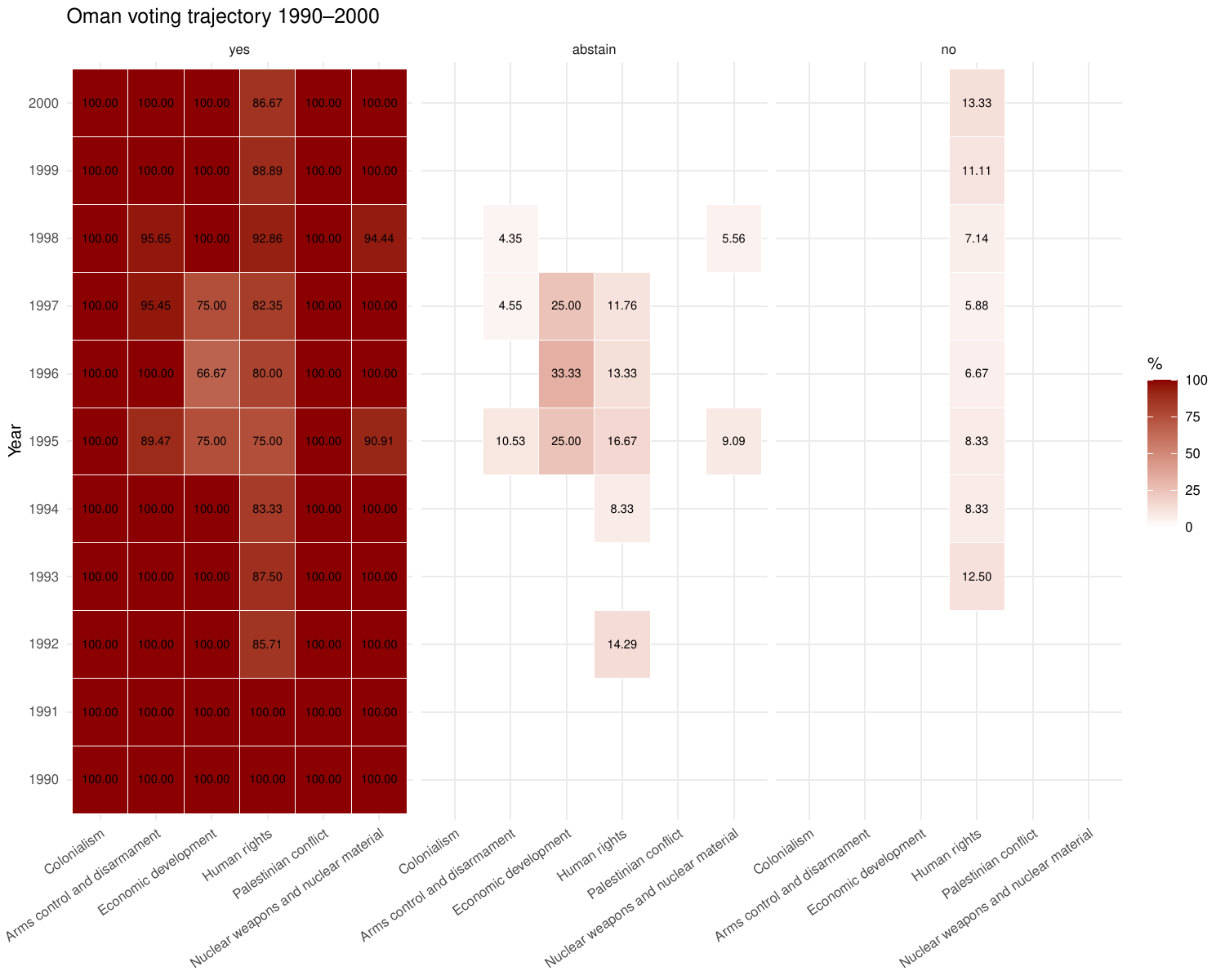}
    \caption{Oman}
    \label{fig:om_traj}
\end{subfigure}
\hfill
\begin{subfigure}[b]{0.3\textwidth}
    \centering
    \includegraphics[width=\linewidth]{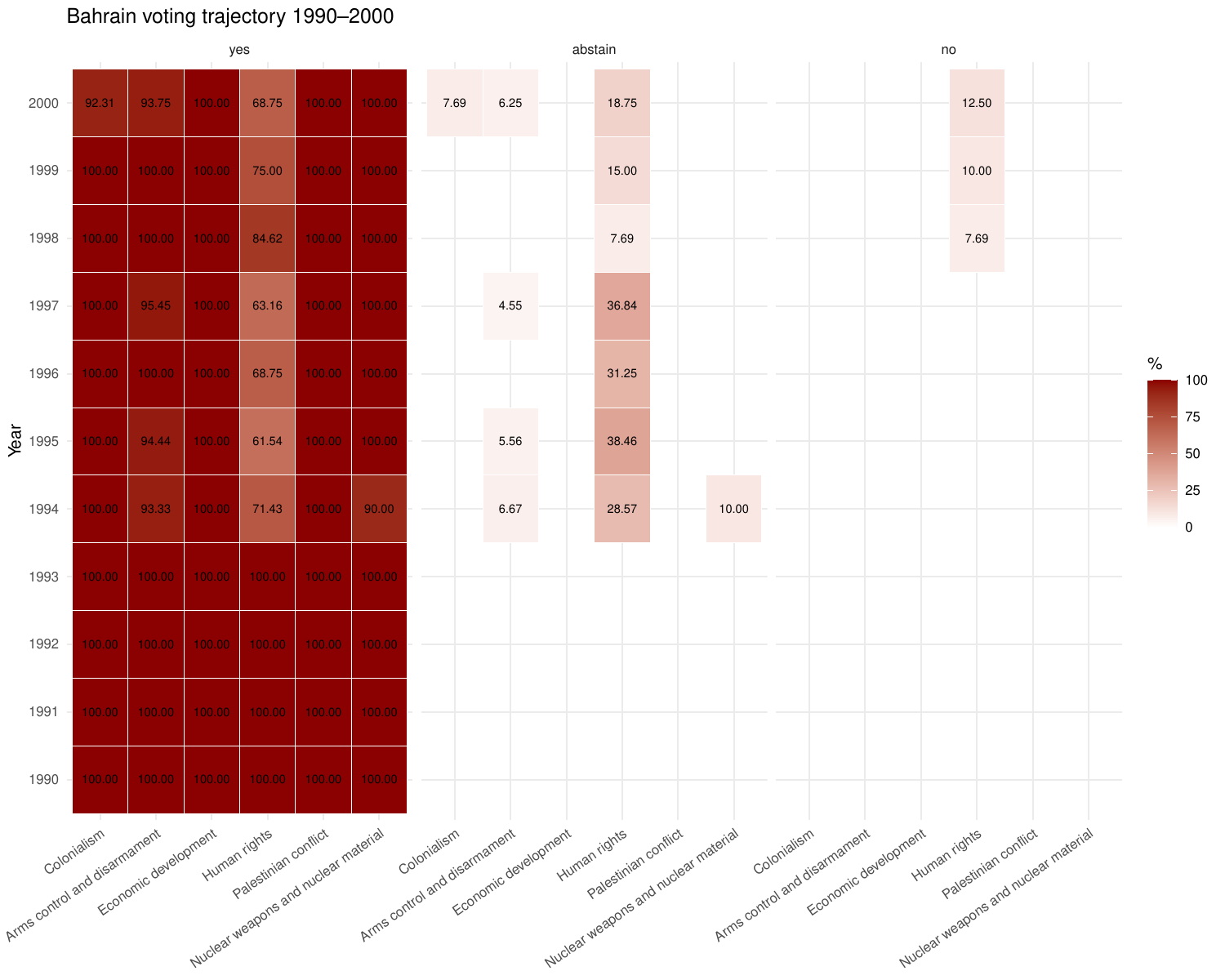}
    \caption{Bahrain}
    \label{fig:bah_traj}
\end{subfigure}
\hfill
\begin{subfigure}[b]{0.3\textwidth}
    \centering
    \includegraphics[width=\linewidth]{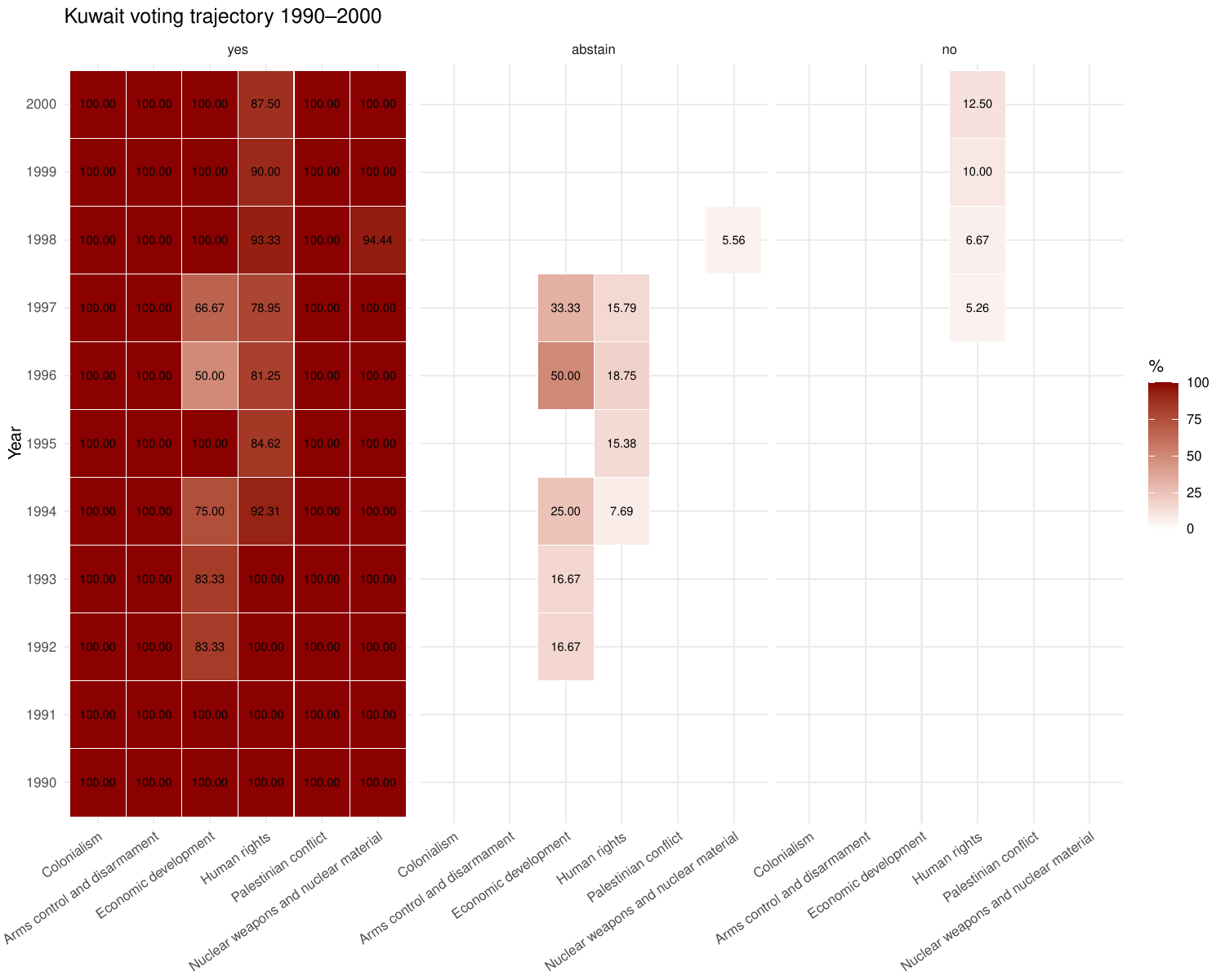}
    \caption{Kuwait}
    \label{fig:ku_traj}
\end{subfigure}

\vspace{0.5cm}

\begin{subfigure}[b]{0.3\textwidth}
    \centering
    \includegraphics[width=\linewidth]{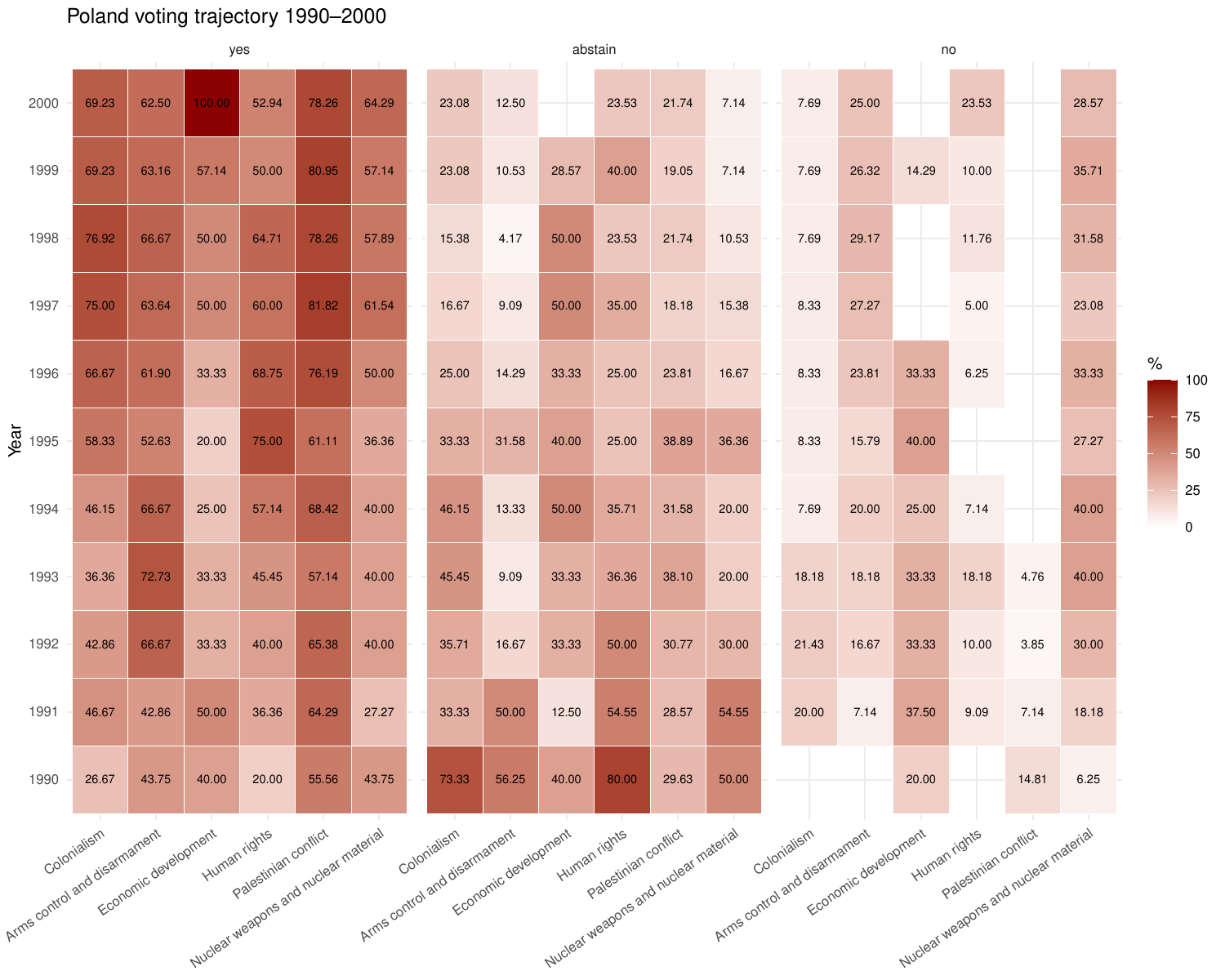}
    \caption{Poland}
    \label{fig:pol_traj}
\end{subfigure}
\hfill
\begin{subfigure}[b]{0.3\textwidth}
    \centering
    \includegraphics[width=\linewidth]{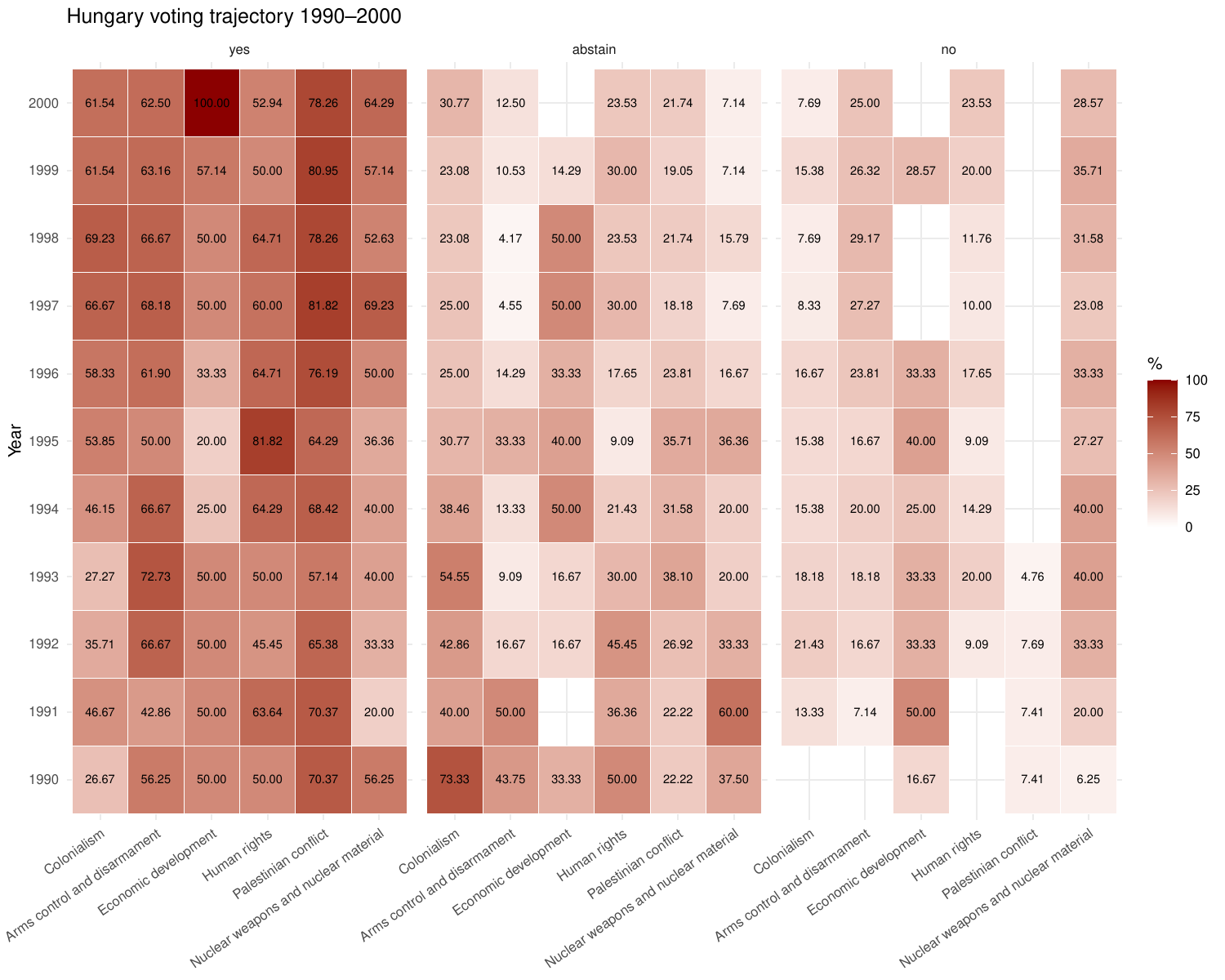}
    \caption{Hungary}
    \label{fig:hung_traj}
\end{subfigure}
\hfill
\begin{subfigure}[b]{0.3\textwidth}
    \centering
    \includegraphics[width=\linewidth]{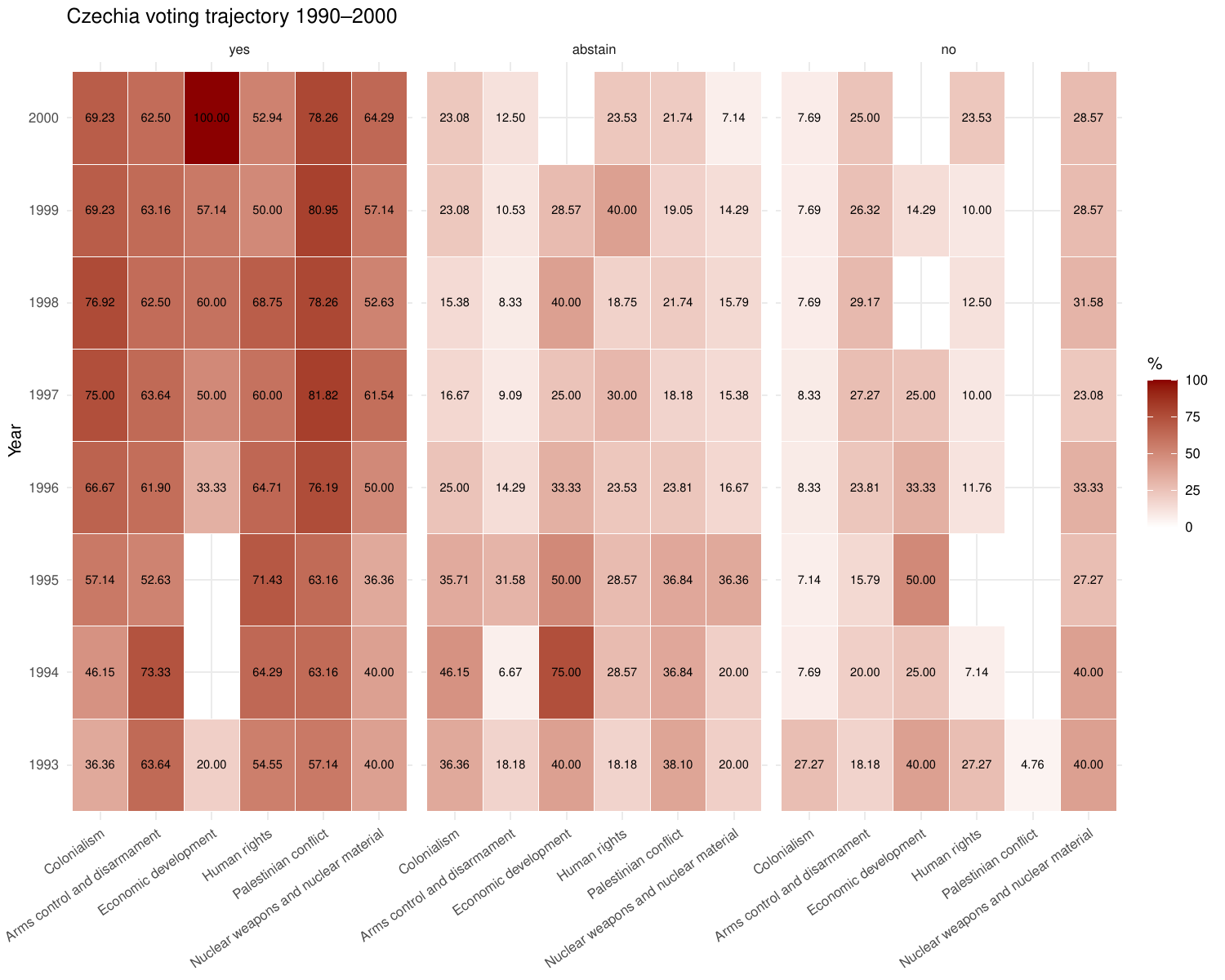}
    \caption{Czechia}
    \label{fig:cz_traj}
\end{subfigure}

\vspace{0.5cm}

\begin{subfigure}[b]{0.3\textwidth}
    \centering
    \includegraphics[width=\linewidth]{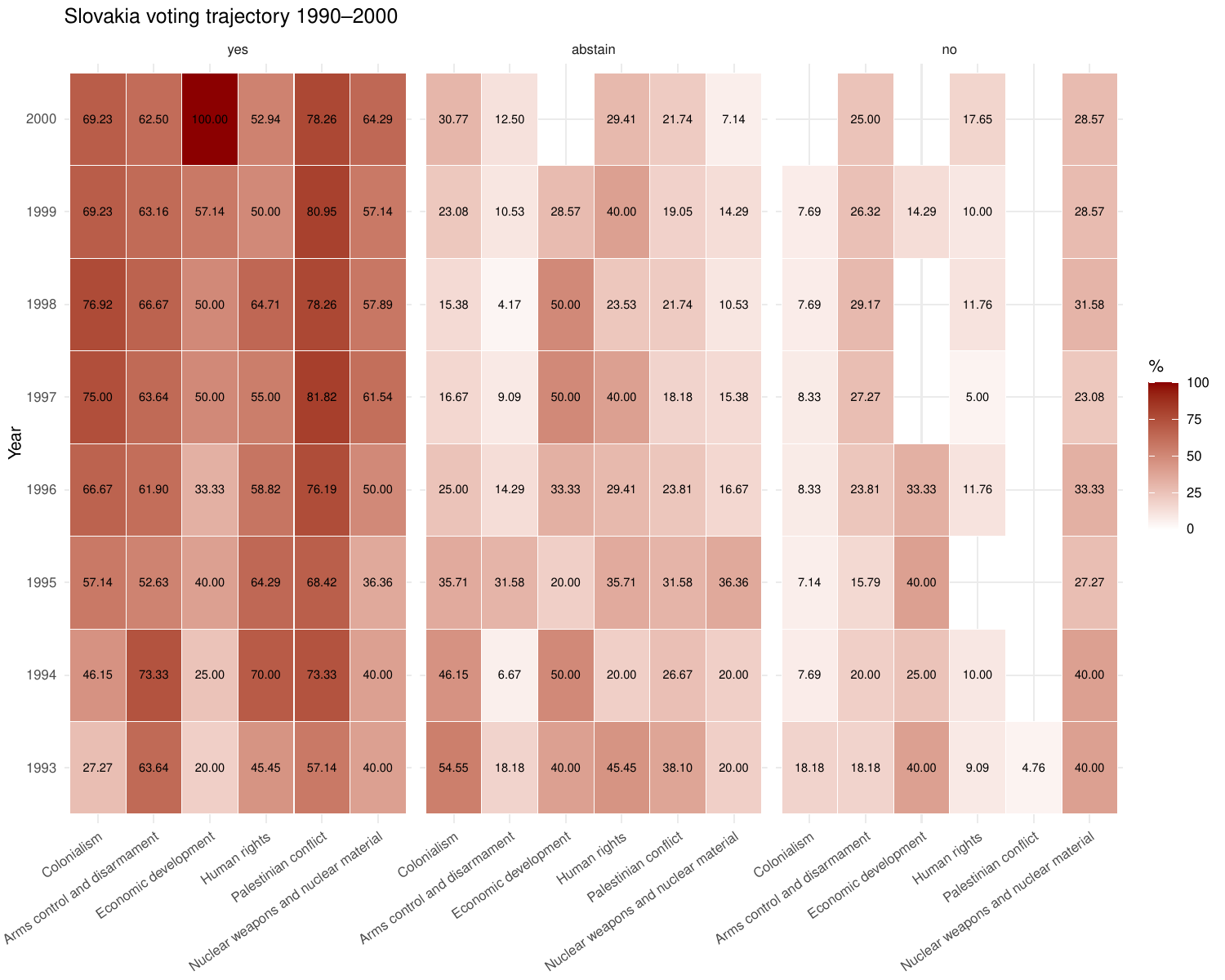}
    \caption{Slovakia}
    \label{fig:slovakia_traj}
\end{subfigure}
\hfill
\begin{subfigure}[b]{0.3\textwidth}
    \centering
    \includegraphics[width=\linewidth]{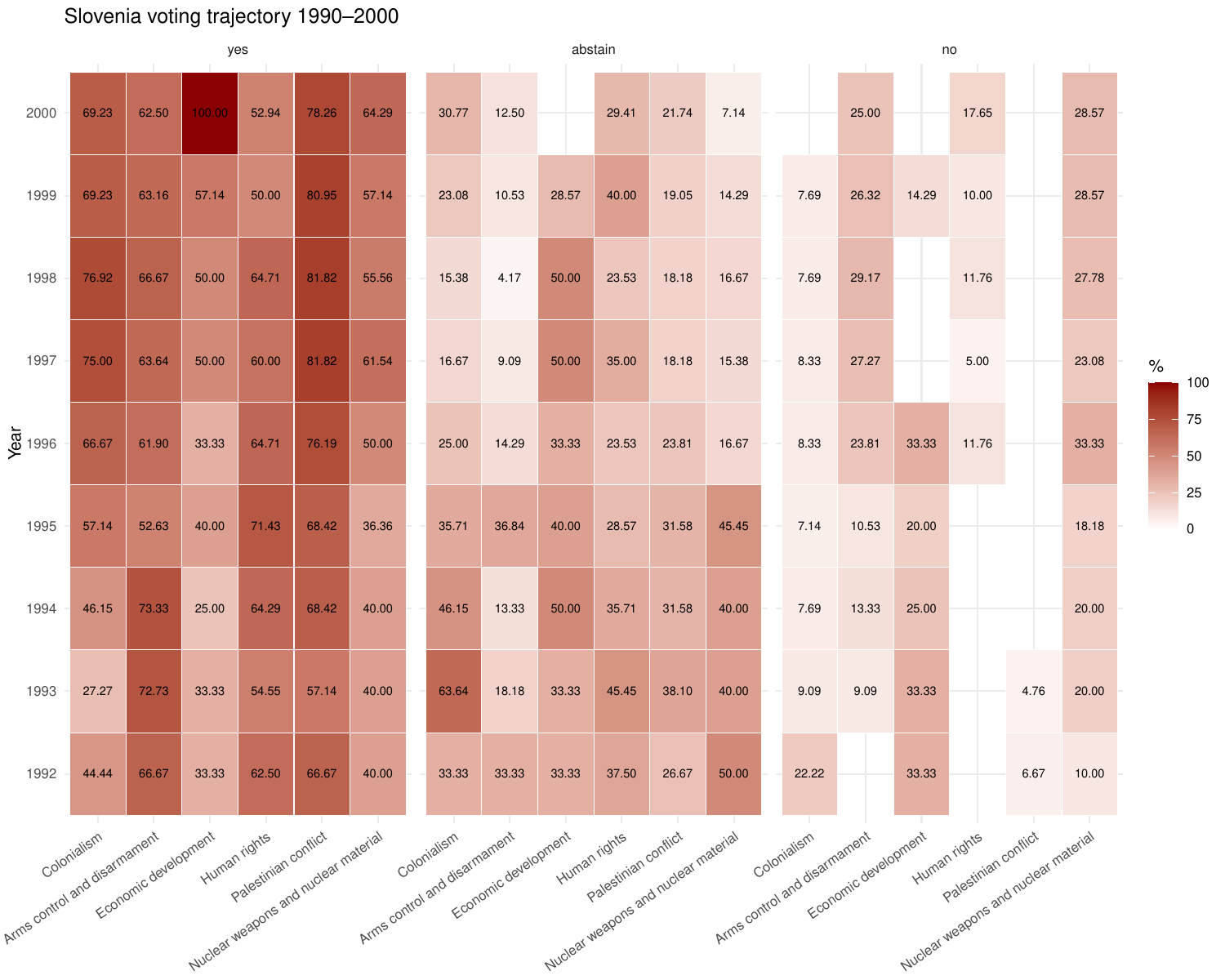}
    \caption{Slovenia}
    \label{fig:slovenia_traj}
\end{subfigure}
\hfill
\begin{subfigure}[b]{0.3\textwidth}
    \centering
    \includegraphics[width=\linewidth]{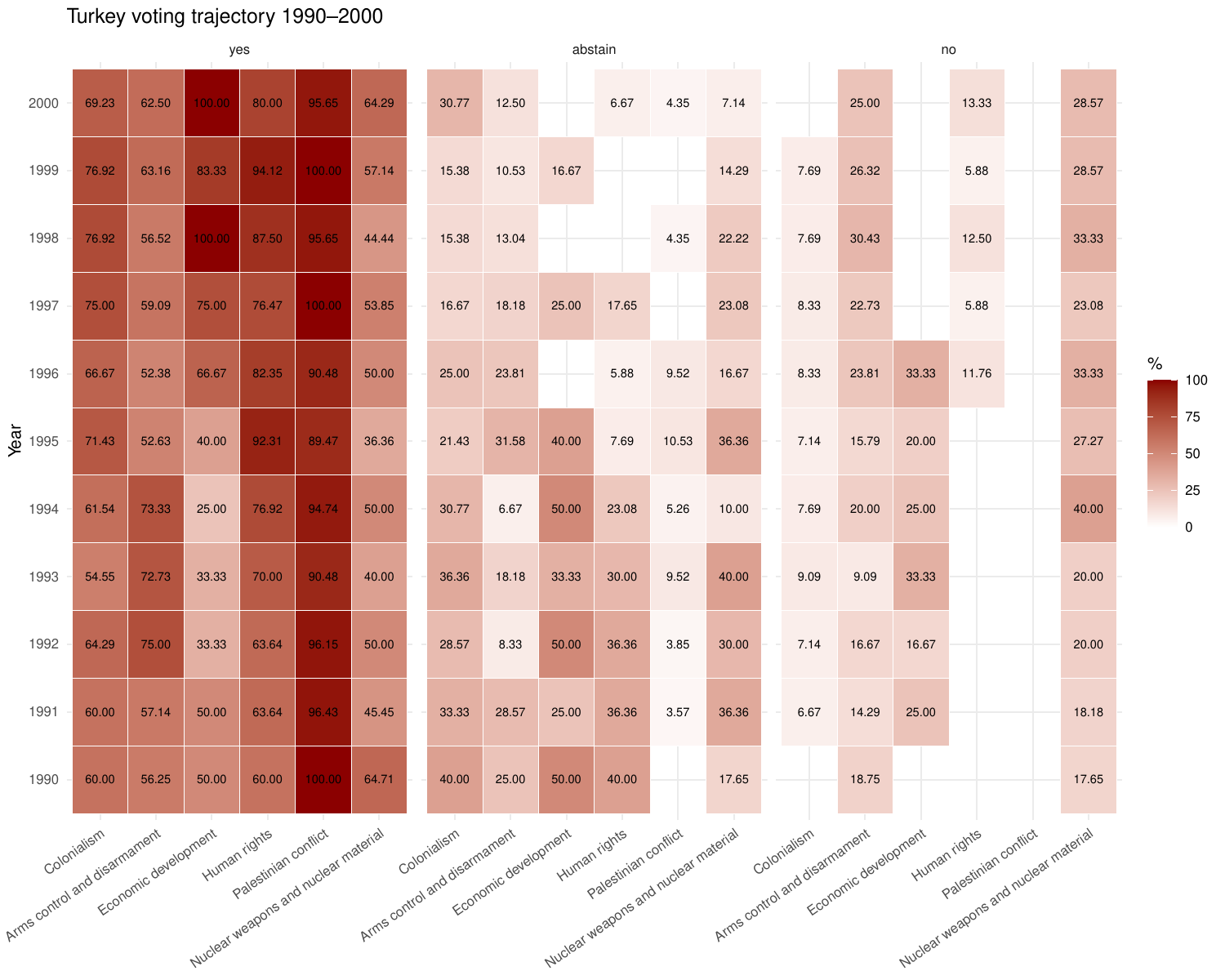}
    \caption{Turkey}
    \label{fig:turkey_traj}
\end{subfigure}
\captionsetup{font=small}
\vspace{-0.02cm}
\captionof{figure}{\footnotesize Heatmap for the trajectory of voting on different issues for selected countries. First row corresponds to USA, France and Israel. Second row contains China, Russia and Cuba. The third and fourth rows are dominated by Gulf countries such as Saudi Arabia, UAE, and Qatar, whereas the last two rows consist primarily of East European countries.}
\label{fig:traj}
\end{center}


\begin{center}
\captionsetup{
type=figure,
font=small,
skip=2pt
}
\centering
\begin{subfigure}[b]{0.3\textwidth}
    \centering
    \includegraphics[width=\linewidth]{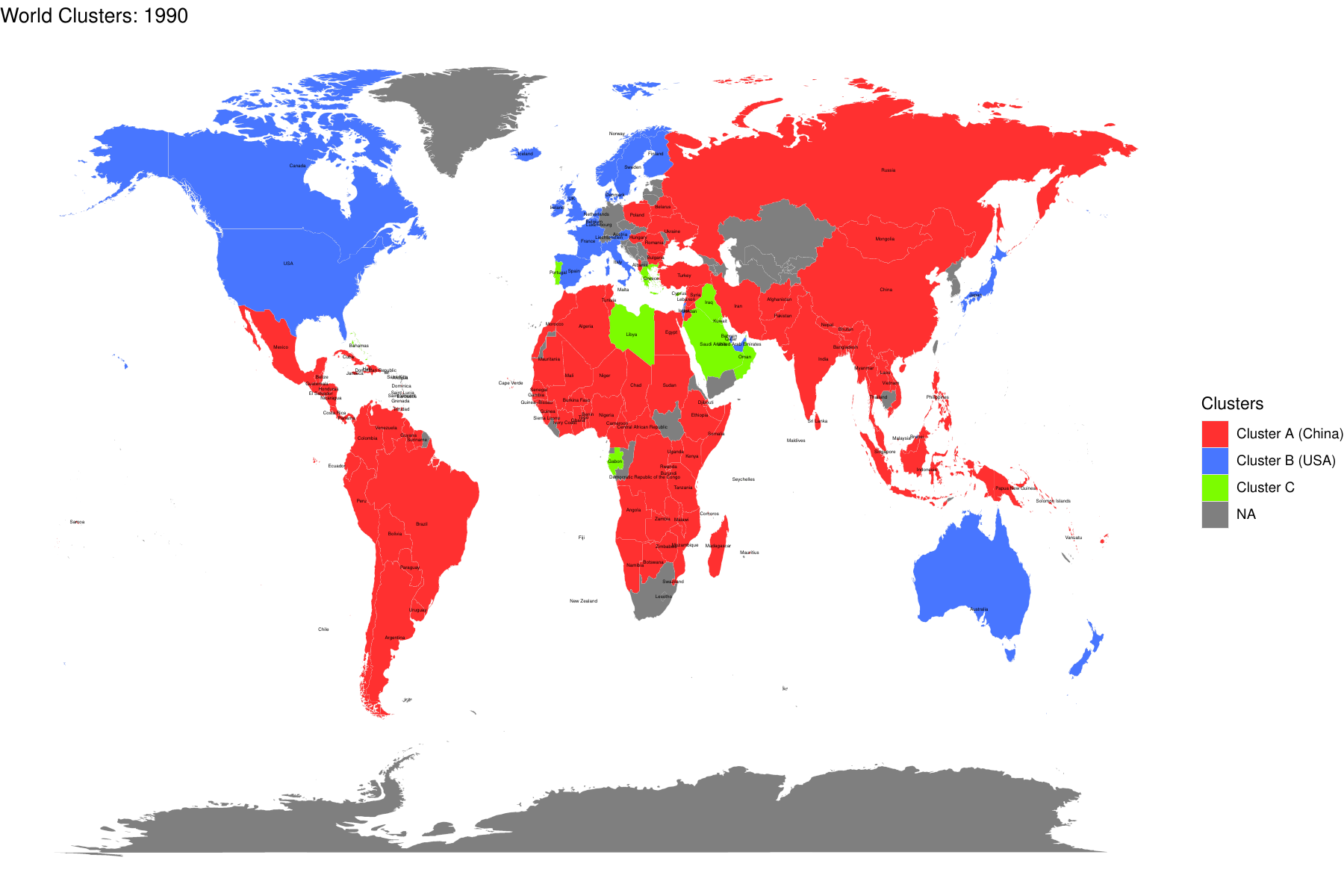}
    \caption{1990}
    \label{fig:wc1990}
\end{subfigure}
\hfill
\begin{subfigure}[b]{0.3\textwidth}
    \centering
    \includegraphics[width=\linewidth]{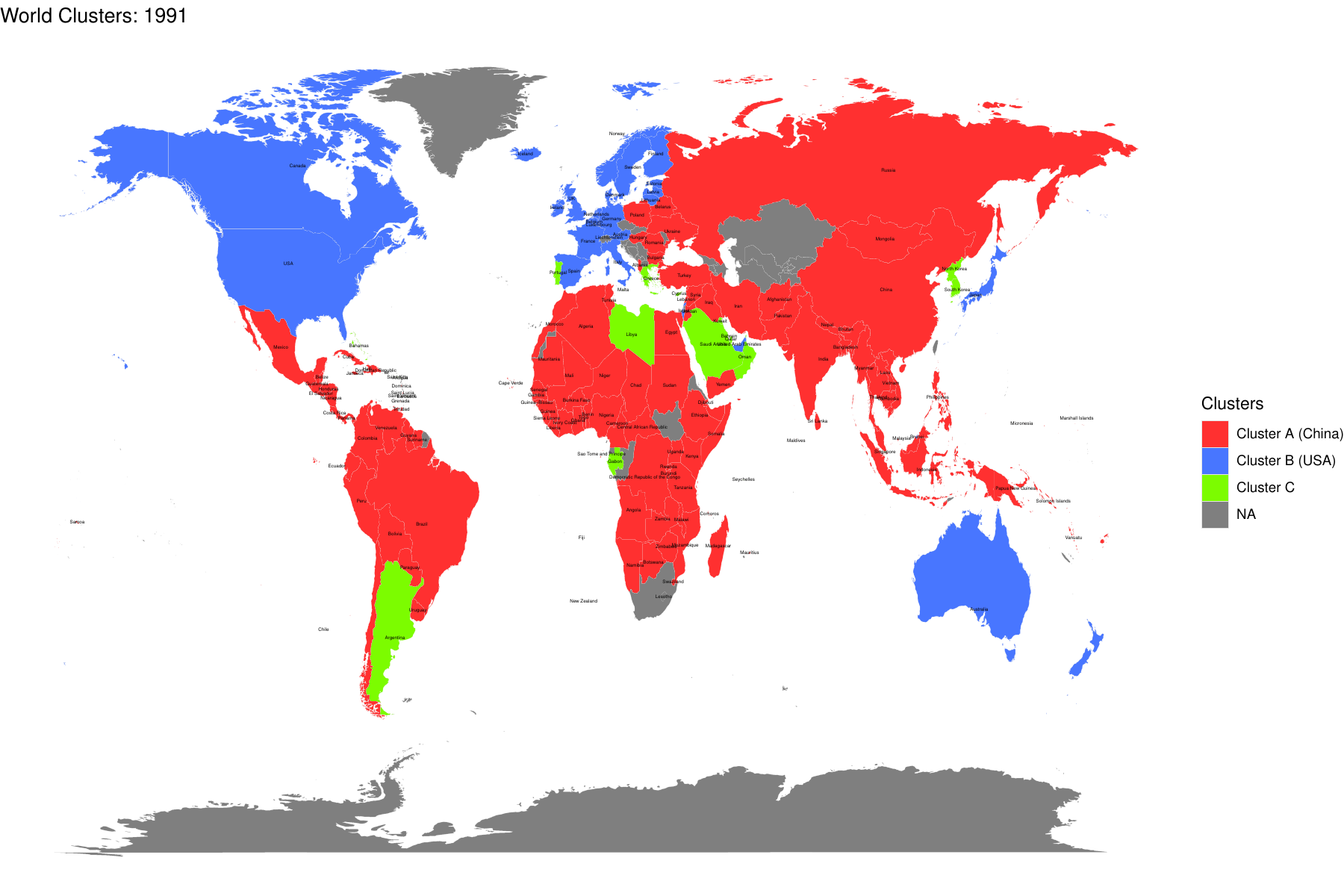}
    \caption{1991}
    \label{fig:wc1991}
\end{subfigure}
\hfill
\begin{subfigure}[b]{0.3\textwidth}
    \centering
    \includegraphics[width=\linewidth]{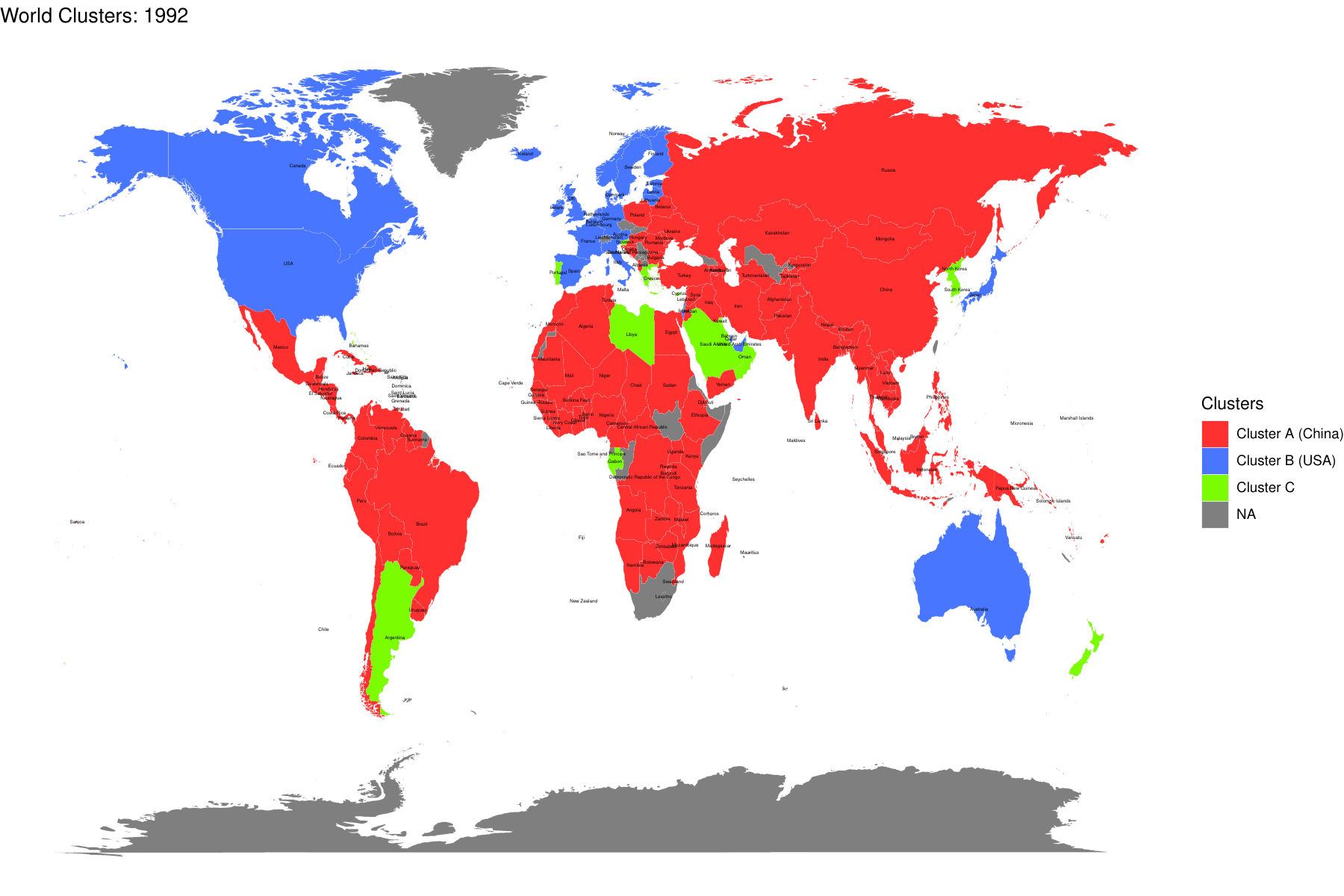}
    \caption{1992}
    \label{fig:wc1992}
\end{subfigure}

\vspace{0.3cm}

\begin{subfigure}[b]{0.3\textwidth}
    \centering
    \includegraphics[width=\linewidth]{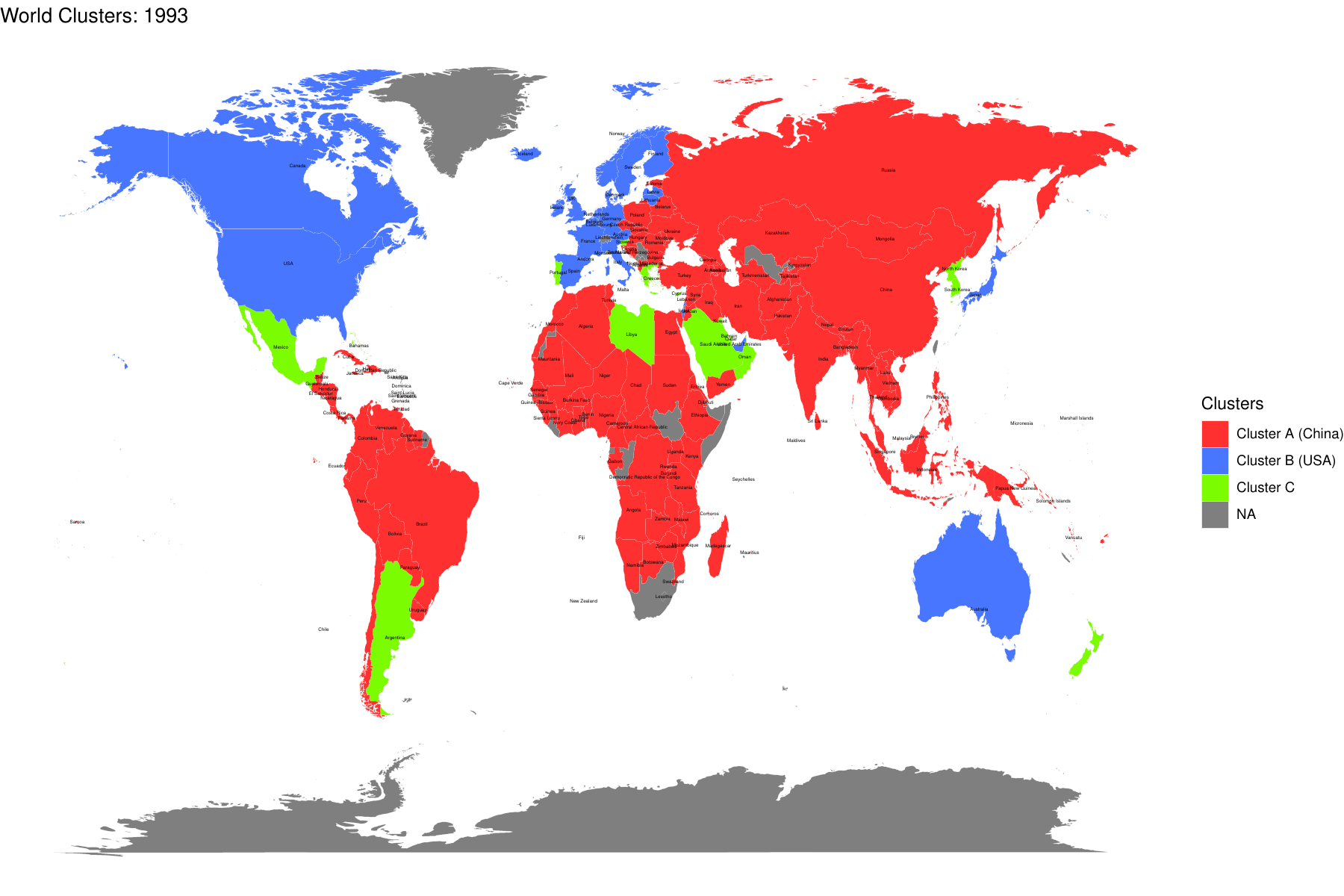}
    \caption{1993}
    \label{fig:wc1993}
\end{subfigure}
\hfill
\begin{subfigure}[b]{0.3\textwidth}
    \centering
    \includegraphics[width=\linewidth]{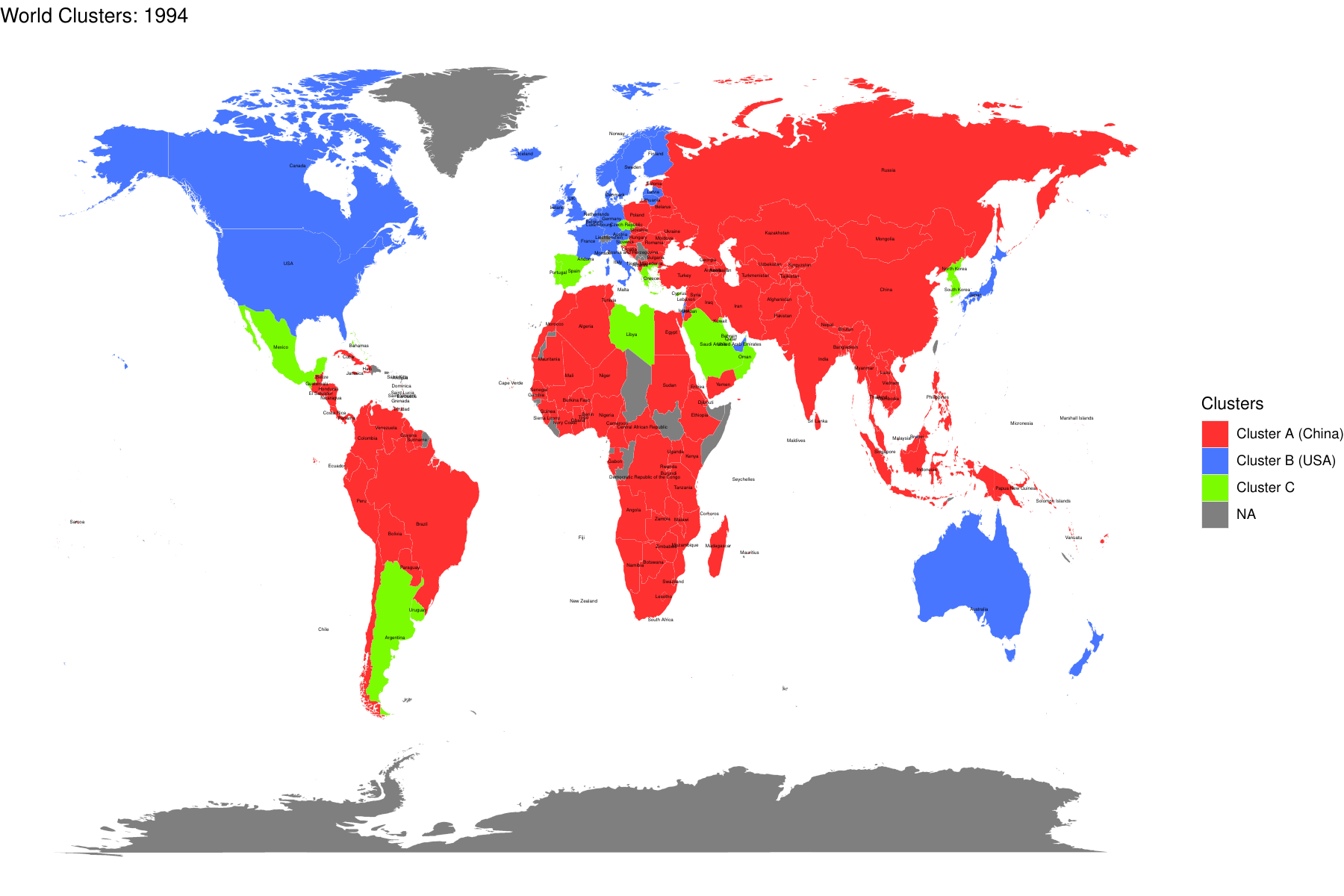}
    \caption{1994}
    \label{fig:wc1994}
\end{subfigure}
\hfill
\begin{subfigure}[b]{0.3\textwidth}
    \centering
    \includegraphics[width=\linewidth]{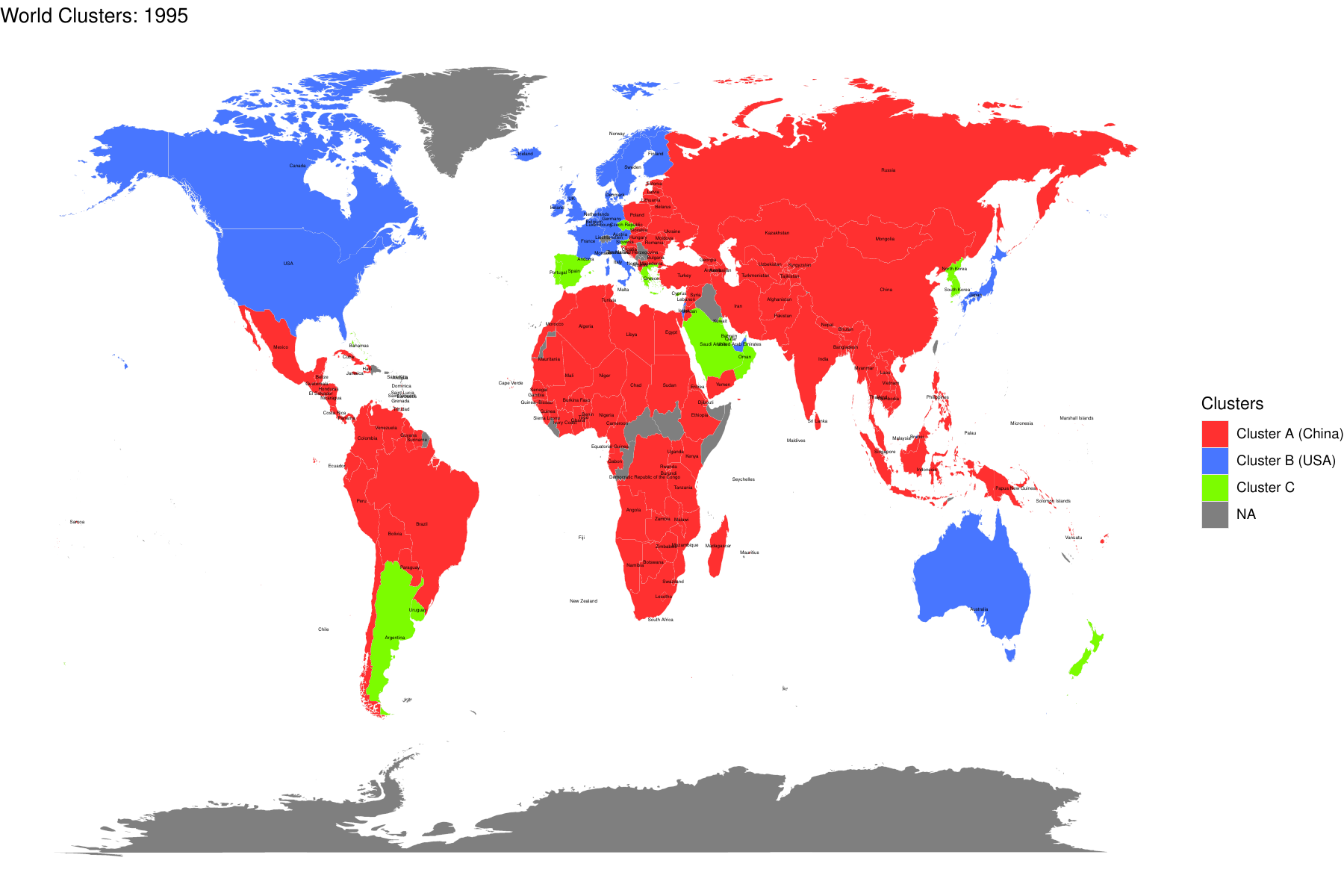}
    \caption{1995}
    \label{fig:wc1995}
\end{subfigure}

\vspace{0.3cm}

\begin{subfigure}[b]{0.3\textwidth}
    \centering
    \includegraphics[width=\linewidth]{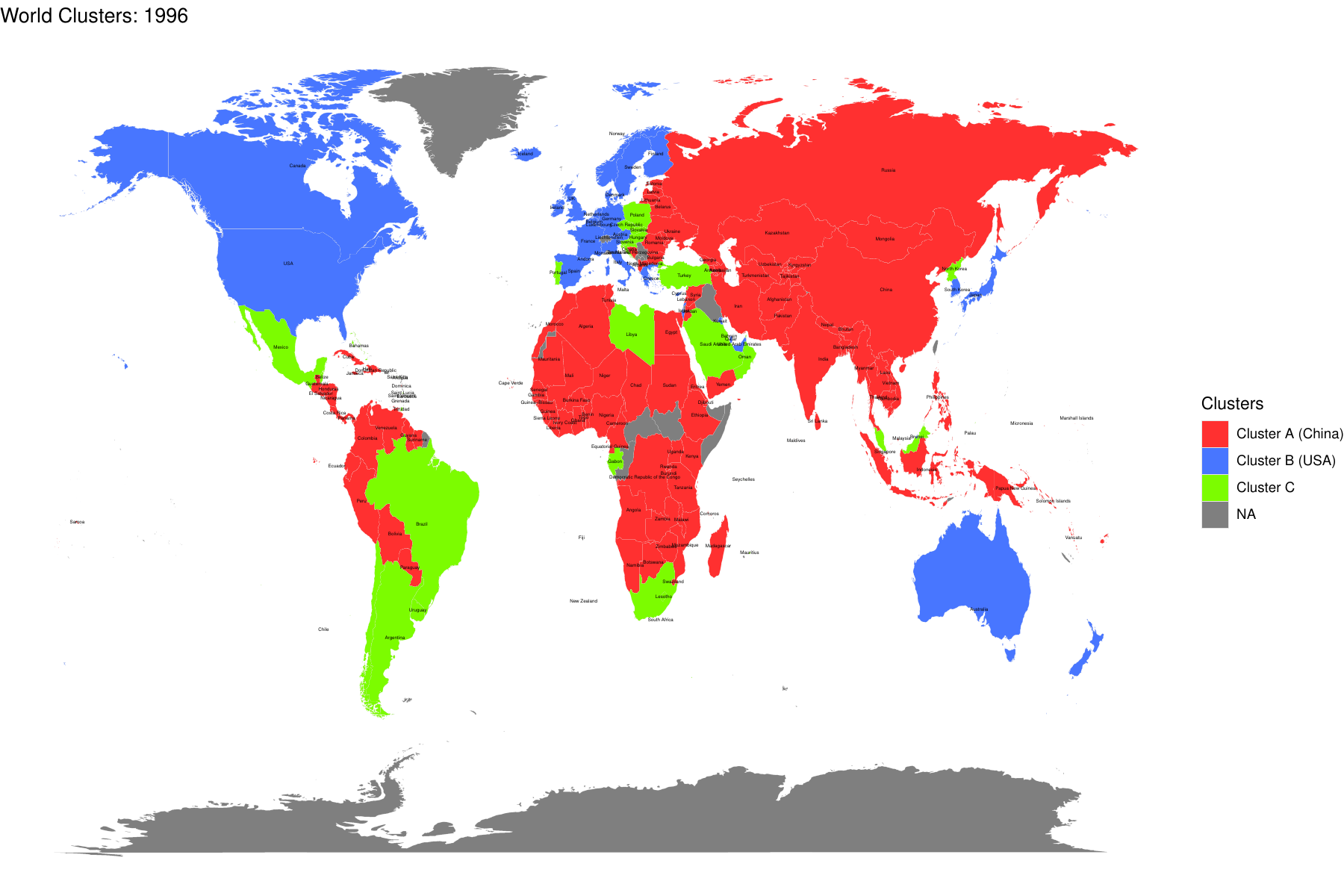}
    \caption{1996}
    \label{fig:wc1996}
\end{subfigure}
\hfill
\begin{subfigure}[b]{0.3\textwidth}
    \centering
    \includegraphics[width=\linewidth]{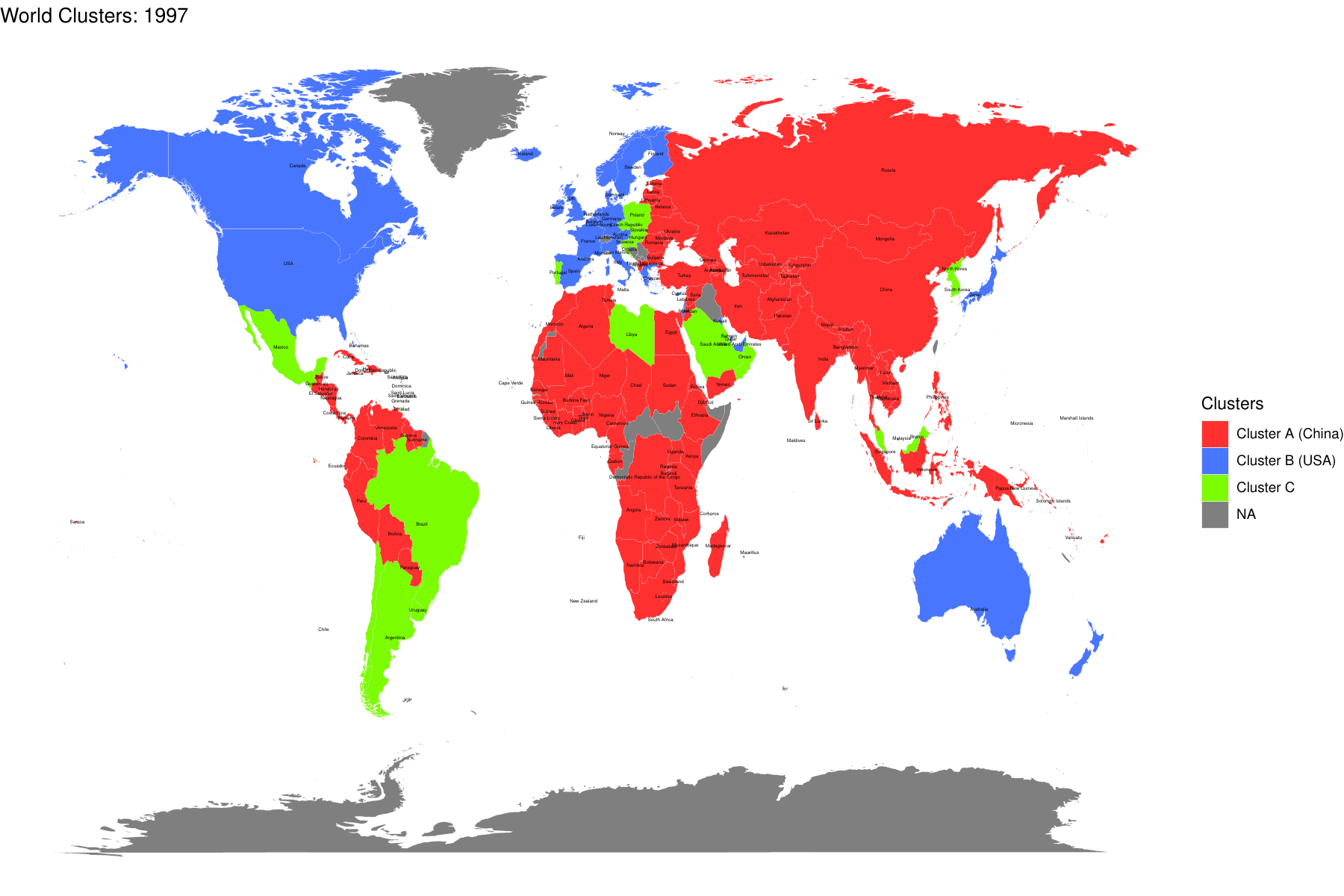}
    \caption{1997}
    \label{fig:wc1997}
\end{subfigure}
\hfill
\begin{subfigure}[b]{0.3\textwidth}
    \centering
    \includegraphics[width=\linewidth]{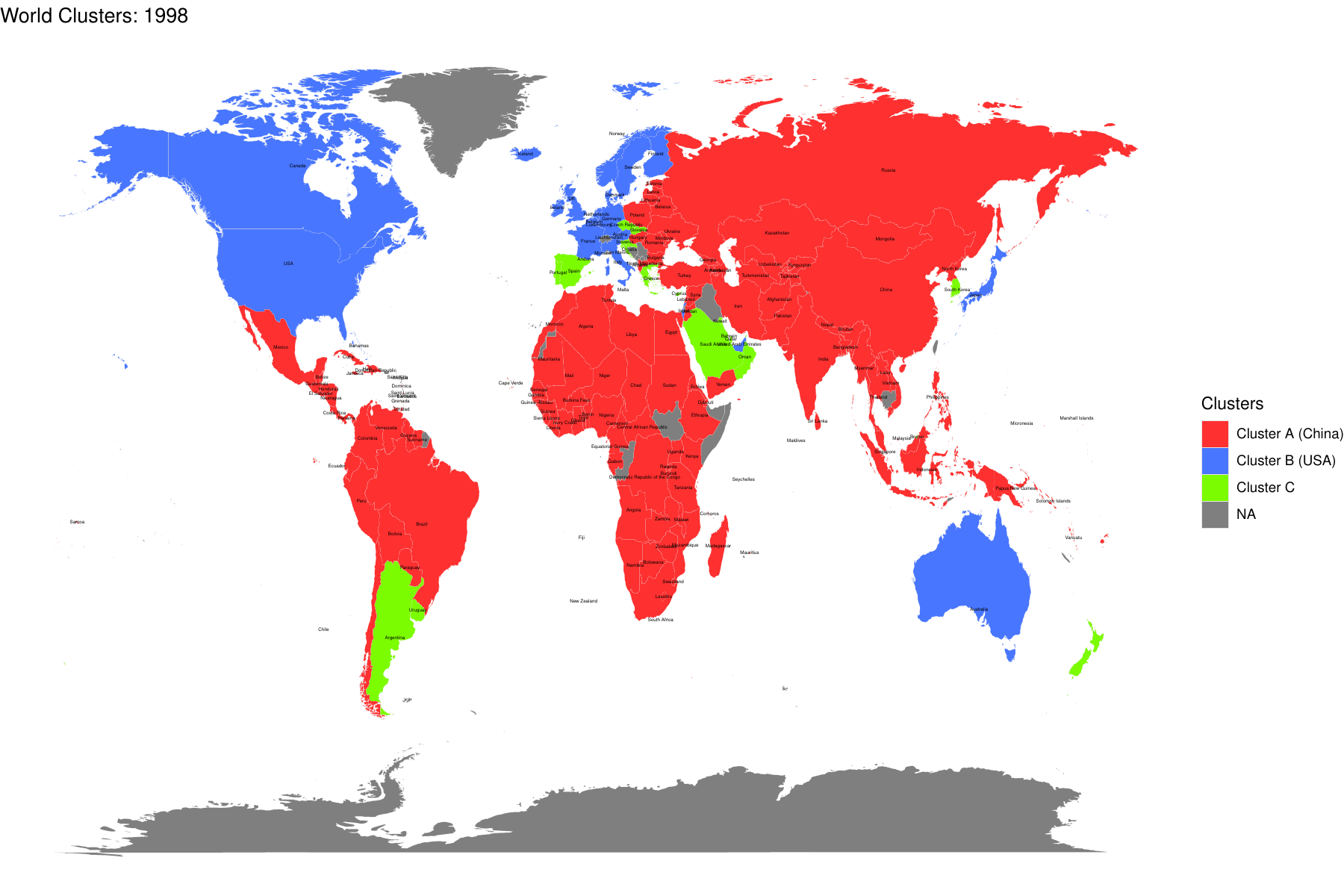}
    \caption{1998}
    \label{fig:wc1998}
\end{subfigure}

\vspace{0.3cm}

\begin{subfigure}[b]{0.3\textwidth}
    \centering
    \includegraphics[width=\linewidth]{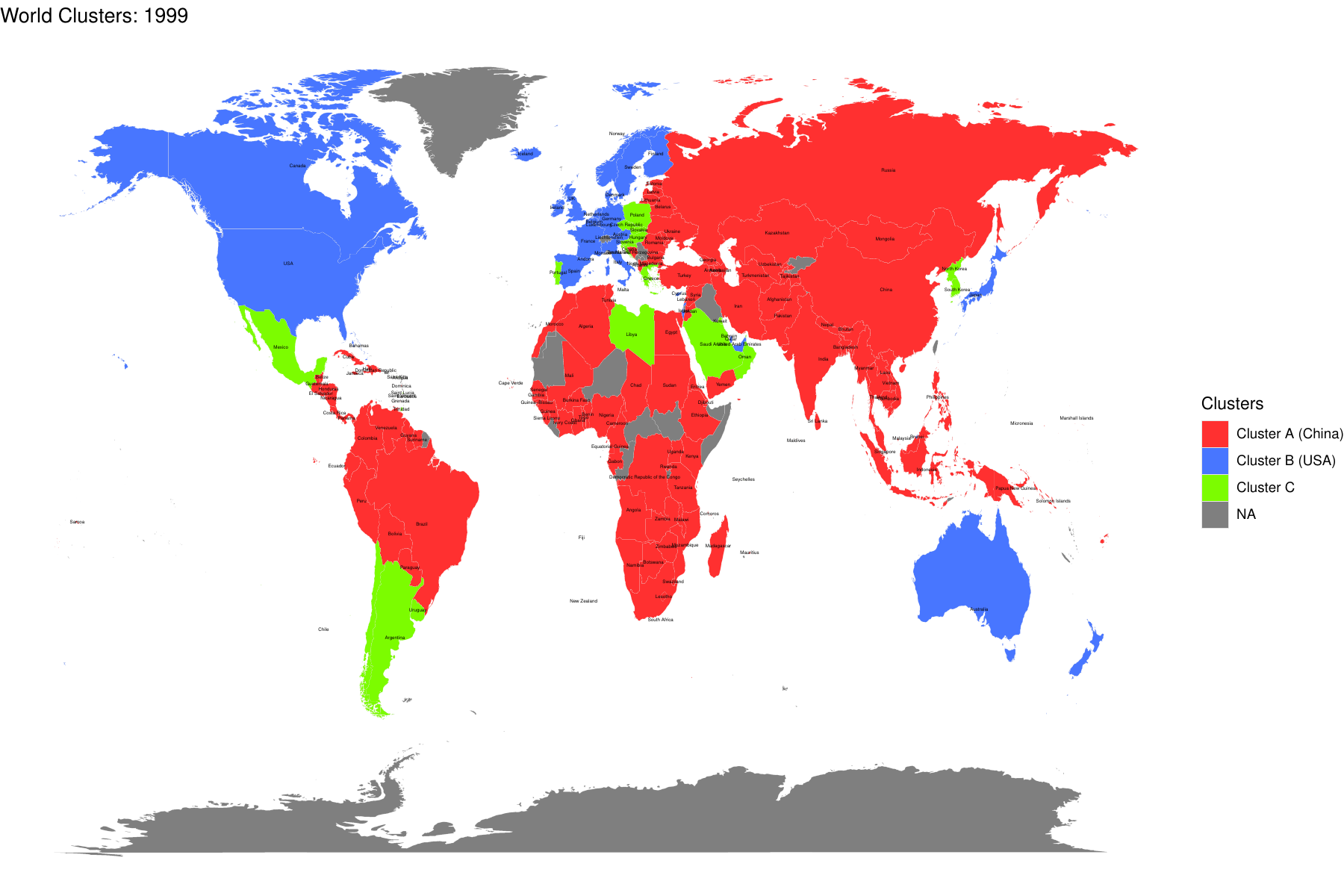}
    \caption{1999}
    \label{fig:wc1999}
\end{subfigure}
\hfill
\begin{subfigure}[b]{0.3\textwidth}
    \centering
    \includegraphics[width=\linewidth]{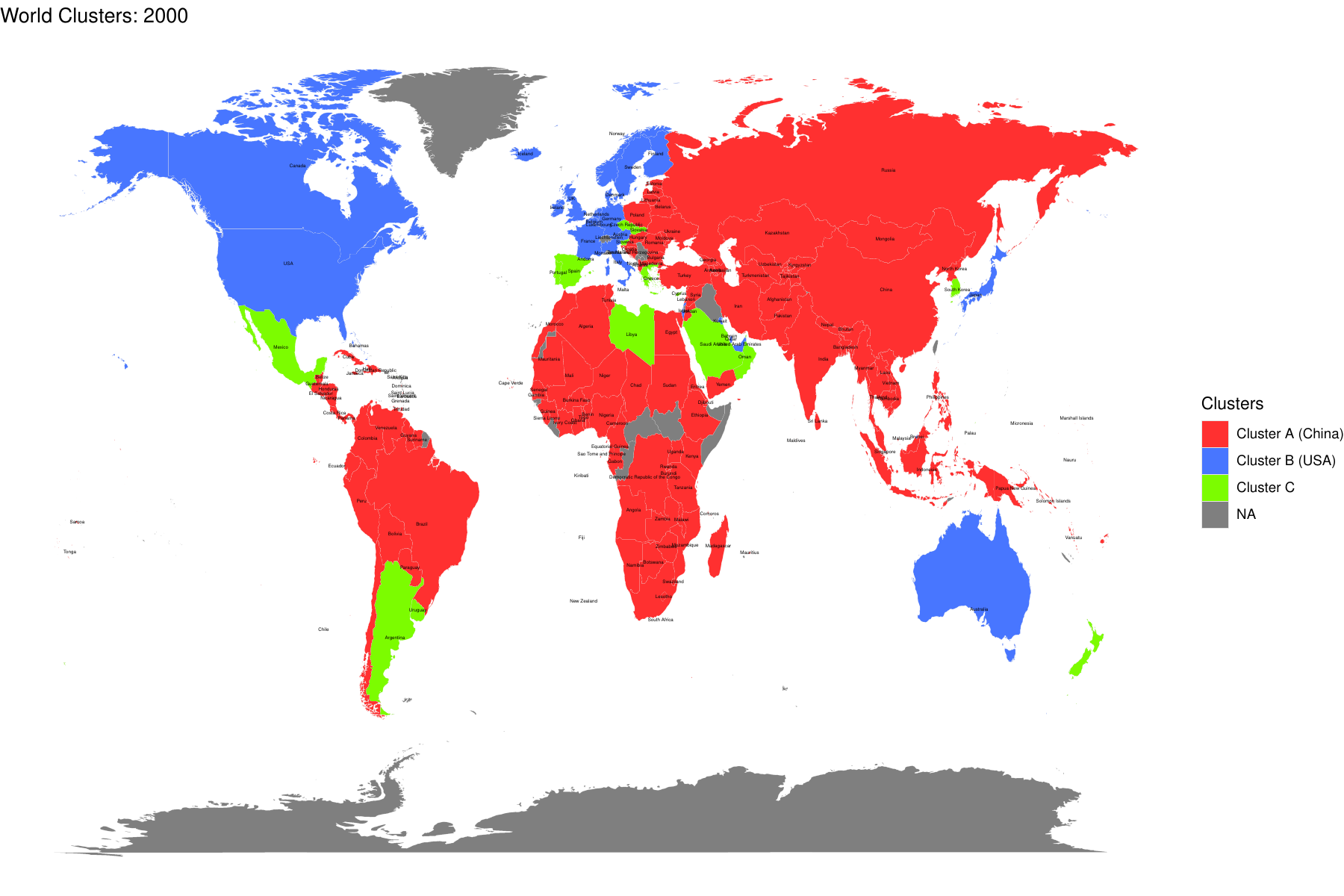}
    \caption{2000}
    \label{fig:wc2000}
\end{subfigure}

\captionof{figure}{\footnotesize Temporal evolution of world clusters from 1990-2000 using Het-Cov clustering with GDP per capita as covariate.}
\label{fig:world_clusters_hetcov}
\end{center}
\begin{center}
\captionsetup{
type=figure,
font=small,
skip=2pt
}
\centering
\begin{subfigure}[b]{0.3\textwidth}
    \centering
    \includegraphics[width=\linewidth]{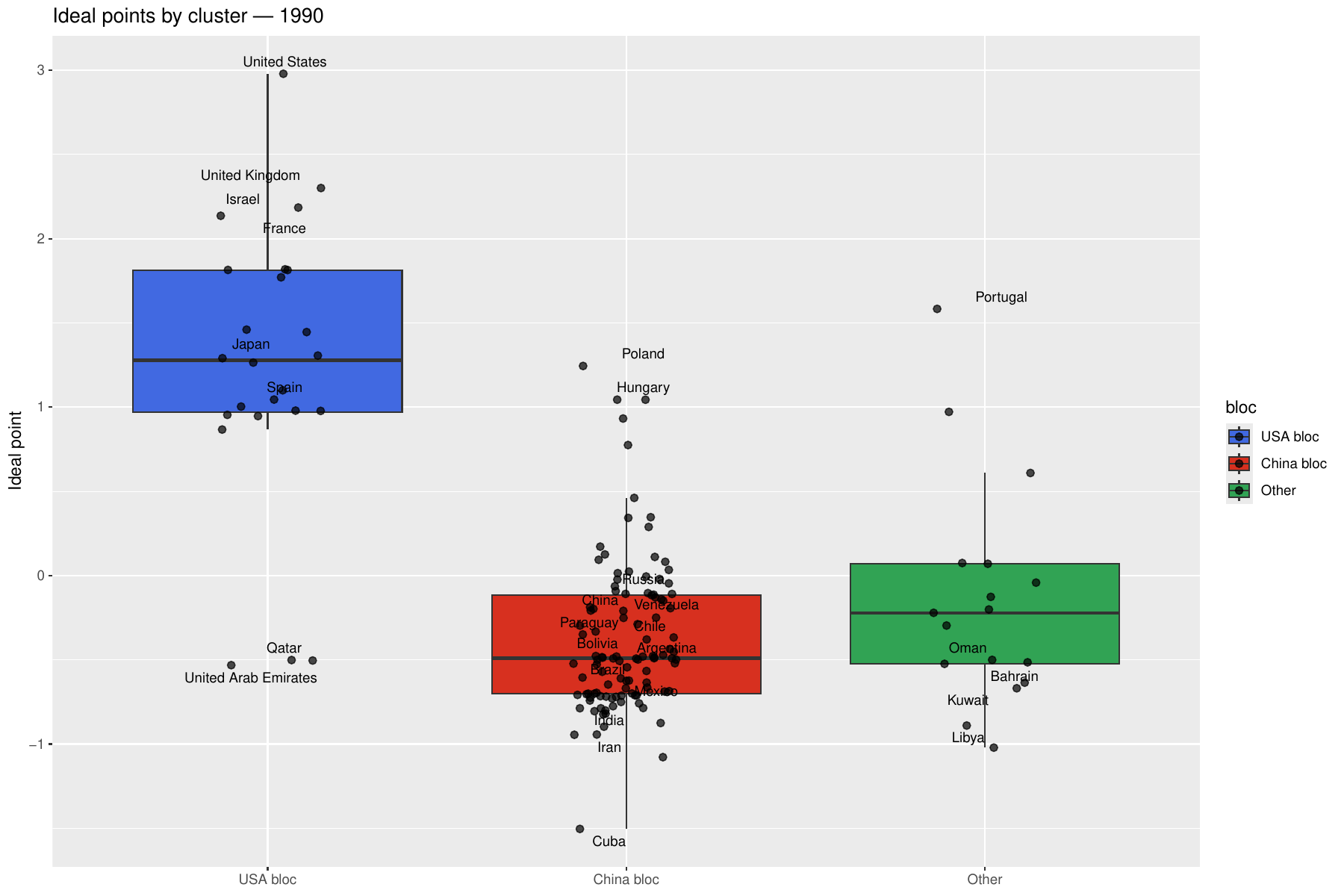}
    \caption{1990}
    \label{fig:bp1990}
\end{subfigure}
\hfill
\begin{subfigure}[b]{0.3\textwidth}
    \centering
    \includegraphics[width=\linewidth]{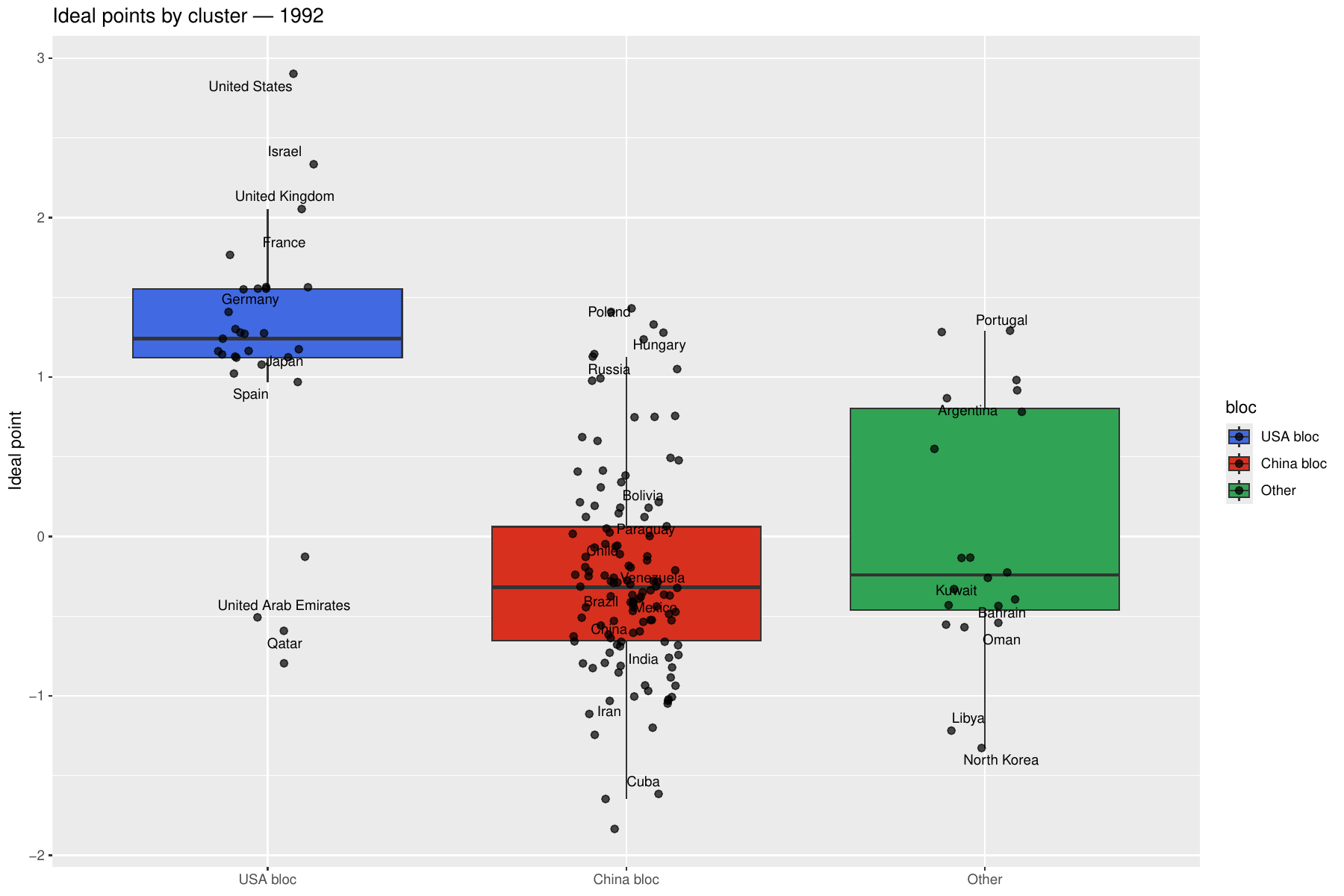}
    \caption{1992}
    \label{fig:bp1992}
\end{subfigure}
\hfill
\begin{subfigure}[b]{0.3\textwidth}
    \centering
    \includegraphics[width=\linewidth]{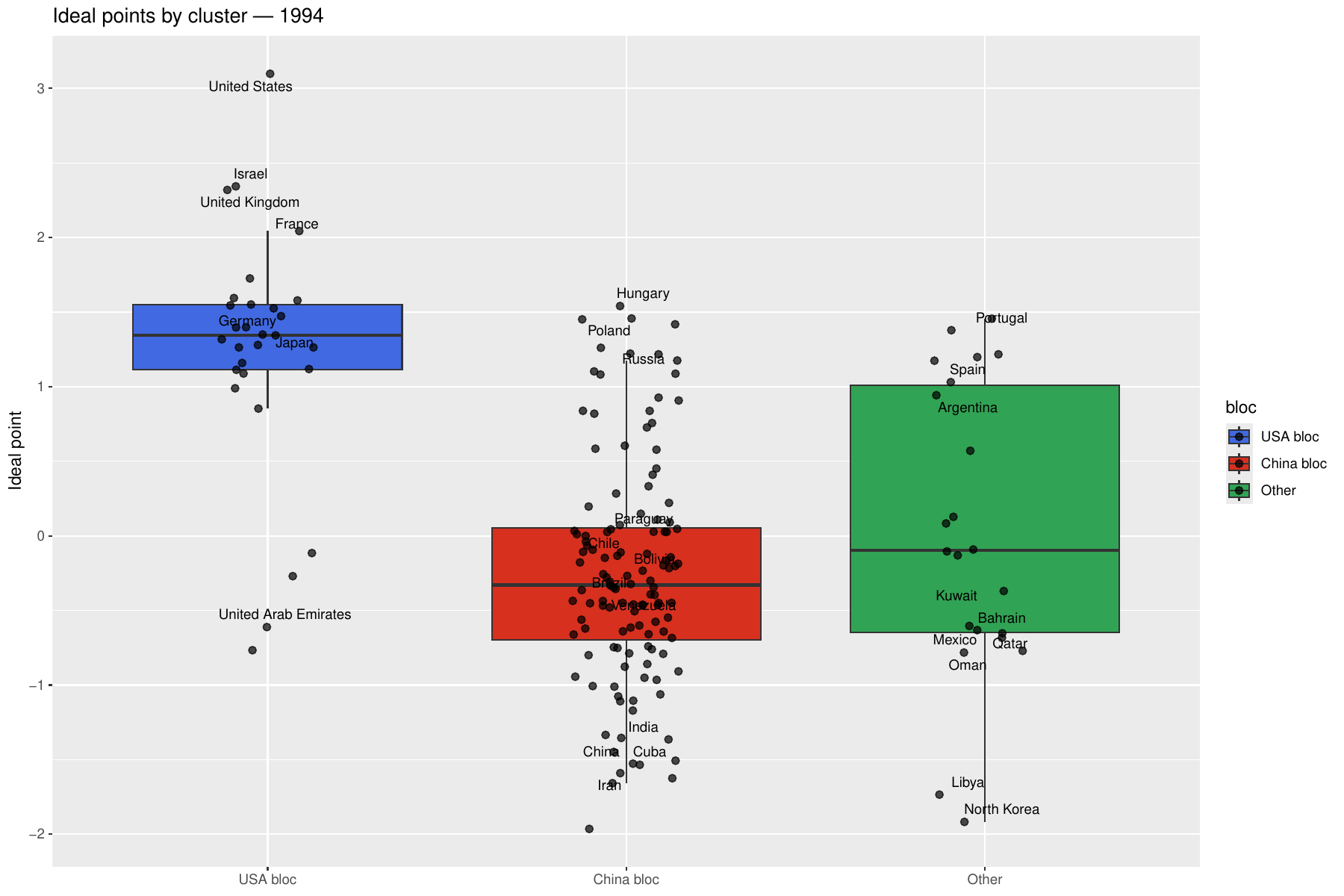}
    \caption{1994}
    \label{fig:bp1994}
\end{subfigure}

\vspace{0.3cm}

\begin{subfigure}[b]{0.3\textwidth}
    \centering
    \includegraphics[width=\linewidth]{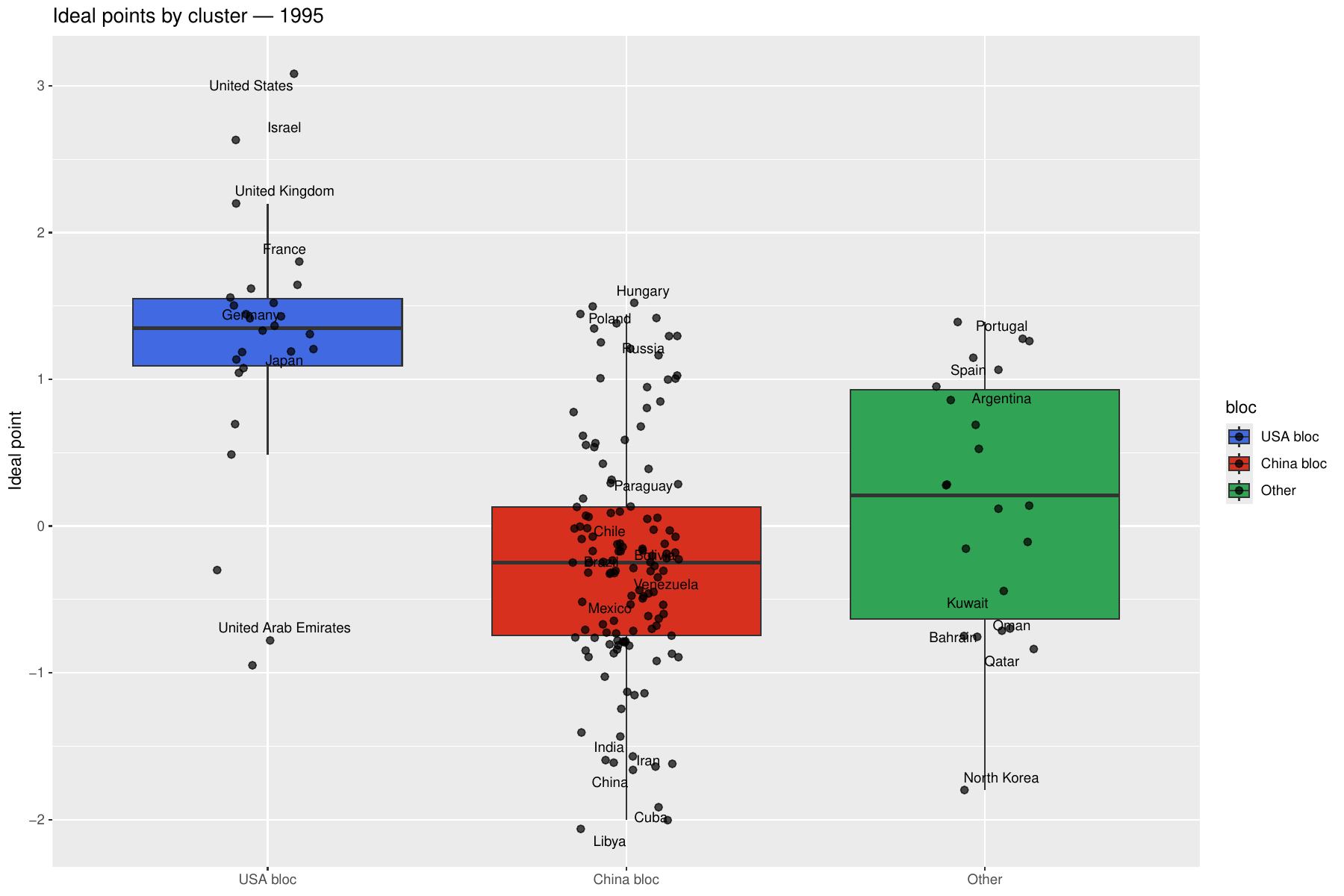}
    \caption{1995}
    \label{fig:bp1995}
\end{subfigure}
\hfill
\begin{subfigure}[b]{0.3\textwidth}
    \centering
    \includegraphics[width=\linewidth]{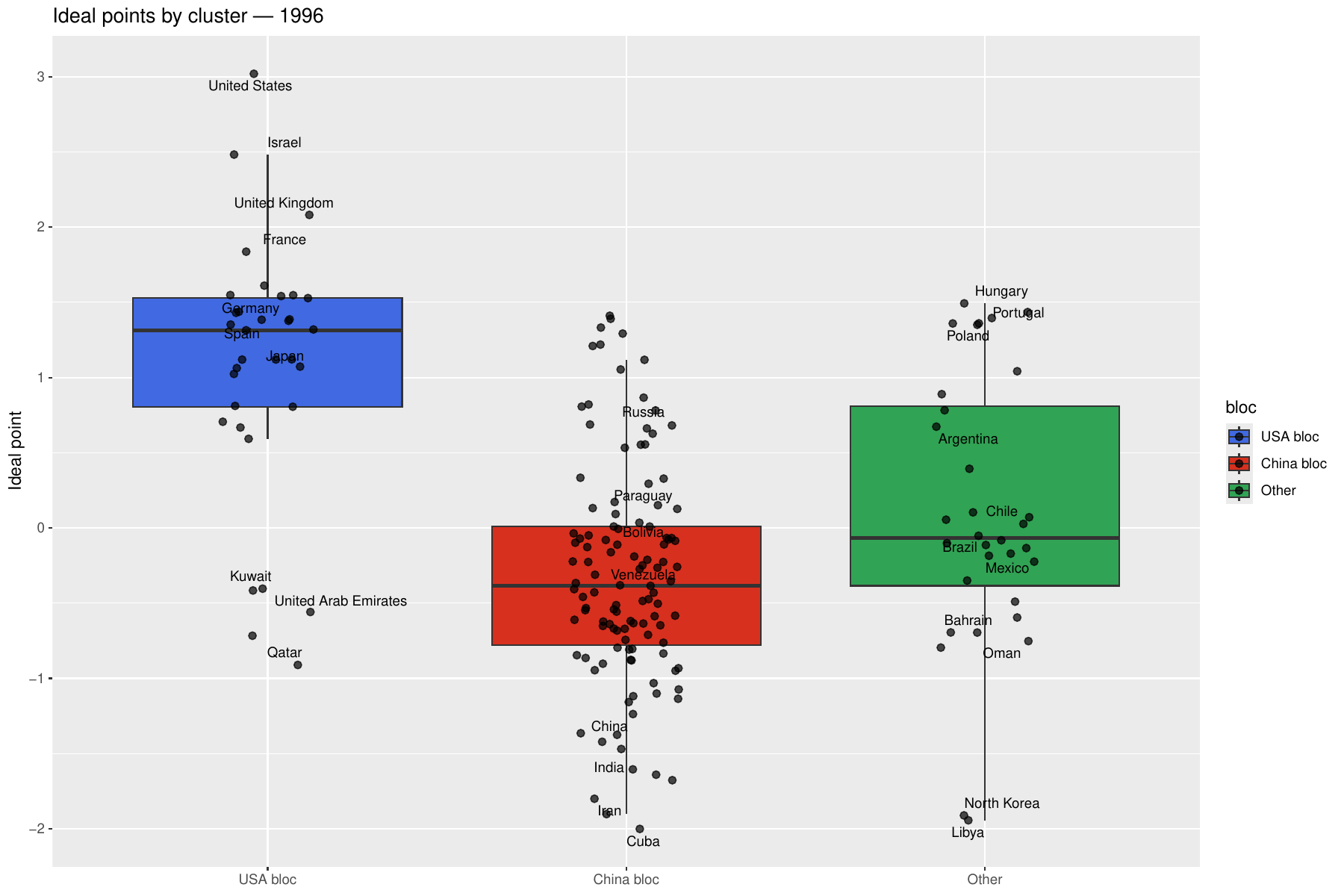}
    \caption{1996}
    \label{fig:bp1996}
\end{subfigure}
\hfill
\begin{subfigure}[b]{0.3\textwidth}
    \centering
    \includegraphics[width=\linewidth]{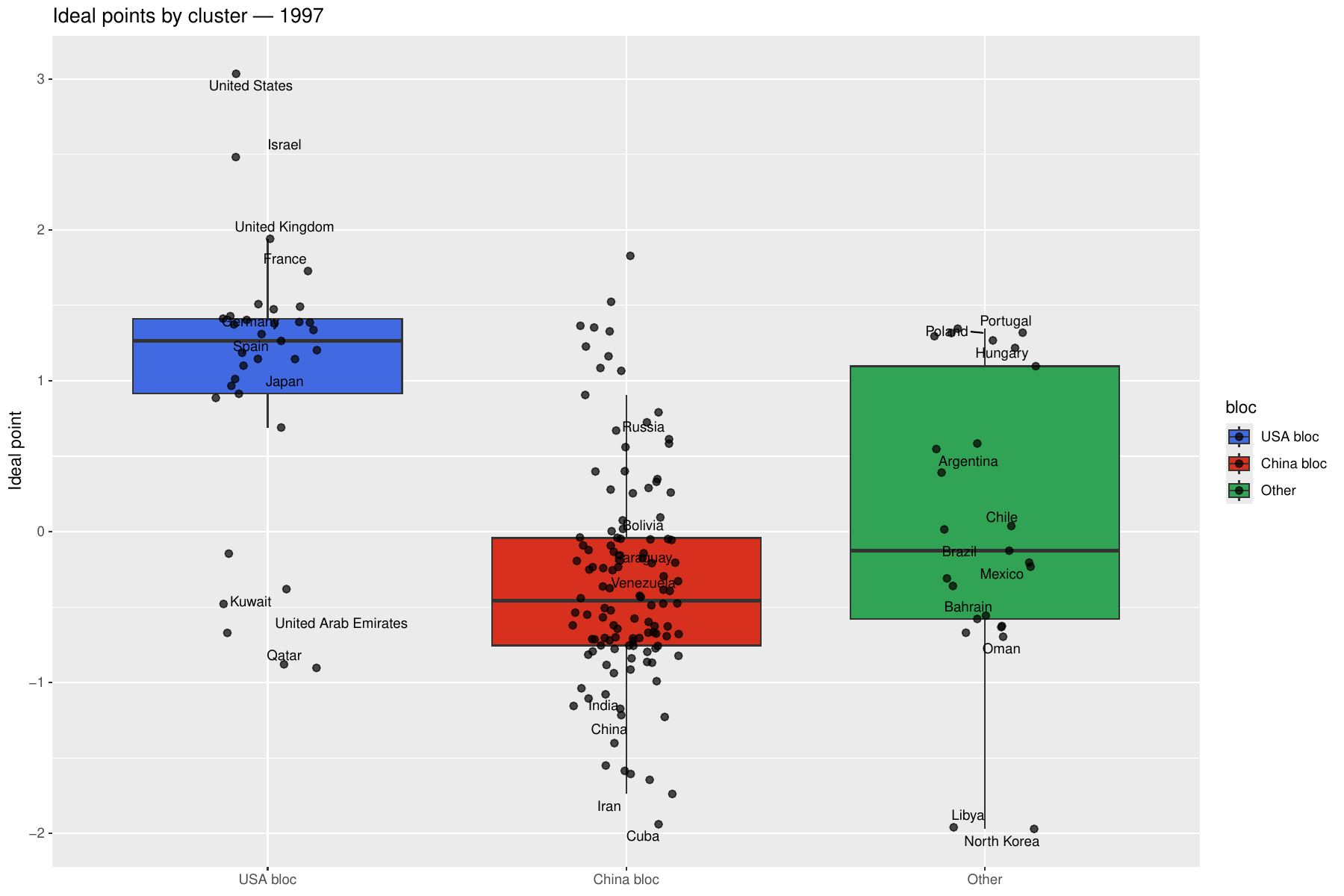}
    \caption{1997}
    \label{fig:bp1997}
\end{subfigure}

\vspace{0.3cm}
\begin{subfigure}[b]{0.3\textwidth}
    \centering
    \includegraphics[width=\linewidth]{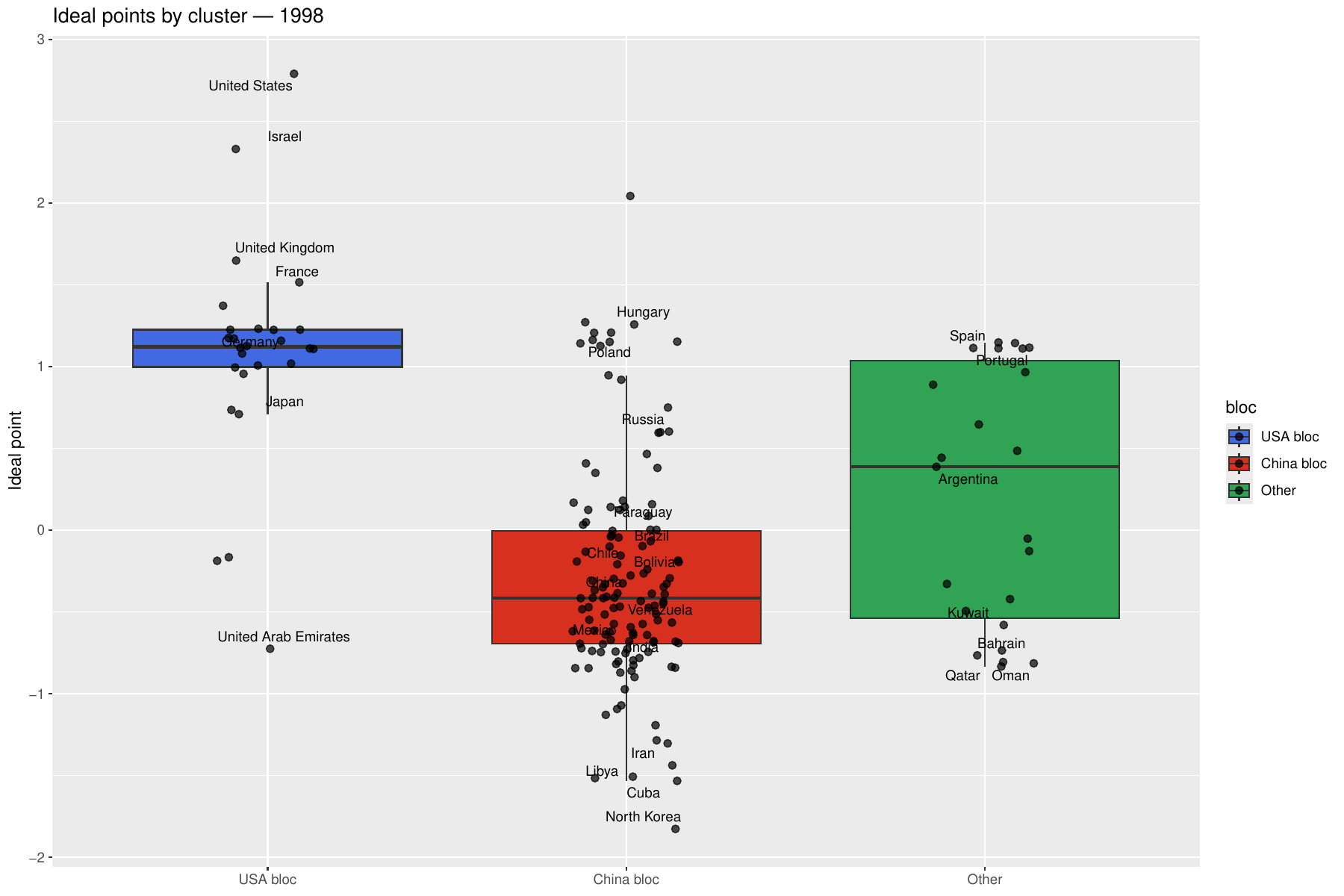}
    \caption{1998}
    \label{fig:bp1998}
\end{subfigure}
\hfill
\begin{subfigure}[b]{0.3\textwidth}
    \centering
    \includegraphics[width=\linewidth]{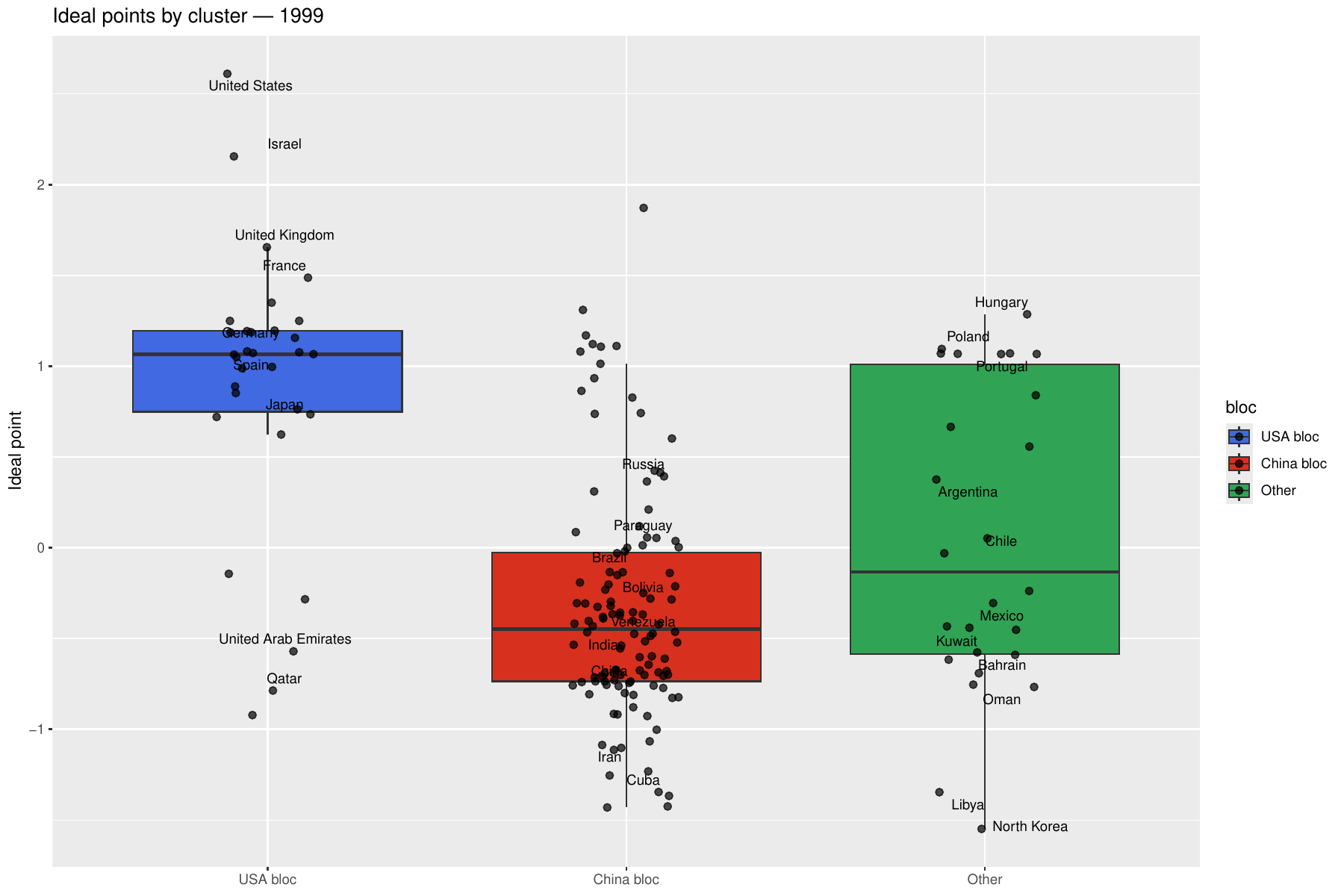}
    \caption{1999}
    \label{fig:bp1999}
\end{subfigure}
\hfill
\begin{subfigure}[b]{0.3\textwidth}
    \centering
    \includegraphics[width=\linewidth]{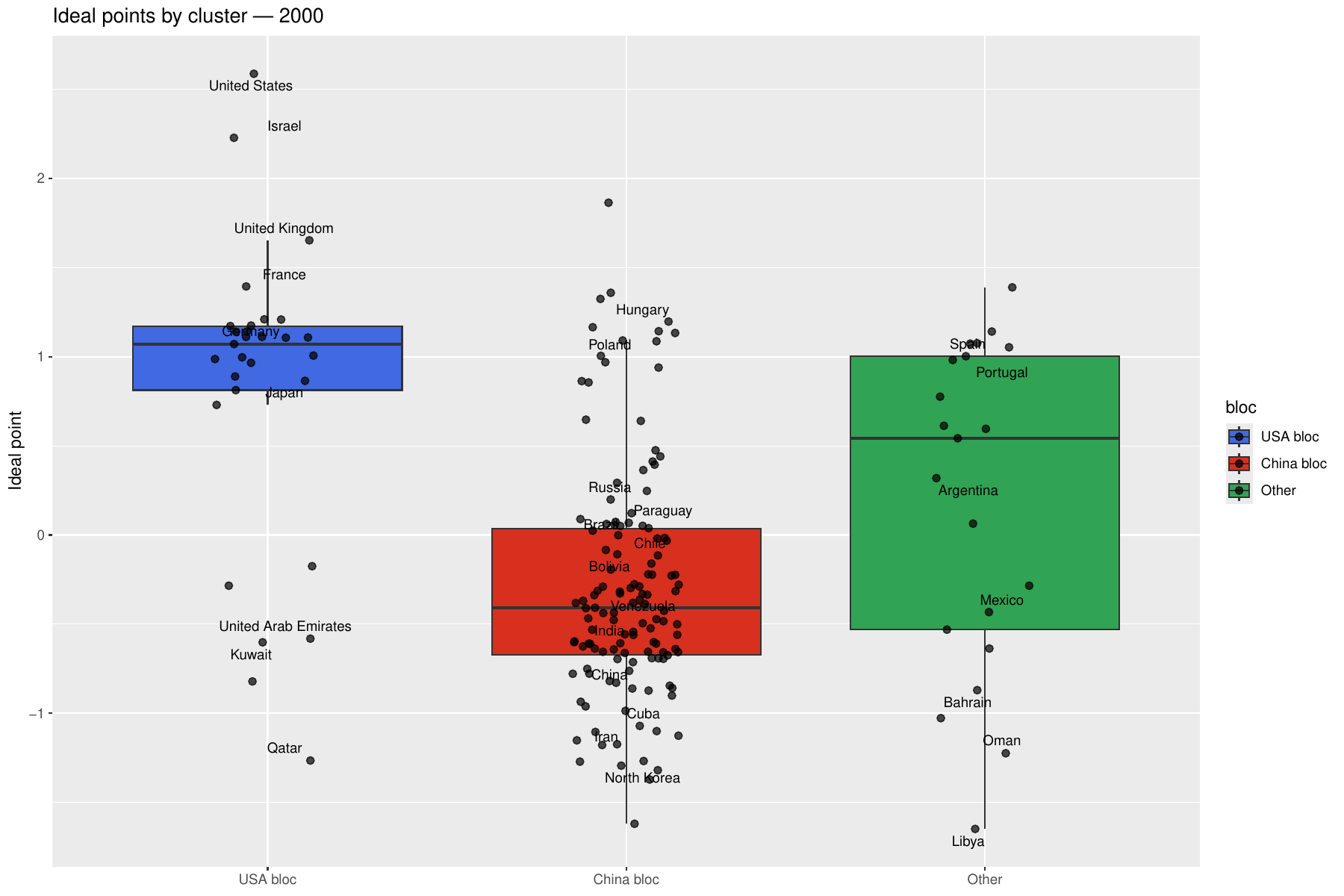}
    \caption{2000}
    \label{fig:bp2000}
\end{subfigure}

\vspace{0.05cm}

\captionof{figure}{\footnotesize Distribution of ideal points within each cluster labelled by anchor countries (USA bloc (blue), China bloc (red), Other bloc (green)) of the Het-Cov algorithm $(K=3)$, selected years 1990–2000. }
\label{fig:boxplots}
\end{center}

\section{Conclusion}

This work develops a methodology for community detection in heterogeneous attributed networks by jointly leveraging both network structure and nodal covariates. We establish weak consistency guarantees for the misclustering rate of individual nodes under the heterogeneous stochastic block model with covariates. Thus the method theoretically justifies the role of heterogeneity and node level covariates in community recovery.

The simulation experiments confirm that the method substantially outperforms competing baseline approaches when both network heterogeneity and covariate signals are strong. This result demonstrates that the joint utilization of network structure and nodal covariates yields improvements that neither source of information alone could provide. The magnitude of improvement which was measured through average misclustering error rate, increases with the strength of both heterogeneity and covariate effects.

The data analysis of voting behaviour in the United Nations General Assembly together with economic indicator like GDP per capita illustrates the method's capacity to recover meaningful geopolitical structure from real data. Specifically, the framework successfully identifies the emergence of regional and ideological coalitions—such as the Shanghai Cooperation Organization (SCO), BRICS etc. - long time before their formal establishment, and captures substantive international tensions during the turbulent 1990 - 2000. These findings suggest that the method provides a reliable account of international alignments during complex geopolitical transitions.

The approach offers a complementary perspective to existing approaches for understanding voting behaviour. Whereas the ideal point model (\cite{voeten_2017}), derives a single  dimension scale summarizing each country's position with respect to US global order, our method explicitly explains community structure through the joint effects of economic and geopolitical factors. This is not a replacement for existing methods, rather a distinct lens that interprets the clustering of nations in coherence with global transitions happened during the last years before 2000. Together, these approaches provide a more complete picture of the mechanisms underlying international dynamics of countries. 
As a direction for future research, we aim to develop a fully dynamic framework that accounts for the temporal evolution of the network, evolving community memberships, and shifts in the covariate distribution over time, instead of performing yearwise network clustering under block model or more general set up.


\section{Supplementary}

\subsection{Proof of Lemma 1}
\label{lemma1_proof}
The argument closely follows that of Lemma~1 in \cite{cov_sc_bink}. For completeness, we provide a brief proof and omit routine intermediate calculations.

Recall that
\[
c_l=\sum_{i}\operatorname{Var}(X_{il}\mid Z_i=l),
\qquad
\bar c=\frac{1}{TK}\sum_{l=1}^{TK} c_l .
\]
Let $\tilde C$ denote the diagonal matrix whose $m^{th}$ diagonal entry is $c_m$. Define another diagonal matrix $C$ through the relation
$CZ = Z\tilde C.$
Further, let
$
\tilde W = Z^{T}Z .
$

    Since
    \[
    \mathcal D_{P}
    =
    \operatorname{diag}\!\left(PZ^{T}\mathbf 1_{N}+\tau\right),
    \]
    it follows that
    \[
    \tilde{\mathcal L}
    =
    Z\tilde P Z^{T}
    +
    \alpha C,
    \]
    where,
    $ \Tilde{P}
= \mathcal{D}_{P}^{-1/2}
PZ^{T}\mathcal{D}_{\tau}^{-1}
Z P \mathcal{D}_{P}^{-1/2}
+ \alpha MM^{T}.$
    
We first note that $\tilde P$ is symmetric and positive definite for every $\alpha \geq 0$. Since $\tilde W=Z^{T}Z$ is diagonal with strictly positive entries, it follows that
\[
\det(\tilde P \tilde W)
=
\det(\tilde P)\det(\tilde W)
>0.
\]
Consequently, the matrix $\tilde P\tilde W+\alpha\tilde C$ is diagonalizable with real eigenvalues. Let its spectral decomposition be
\[
\tilde P\tilde W+\alpha\tilde C
=
\nu \Lambda \nu^{T},
\]
where $\Lambda$ is the diagonal matrix of eigenvalues and $\nu$ is the corresponding orthonormal eigenvector matrix.

Using the identity
\[
\tilde{\mathcal L}
=
Z\tilde P Z^{T}
+
\alpha C,
\]
together with $CZ=Z\tilde C$, we obtain
\[
\tilde{\mathcal L}Z\nu
=
\left(Z\tilde P Z^{T}+\alpha C\right)Z\nu
=
Z(\tilde P\tilde W+\alpha\tilde C)\nu
=
Z\nu\Lambda.
\]
Therefore, the columns of $Z\nu$ are eigenvectors of $\tilde{\mathcal L}$, and the corresponding eigenvalues are precisely the diagonal entries of $\Lambda$, i.e., the eigenvalues of $\tilde P\tilde W+\alpha\tilde C$.

However, this argument only establishes that these eigenvalues belong to the spectrum of $\tilde{\mathcal L}$. It does not guarantee that they correspond to the largest $TK$ eigenvalues of $\tilde{\mathcal L}$. To identify $Z\nu$ with the population eigenspace used in spectral clustering, an additional condition is required ensuring that the eigenvalues of $\tilde P\tilde W+\alpha\tilde C$ are separated from the remaining $N-TK$ eigenvalues of $\tilde{\mathcal L}$. We derive such a condition below.

Let $y\in\mathbb{R}^{N}$ satisfy $\|y\|_2=1$ and suppose that
\[
y \perp \operatorname{span}(Z\nu).
\]
Since $\nu$ is nonsingular, this implies
\[
y^{T}Z=0.
\]
Therefore,
\[
y^{T}\tilde{\mathcal L}y
=
y^{T}(Z\tilde P Z^{T}+\alpha C)y
=
\alpha y^{T}Cy.
\]
Using the decomposition
\[
C=\bar c\,I_N+(C-\bar c\,I_N),
\]
we obtain
\[
y^{T}\tilde{\mathcal L}y
=
\alpha\bar c
+
\alpha y^{T}(C-\bar c\,I_N)y.
\]
Hence,
\[
y^{T}\tilde{\mathcal L}y
\le
\alpha\bar c
+
\alpha \|C-\bar c\,I_N\|
=
\alpha(\bar c+\mathfrak X),
\]
where
\[
\mathfrak X
=
\max_{1\le l\le TK}|c_l-\bar c|.
\]

Consequently, every eigenvalue of $\tilde{\mathcal L}$ associated with the orthogonal complement of $\operatorname{span}(Z\nu)$ is bounded above by $\alpha(\bar c+\mathfrak X)$.

Next, consider the smallest eigenvalue of
$\tilde P\tilde W+\alpha\tilde C$. By the Courant--Fischer characterization,
\[
\lambda_{TK}(\tilde P\tilde W+\alpha\tilde C)
=
\min_{\|l\|_2=1}
l^{T}(\tilde P\tilde W+\alpha\tilde C)l .
\]
Using
\[
\tilde C
=
\bar c\,I_{TK}
+
(\tilde C-\bar c\,I_{TK}),
\]
we obtain
\[
\begin{aligned}
\lambda_{TK}(\tilde P\tilde W+\alpha\tilde C)
&=
\min_{\|l\|_2=1}
\Bigl[
l^{T}(\tilde P\tilde W)l
+
\alpha\bar c
+
\alpha l^{T}(\tilde C-\bar c\,I_{TK})l
\Bigr] \\
&\ge
\lambda_{TK}(\tilde P\tilde W)
+
\alpha\bar c
-
\alpha\|\tilde C-\bar c\,I_{TK}\| \\
&=
\lambda_{TK}(\tilde P\tilde W)
+
\alpha(\bar c-\mathfrak X).
\end{aligned}
\]

Hence, the eigenvectors in $Z\nu$ correspond to the top $TK$ eigenvalues of $\tilde{\mathcal L}$ provided that the smallest eigenvalue associated with $\operatorname{span}(Z\nu)$ exceeds the largest eigenvalue corresponding to its orthogonal complement. Using the bounds derived above, it suffices to require that
\[
\lambda_{TK}(\tilde P\tilde W+\alpha\tilde C)
>
\max_{\substack{x\perp \operatorname{span}(Z\nu)\\ \|x\|_2=1}}
x^{T}\tilde{\mathcal L}x.
\]
Since
\[
\lambda_{TK}(\tilde P\tilde W+\alpha\tilde C)
\ge
\lambda_{TK}(\tilde P\tilde W)
+
\alpha(\bar c-\mathfrak X),
\]
and
\[
\max_{\substack{x\perp \operatorname{span}(Z\nu)\\ \|x\|_2=1}}
x^{T}\tilde{\mathcal L}x
\le
\alpha(\bar c+\mathfrak X),
\]
a sufficient condition is
\[
\lambda_{TK}(\tilde P\tilde W)
+
\alpha(\bar c-\mathfrak X)
>
\alpha(\bar c+\mathfrak X),
\]
which simplifies to
\[
\lambda_{TK}(\tilde P\tilde W)
>
2\alpha\mathfrak X.
\]
Therefore, the condition
\[
\lambda_{TK}(\tilde P\tilde W) > 2\alpha\mathfrak X
\]
ensures that the top $TK$ eigenvectors of $\tilde{\mathcal L}$ are precisely the columns of $Z\nu$.

Finally, since $\nu$ is nonsingular,
\[
Z_i\nu = Z_j\nu
\iff
Z_i = Z_j.
\]
Hence, the population eigenspace uniquely identifies the type--block memberships. \hfill $\square$
\subsection*{Proof of Theorem 2} 

Recall that the covariate assisted sample Laplacian and its population counterpart are defined as
\[
\tilde{L}(\alpha)
=
D_{\tau}^{-1/2} A D_{\tau}^{-1} A D_{\tau}^{-1/2}
+
\alpha XX^{\mathrm T},
\]
and
\[
\tilde{\mathcal L}(\alpha)
=
\mathcal D_{\tau}^{-1/2}
\mathcal A
\mathcal D_{\tau}^{-1}
\mathcal A
\mathcal D_{\tau}^{-1/2}
+
\alpha \mathbb E\!\left(XX^{\mathrm T}\right),
\]
respectively.

To facilitate the analysis, define
\[
H
=
\mathcal D_{\tau}^{-1/2}
A
\mathcal D_{\tau}^{-1}
A
\mathcal D_{\tau}^{-1/2}.
\]
Our objective is to derive a high probability bound on the spectral norm deviation between the sample covariate assisted Laplacian and its population counterpart. To facilitate the analysis, define
\[
H
=
\mathcal D_{\tau}^{-1/2}
A
\mathcal D_{\tau}^{-1}
A
\mathcal D_{\tau}^{-1/2}.
\]
Applying the triangle inequality yields
\begin{align*}
\bigl\|
\tilde{L}(\alpha)-\tilde{\mathcal L}(\alpha)
\bigr\|
&=
\Bigl\|
\alpha\bigl(
XX^{\mathrm T}
-
\mathbb E(XX^{\mathrm T})
\bigr)
+
\Bigl(
D_{\tau}^{-1/2} A D_{\tau}^{-1} A D_{\tau}^{-1/2}
-
\mathcal D_{\tau}^{-1/2}
\mathcal A
\mathcal D_{\tau}^{-1}
\mathcal A
\mathcal D_{\tau}^{-1/2}
\Bigr)
\Bigr\| \\
&\le
\alpha
\bigl\|
XX^{\mathrm T}
-
\mathbb E(XX^{\mathrm T})
\bigr\|
+
\bigl\|
H
-
\mathcal D_{\tau}^{-1/2}
\mathcal A
\mathcal D_{\tau}^{-1}
\mathcal A
\mathcal D_{\tau}^{-1/2}
\bigr\| \\
&\qquad
+
\bigl\|
D_{\tau}^{-1/2}
A
D_{\tau}^{-1}
A
D_{\tau}^{-1/2}
-
H
\bigr\|.
\end{align*}

Accordingly, our analysis reduces to establishing suitable upper bounds for the three terms on the right-hand side of the above decomposition. We treat these terms separately in the following three steps. Combining the resulting bounds then yields the desired concentration inequality for
$\bigl\|
\tilde{L}(\alpha)-\tilde{\mathcal L}(\alpha)
\bigr\|.$

\paragraph{Step 1: Bounding the covariate term.}
We begin by deriving a concentration bound for the covariate component. Observe that
\[
\alpha XX^{\mathrm T}
=
\alpha \sum_{k=1}^{R} X_k X_k^{\mathrm T},
\]
where $X_k$ denotes the $k$th column of the covariate matrix
$X\in \mathbb{R}^{N\times R}$.
Consequently,
\[
\alpha\!\left(XX^{\mathrm T}
-
\mathbb E(XX^{\mathrm T})\right)
=
\alpha \sum_{k=1}^{R}
\left(
X_kX_k^{\mathrm T}
-
\mathbb E(X_kX_k^{\mathrm T})
\right).
\]
Since the above expression is a sum of independent random matrices, we shall apply the Matrix Bernstein inequality to obtain a concentration bound.

To this end, consider
\[
\alpha
\left(
X_kX_k^{\mathrm T}
-
\mathbb E(X_kX_k^{\mathrm T})
\right).
\]
Using the identity
\[
\mathbb E(X_kX_k^{\mathrm T})
=
\mathcal X_k \mathcal X_k^{\mathrm T}
+
\operatorname{diag}
\!\left(
\mathcal X_k^{(2)}
-
\mathcal X_k^{2}
\right),
\]
where
$\mathcal X_k=\mathbb E(X_k),$
we obtain
\begin{align*}
&\left\|
\alpha
\left(
X_kX_k^{\mathrm T}
-
\mathbb E(X_kX_k^{\mathrm T})
\right)
\right\| \\
&=
\alpha
\left\|
X_kX_k^{\mathrm T}
-
\mathcal X_k\mathcal X_k^{\mathrm T}
-
\operatorname{diag}
\!\left(
\mathcal X_k^{(2)}
-
\mathcal X_k^{2}
\right)
\right\| \\
&\le
\alpha
\Bigl(
\|X_kX_k^{\mathrm T}\|
+
\|\mathcal X_k\mathcal X_k^{\mathrm T}\|
+
\bigl\|
\operatorname{diag}
(
\mathcal X_k^{(2)}
-
\mathcal X_k^{2}
)
\bigr\|
\Bigr).
\end{align*}

Assuming that the covariates are uniformly bounded by $J$, we have
\[
\|X_kX_k^{\mathrm T}\|
\le NJ^2,
\qquad
\|\mathcal X_k\mathcal X_k^{\mathrm T}\|
\le NJ^2,
\]
and
\[
\bigl\|
\operatorname{diag}
(
\mathcal X_k^{(2)}
-
\mathcal X_k^{2}
)
\bigr\|
=
\max_{1\le i\le N}
\left|
\mathbb E(X_{ik}^{2})
-
\mathbb E(X_{ik})^{2}
\right|
\le J^2.
\]
Therefore,
\[
\left\|
\alpha
\left(
X_kX_k^{\mathrm T}
-
\mathbb E(X_kX_k^{\mathrm T})
\right)
\right\|
\le
\alpha(2NJ^2+J^2)
\le
3\alpha NJ^2.
\]
Define
\[
S:=3\alpha NJ^2.
\]

Finally, we introduce the notation
\[
\mathcal X_k
=
\begin{pmatrix}
\mathbb E(X_{1k})\\
\mathbb E(X_{2k})\\
\vdots\\
\mathbb E(X_{Nk})
\end{pmatrix},
\qquad
\mathcal X_k^{(2)}
=
\begin{pmatrix}
\mathbb E(X_{1k}^{2})\\
\mathbb E(X_{2k}^{2})\\
\vdots\\
\mathbb E(X_{Nk}^{2})
\end{pmatrix},
\]
and
\[
\mathcal X_k^{2}
=
\begin{pmatrix}
\mathbb E(X_{1k})^{2}\\
\mathbb E(X_{2k})^{2}\\
\vdots\\
\mathbb E(X_{Nk})^{2}
\end{pmatrix}.
\]

The quantity $S$ will serve as the uniform bound required for the application of the Matrix Bernstein inequality.

We next derive a bound on the matrix variance parameter associated with
\[
\alpha\sum_{k=1}^{R}
\left(
X_kX_k^{\mathrm T}
-
\mathbb E(X_kX_k^{\mathrm T})
\right).
\]

Recall that
\[
\mathbb E(X_kX_k^{\mathrm T})
=
\mathcal X_k\mathcal X_k^{\mathrm T}
+
\operatorname{diag}
\!\left(
\mathcal X_k^{(2)}
-
\mathcal X_k^{2}
\right).
\]
A straightforward calculation yields
\begin{align*}
\mathbb E(X_kX_k^{\mathrm T})
\mathbb E(X_kX_k^{\mathrm T})
=
&
\left(
\sum_{i=1}^{N}\mathcal X_{ik}^{2}
\right)
\mathcal X_k\mathcal X_k^{\mathrm T}
-
\mathcal X_k
\Bigl\{
\mathcal X_k
\bigl(
\mathcal X_k^{2}
-
\mathcal X_k^{(2)}
\bigr)
\Bigr\}^{\!\mathrm T}
\\
&
-
\Bigl\{
\mathcal X_k
\bigl(
\mathcal X_k^{2}
-
\mathcal X_k^{(2)}
\bigr)
\Bigr\}
\mathcal X_k^{\mathrm T}
+
\operatorname{diag}
\!\left\{
\bigl(
\mathcal X_k^{2}
-
\mathcal X_k^{(2)}
\bigr)^2
\right\}.
\end{align*}

Similarly,
\[
X_kX_k^{\mathrm T}X_kX_k^{\mathrm T}
=
\left(
\sum_{i=1}^{N}X_{ik}^{2}
\right)
X_kX_k^{\mathrm T},
\]
and hence, after taking expectations and collecting terms,
\begin{align*}
\operatorname{Var}(X_kX_k^{\mathrm T})
=
&
\mathcal X_k\mathcal X_k^{\mathrm T}
\sum_{i=1}^{N}
\Bigl(
\mathcal X_{ik}^{(2)}
-
\mathcal X_{ik}^{2}
\Bigr)
\\
&
+
\mathcal X_k
\Bigl\{
\mathcal X_k
\bigl(
\mathcal X_k^{2}
-
2\mathcal X_k^{(2)}
\bigr)
+
\mathcal X_k^{(3)}
\Bigr\}^{\!\mathrm T}
\\
&
+
\Bigl\{
\mathcal X_k
\bigl(
\mathcal X_k^{2}
-
2\mathcal X_k^{(2)}
\bigr)
+
\mathcal X_k^{(3)}
\Bigr\}
\mathcal X_k^{\mathrm T}
\\
&
+
\operatorname{diag}
\!\Bigl[
\bigl(
\mathcal X_k^{(2)}
-
\mathcal X_k^{2}
\bigr)
\Bigl(
\sum_{i=1}^{N}\mathcal X_{ik}^{(2)}
\Bigr)
-
\mathcal X_k^{(2)2}
+
2\mathcal X_k^{2}\mathcal X_k^{(2)}
\\
&
\hspace{2.5cm}
-
2\mathcal X_k\mathcal X_k^{(3)}
+
\mathcal X_k^{(4)}
-
\bigl(
\mathcal X_k^{(2)}
-
\mathcal X_k^{2}
\bigr)^2
\Bigr].
\end{align*}

Using the boundedness assumption on the covariates and repeatedly applying the triangle inequality, we obtain
\[
\left\|
\sum_{k=1}^{R}
\operatorname{Var}(X_kX_k^{\mathrm T})
\right\|
\le
8
\sum_{k=1}^{R}
\left\{
\sum_{i=1}^{N}
\mathcal X_{ik}^{(2)}
\sum_{l=1}^{N}
\left(
\mathcal X_{lk}^{(2)}
-
\mathcal X_{lk}^{2}
\right)
+
\mathcal X_{ik}^{(4)}
\right\}.
\]

Consequently,
\[
\left\|
\sum_{k=1}^{R}
\operatorname{Var}
\!\left(
\alpha X_kX_k^{\mathrm T}
\right)
\right\|
\le
8\alpha^2
\sum_{k=1}^{R}
\left\{
\sum_{i=1}^{N}
\mathcal X_{ik}^{(2)}
\sum_{l=1}^{N}
\left(
\mathcal X_{lk}^{(2)}
-
\mathcal X_{lk}^{2}
\right)
+
\mathcal X_{ik}^{(4)}
\right\}
\equiv \varpi.
\]

Let
\[
b=
\Bigl\{
3\varpi
\log\!\left(\frac{8N}{\epsilon}\right)
\Bigr\}^{1/2},
\]
and assume that
\[
\frac{\varpi}{S^2}
>
3\log\!\left(\frac{8N}{\epsilon}\right).
\]
Under this condition, \(b<\varpi/S\). In particular, the above assumption requires
\[
R \ge C\log N
\]
for some positive constant \(C\).

Applying the Matrix Bernstein inequality yields
\[
\Pr\!\left(
\left\|
\alpha XX^{\mathrm T}
-
\mathbb E(\alpha XX^{\mathrm T})
\right\|
>b
\right)
\le
2N
\exp\!\left(
-\frac{b^2}
{2\varpi+\frac{2Sb}{3}}
\right)
\le
\frac{\epsilon}{4}.
\]

Therefore, with probability at least \(1-\epsilon/4\),
\[
\left\|
\alpha XX^{\mathrm T}
-
\mathbb E(\alpha XX^{\mathrm T})
\right\|
\le
\Bigl\{
3\varpi
\log\!\left(\frac{8N}{\epsilon}\right)
\Bigr\}^{1/2}.
\]

\paragraph{Step 2: Bounding the graph concentration term.}

We next derive a bound for the second term appearing in the decomposition of
\(
\|\tilde{L}(\alpha)-\tilde{\mathcal L}(\alpha)\|.
\)
Using the submultiplicativity of the spectral norm, we obtain
\begin{align*}
&
\left\|
\mathcal D_{\tau}^{-1/2}
A
\mathcal D_{\tau}^{-1}
A
\mathcal D_{\tau}^{-1/2}
-
\mathcal D_{\tau}^{-1/2}
\mathcal A
\mathcal D_{\tau}^{-1}
\mathcal A
\mathcal D_{\tau}^{-1/2}
\right\| \\
&=
\left\|
\Bigl(
\mathcal D_{\tau}^{-1/2}
A
\mathcal D_{\tau}^{-1/2}
\Bigr)^2
-
\Bigl(
E(\mathcal D_{\tau}^{-1/2}
A
\mathcal D_{\tau}^{-1/2})
\Bigr)^2
\right\| \\
&\le
\left\|
\mathcal D_{\tau}^{-1/2}
A
\mathcal D_{\tau}^{-1/2}
-
E\!\left(
\mathcal D_{\tau}^{-1/2}
A
\mathcal D_{\tau}^{-1/2}
\right)
\right\|
\\
&\hspace{1cm}\times
\left\|
\mathcal D_{\tau}^{-1/2}
A
\mathcal D_{\tau}^{-1/2}
+
E\!\left(
\mathcal D_{\tau}^{-1/2}
A
\mathcal D_{\tau}^{-1/2}
\right)
\right\|.
\end{align*}

The first factor can be bounded using the concentration result established in \cite{qin_rohe_2013}. Assume that
\[
d+\tau > 3\log\!\left(\frac{8N}{\epsilon}\right),
\]
where
\(
d=\min_i \mathcal D_{ii}
\),
and define
\[
a=
\left\{
\frac{3\log(8N/\epsilon)}
     {d+\tau}
\right\}^{1/2}.
\]
Under this assumption, \(a<1\), and with probability at least \(1-\epsilon/4\),
\[
\left\|
\mathcal D_{\tau}^{-1/2}
A
\mathcal D_{\tau}^{-1/2}
-
E\!\left(
\mathcal D_{\tau}^{-1/2}
A
\mathcal D_{\tau}^{-1/2}
\right)
\right\|
\le a.
\]

To bound the second factor, we use the facts that
\(
\|\mathcal L_{\tau}\|\le 1
\),
\(
\|L_{\tau}\|\le 1
\),
and
\(
\|\mathcal D_{\tau}^{-1/2}D_{\tau}^{1/2}\|
\le a+1
\),
which follow from \cite{qin_rohe_2013}. Consequently, with probability at least \(1-\epsilon/4\),
\begin{align*}
&
\left\|
\mathcal D_{\tau}^{-1/2}
A
\mathcal D_{\tau}^{-1/2}
+
E\!\left(
\mathcal D_{\tau}^{-1/2}
A
\mathcal D_{\tau}^{-1/2}
\right)
\right\| \\
&\le
\left\|
\mathcal D_{\tau}^{-1/2}
D_{\tau}^{1/2}
L_{\tau}
D_{\tau}^{1/2}
\mathcal D_{\tau}^{-1/2}
\right\|
+
\|\mathcal L_{\tau}\| \\
&\le
\|\mathcal D_{\tau}^{-1/2}D_{\tau}^{1/2}\|
\,
\|L_{\tau}\|
\,
\|D_{\tau}^{1/2}\mathcal D_{\tau}^{-1/2}\|
+
1 \\
&\le
(a+1)^2+1.
\end{align*}

Combining the above bounds yields that, with probability at least \(1-\epsilon/4\),
\[
\left\|
\mathcal D_{\tau}^{-1/2}
A
\mathcal D_{\tau}^{-1}
A
\mathcal D_{\tau}^{-1/2}
-
E\!\left(
\mathcal D_{\tau}^{-1/2}
A
\mathcal D_{\tau}^{-1}
A
\mathcal D_{\tau}^{-1/2}
\right)
\right\|
\le
a\bigl\{(a+1)^2+1\bigr\}.
\]

\paragraph{Step 3: Bounding the degree normalization error.}

We now derive a bound for the third term in the decomposition of
\(
\|\tilde{L}(\alpha)-\tilde{\mathcal L}(\alpha)\|.
\)
By the arguments of \cite{qin_rohe_2013}, we have
\[
\left\|
\mathcal D_{\tau}^{-1/2}D_{\tau}^{1/2}-I
\right\|
\le a
\]
with probability at least \(1-\epsilon/4\). By an analogous argument,
\[
\left\|
D_{\tau}^{1/2}\mathcal D_{\tau}^{-1}D_{\tau}^{1/2}-I
\right\|
\le a
\]
with probability at least \(1-\epsilon/4\). Using these bounds, we control the discrepancy arising from replacing the sample degree matrix by its population counterpart.

We have,
\begin{align*}
&
\left\|
D_{\tau}^{-1/2}
A
D_{\tau}^{-1}
A
D_{\tau}^{-1/2}
-
\mathcal D_{\tau}^{-1/2}
A
\mathcal D_{\tau}^{-1}
A
\mathcal D_{\tau}^{-1/2}
\right\| \\
&=
\left\|
L_{\tau}^{2}
-
\mathcal D_{\tau}^{-1/2}
D_{\tau}^{1/2}
L_{\tau}
D_{\tau}^{1/2}
\mathcal D_{\tau}^{-1}
D_{\tau}^{1/2}
L_{\tau}
D_{\tau}^{1/2}
\mathcal D_{\tau}^{-1/2}
\right\| \\
&=
\Bigl\|
L_{\tau}^{2}
-
L_{\tau}
D_{\tau}^{1/2}
\mathcal D_{\tau}^{-1}
D_{\tau}^{1/2}
L_{\tau}
D_{\tau}^{1/2}
\mathcal D_{\tau}^{-1/2}
\\
&\qquad\qquad
+
\left(
I-\mathcal D_{\tau}^{-1/2}D_{\tau}^{1/2}
\right)
L_{\tau}
D_{\tau}^{1/2}
\mathcal D_{\tau}^{-1}
D_{\tau}^{1/2}
L_{\tau}
D_{\tau}^{1/2}
\mathcal D_{\tau}^{-1/2}
\Bigr\| \\
&\le
\left\|
L_{\tau}
\Bigl(
L_{\tau}
-
D_{\tau}^{1/2}
\mathcal D_{\tau}^{-1}
D_{\tau}^{1/2}
L_{\tau}
D_{\tau}^{1/2}
\mathcal D_{\tau}^{-1/2}
\Bigr)
\right\|
+
a(a+1)^2.
\end{align*}

Applying the triangle inequality together with the bounds
\(
\|L_{\tau}\|\le 1
\),
\(
\|D_{\tau}^{1/2}\mathcal D_{\tau}^{-1/2}\|\le a+1
\),
and
\(
\|D_{\tau}^{1/2}\mathcal D_{\tau}^{-1}D_{\tau}^{1/2}-I\|\le a
\),
we obtain
\begin{align*}
&
\left\|
D_{\tau}^{-1/2}
A
D_{\tau}^{-1}
A
D_{\tau}^{-1/2}
-
\mathcal D_{\tau}^{-1/2}
A
\mathcal D_{\tau}^{-1}
A
\mathcal D_{\tau}^{-1/2}
\right\| \\
&\le
\left\|
D_{\tau}^{1/2}
\mathcal D_{\tau}^{-1}
D_{\tau}^{1/2}
L_{\tau}
\left(
D_{\tau}^{1/2}
\mathcal D_{\tau}^{-1/2}
-I
\right)
\right\|
\\
&\qquad
+
\left\|
\left(
D_{\tau}^{1/2}
\mathcal D_{\tau}^{-1}
D_{\tau}^{1/2}
-I
\right)
L_{\tau}
\right\|
+
a(a+1)^2 \\
&\le
a(a+1)+a+a(a+1)^2 .
\end{align*}

Hence, with probability at least \(1-\epsilon/2\),
\[
\left\|
D_{\tau}^{-1/2}
A
D_{\tau}^{-1}
A
D_{\tau}^{-1/2}
-
\mathcal D_{\tau}^{-1/2}
A
\mathcal D_{\tau}^{-1}
A
\mathcal D_{\tau}^{-1/2}
\right\|
\le
a(a+1)+a+a(a+1)^2.
\]

Combining the bounds obtained for the three terms in the decomposition and applying a union bound yields that, with probability at least \(1-\epsilon\),
\begin{align*}
\|
\tilde L(\alpha)-\tilde{\mathcal L}(\alpha)
\|
&\le
2a^{3}+5a^{2}+5a+b \\
&\le
12a+b \\
&=
\left\{
\varpi^{1/2}
+
12(d+\tau)^{-1/2}
\right\}
\left\{
3\log\!\left(\frac{8N}{\epsilon}\right)
\right\}^{1/2}.
\end{align*}

Defining
\[
\delta
=
\varpi^{1/2}
+
12(d+\tau)^{-1/2},
\]
the preceding result can be expressed compactly as
\[
\|
\tilde L(\alpha)-\tilde{\mathcal L}(\alpha)
\|
\le
\delta
\left\{
3\log\!\left(\frac{8N}{\epsilon}\right)
\right\}^{1/2}.
\]

\subsection*{Proof of Theorem~3}

The proof follows the argument of Theorem~3 in \cite{cov_sc_bink}, which itself builds upon the perturbation analysis developed in Theorem~4.2 of \cite{qin_rohe_2013}.

Let \(P_{\tilde L}\) denote the orthogonal projection onto the subspace spanned by the leading \(TK\) left singular vectors of \(\tilde L\). By Lemma~9 of \cite{McSherry_2001}, \(P_{\tilde L}\) provides the optimal rank-\(TK\) approximation to \(\tilde L\), and therefore
\[
\left\|
P_{\tilde L}\tilde L-\tilde{\mathcal L}
\right\|_F^2
\le
8TK
\left\|
\tilde L-\tilde{\mathcal L}
\right\|^2 .
\]

We now apply the Davis--Kahan theorem \cite{davis_kahan_1970} to compare the eigenspaces of \(\tilde L\) and \(\tilde{\mathcal L}\). Let \(W\subset \mathbb R\) be an interval and define the eigengap
\[
\Lambda
=
\min
\Bigl\{
|\lambda-r|:
\lambda \in \sigma(\tilde{\mathcal L})\setminus W,
\,
r\in W
\Bigr\},
\]
where \(\sigma(\tilde{\mathcal L})\) denotes the spectrum of \(\tilde{\mathcal L}\).

Since \(\operatorname{rank}(\tilde{\mathcal L})=TK\), let \(\lambda_{TK}\) denote the smallest nonzero eigenvalue of \(\tilde{\mathcal L}\), and choose
\[
W=
\left(
\frac{\lambda_{TK}}{2},
\infty
\right).
\]
The nearest eigenvalue of \(\tilde{\mathcal L}\) lying outside \(W\) is \(0\), while the nearest point in \(W\) is \(\lambda_{TK}/2\). Consequently,
\[
\Lambda
=
\frac{\lambda_{TK}}{2}.
\]

Let \(\omega_{TK}\) denote the \(TK\)th largest eigenvalue of \(\tilde L\). By Weyl's inequality,
\[
\bigl|
\lambda_{TK}-\omega_{TK}
\bigr|
\le
\|
\tilde L-\tilde{\mathcal L}
\|.
\]
Combining this with the concentration result established in Theorem~2 and the assumption
\[
\delta
\left\{
3\log\!\left(\frac{8N}{\epsilon}\right)
\right\}^{1/2}
\le
\frac{\lambda_{TK}}{2},
\]
yields
\[
\bigl|
\lambda_{TK}-\omega_{TK}
\bigr|
\le
\|
\tilde L-\tilde{\mathcal L}
\|
\le
\delta
\left\{
3\log\!\left(\frac{8N}{\epsilon}\right)
\right\}^{1/2}
\le
\frac{\lambda_{TK}}{2}.
\]
Therefore,
\[
\omega_{TK}
\ge
\frac{\lambda_{TK}}{2},
\]
which implies that \(\omega_{TK}\in W\).

To verify that the dimensions of the corresponding eigenspaces agree, consider the \((TK+1)\)st largest eigenvalue of \(\tilde L\), denoted by \(\omega_{TK+1}\). Since \(\lambda_{TK+1}=0\), Weyl's inequality again gives
\[
\omega_{TK+1}
\le
\lambda_{TK+1}
+
\|
\tilde L-\tilde{\mathcal L}
\|
\le
\frac{\lambda_{TK}}{2}.
\]
Hence,
\[
\omega_{TK+1}\notin W.
\]
It follows that exactly \(TK\) eigenvalues of \(\tilde L\) lie in \(W\), and therefore the eigenspaces associated with \(\tilde L\) and \(\tilde{\mathcal L}\) have the same dimension.

Let \(U\) and \(\mathcal U\) denote the matrices whose columns consist of the leading \(TK\) eigenvectors of \(\tilde L\) and \(\tilde{\mathcal L}\), respectively. By the Davis--Kahan theorem, there exists an orthogonal matrix \(O\) such that
\begin{align*}
\|U-\mathcal UO\|_F
&\le
\frac{\sqrt{2}\,
\|P_{\tilde L}\tilde L-\tilde{\mathcal L}\|_F}
{\Lambda} \\
&\le
\frac{\sqrt{8}\,
\|P_{\tilde L}\tilde L-\tilde{\mathcal L}\|_F}
{\lambda_{TK}} \\
&\le
\frac{8\sqrt{TK}\,
\|\tilde L-\tilde{\mathcal L}\|}
{\lambda_{TK}} \\
&\le
\frac{
8\delta
\left\{
3TK\log(8N/\epsilon)
\right\}^{1/2}
}
{\lambda_{TK}} .
\end{align*}

Therefore, with probability at least \(1-\epsilon\),
\[
\|U-\mathcal UO\|_F
\le
\frac{
8\delta
\left\{
3TK\log(8N/\epsilon)
\right\}^{1/2}
}
{\lambda_{TK}},
\]
which establishes the desired eigenspace perturbation bound.

\subsection*{Proof of Theorem~4}

For notational simplicity, we present the proof for the case \(T=2\), corresponding to a bipartite  network. The argument for general \(T>2\) follows analogously with only minor modifications to the notation.

In this setting, the membership matrix \(Z\) is of dimension \(N\times 2K\) and admits the block diagonal representation
\[
Z=
\begin{pmatrix}
Z_{11}^{N_1\times K} & 0 \\
0 & Z_{22}^{N_2\times K}
\end{pmatrix}
=
\begin{pmatrix}
Z^{(1)} \\
Z^{(2)}
\end{pmatrix},
\]
where \(N_1+N_2=N\), and \(Z^{(t)}\) denotes the submatrix corresponding to nodes of type \(t\), \(t=1,2\). Since \(Z_{11}\) and \(Z_{22}\) are community membership matrices, each contains exactly \(K\) distinct rows.

From Theorem~1, we have
\[
\mathcal U
=
Z(Z^{T}Z)^{-1/2}V
=
Z\mu,
\]
where
\[
\mu=(Z^{T}Z)^{-1/2}V
\]
is a nonsingular \(2K\times 2K\) matrix. Partition \(\mu\) conformably as
\[
\mu=
\begin{pmatrix}
\mu_{11}^{K\times K} & \mu_{12}^{K\times K} \\
\mu_{21}^{K\times K} & \mu_{22}^{K\times K}
\end{pmatrix}.
\]

Using the block structure of \(Z\), we obtain
\[
\mathcal U
=
\begin{pmatrix}
Z_{11} & 0\\
0 & Z_{22}
\end{pmatrix}
\begin{pmatrix}
\mu_{11} & \mu_{12}\\
\mu_{21} & \mu_{22}
\end{pmatrix}
=
\begin{pmatrix}
Z_{11}\mu_{11} & Z_{11}\mu_{12}\\
Z_{22}\mu_{21} & Z_{22}\mu_{22}
\end{pmatrix}.
\]

Accordingly, we write
\[
\mathcal U=
\begin{pmatrix}
\mathcal U^{(1)^{N_1 \times K}}\\
\mathcal U^{(2)^{N_2 \times K}}
\end{pmatrix},
\] 
where \(\mathcal U^{(t)}\) contains the rows corresponding to nodes of type \(t\), \(t=1,2\). Since \(Z_{11}\) and \(Z_{22}\) each have exactly \(K\) distinct rows and \(\mu\) is nonsingular, it follows that both \(\mathcal U^{(1)}\) and \(\mathcal U^{(2)}\) contain exactly \(K\) distinct row vectors. These distinct rows correspond to the population centroids of the $K$ communities within each node type; that is, the rows of $\mathcal{U}^t$ represent the type-$t$ population centroids.

Let \(\mathcal O\) be the orthogonal matrix minimizing
\[
\|U\mathcal O^{T}-\mathcal U\|_{F}.
\]
For \(t\in\{1,2\}\), define the set of misclustered nodes of type \(t\) by
\[
\mathcal M_t
=
\Bigl\{
i \in \text{type-}t :
\exists\, j\in \text{type-}t,\; j\neq i,
\text{ such that }
\|C_i^{(t)}\mathcal O^{T}-\mathcal C_i^{(t)}\|_2
>
\|C_i^{(t)}\mathcal O^{T}-\mathcal C_j^{(t)}\|_2
\Bigr\},
\]
where \(C_i^{(t)}\) and \(\mathcal C_i^{(t)}\) denote the empirical and population cluster centroids, respectively, associated with node \(i\) of type \(t\).

It suffices to establish the result for type \(1\), since the proof for type \(2\) is identical.
Define
\[
P=\max_{1\leq i\leq 2K}(Z^{T}Z)_{ii}.
\]
Following the argument of \cite{qin_rohe_2013}, for any two distinct communities \(i\) and \(j\) within type \(1\), we have
\begin{align*}
\|\mathcal C_i^{(1)}-\mathcal C_j^{(1)}\|_2
&=
\|\mathcal U_i^{(1)}-\mathcal U_j^{(1)}\|_2 \\
&=
\|(Z_i-Z_j)(Z^{T}Z)^{-1/2}V\|_2 \\
&=
\|(Z_i-Z_j)(Z^{T}Z)^{-1/2}\|_2 \\
&\geq
\sqrt{\frac{2}{P}}.
\end{align*}
The final inequality follows from the fact that \(Z_i\) and \(Z_j\) correspond to distinct membership vectors and from the bound
\[
\|(Z^{T}Z)^{-1/2}\|_2
=
\lambda_{\min}^{-1/2}(Z^{T}Z)
\geq
P^{-1/2}.
\]

Hence, the population centroids corresponding to distinct communities are separated by at least \(\sqrt{2/P}\), a fact that will be used in the subsequent derivation of the misclustering error bound.

To bound the misclustering rate, we first establish a sufficient condition under which a node is correctly clustered. For any pair of distinct population centroids \(\mathcal C_i^{(1)}\) and \(\mathcal C_j^{(1)}\) corresponding to different communities of type \(1\), a sufficient condition for the empirical centroid \(C_i^{(1)}\mathcal O^T\) to be closer to its true population centroid than to any other population centroid is
\[
\left\|
C_i^{(1)}\mathcal O^T-\mathcal C_i^{(1)}
\right\|_2
<
\frac{1}{\sqrt{2P}}.
\]
Indeed, for any \(Z_j^{(1)}\neq Z_i^{(1)}\), the triangle inequality yields
\begin{align*}
\left\|
C_i^{(1)}\mathcal O^T-\mathcal C_j^{(1)}
\right\|_2
&\ge
\left\|
\mathcal C_i^{(1)}-\mathcal C_j^{(1)}
\right\|_2
-
\left\|
C_i^{(1)}\mathcal O^T-\mathcal C_i^{(1)}
\right\|_2 \\
&\ge
\sqrt{\frac{2}{P}}
-
\sqrt{\frac{1}{2P}}
=
\frac{1}{\sqrt{2P}},
\end{align*}
where the second inequality follows from the lower bound on the separation between distinct population centroids established above. Consequently, whenever
\(
\|C_i^{(1)}\mathcal O^T-\mathcal C_i^{(1)}\|_2
<
(2P)^{-1/2},
\)
the empirical centroid is necessarily closer to its true population centroid than to any other population centroid.

Define the set
\[
\mathcal G^{(1)}
=
\left\{
i:
\left\|
C_i^{(1)}\mathcal O^T-\mathcal C_i^{(1)}
\right\|_2
\ge
\frac{1}{\sqrt{2P}}
\right\}.
\]
By construction, every misclustered node must belong to \(\mathcal G^{(1)}\), and therefore
\[
\mathcal M_1\subseteq \mathcal G^{(1)}.
\]
Let \(Q^{(1)}\in\mathbb R^{N_1 \times 2K}\) denote the matrix whose \(i\)th row is the empirical centroid \(C_i^{(1)}\). Since \(Q^{(1)}\) is obtained via the \(K\)-means procedure, it satisfies the optimality property
\[
\|U^{(1)}-Q^{(1)}\|_F
\le
\|U^{(1)}-\mathcal U^{(1)}\mathcal O\|_F.
\]
Applying the triangle inequality gives
\begin{align*}
\|Q^{(1)}-Z^{(1)}\mu\mathcal O\|_F
&=
\|Q^{(1)}-\mathcal U^{(1)}\mathcal O\|_F \\
&\le
\|U^{(1)}-Q^{(1)}\|_F
+
\|U^{(1)}-\mathcal U^{(1)}\mathcal O\|_F \\
&\le
2\|U^{(1)}-\mathcal U^{(1)}\mathcal O\|_F.
\end{align*}

We now bound the proportion of misclustered type-\(1\) nodes:
\begin{align*}
\frac{|\mathcal M_1|}{N_1}
&\le
\frac{|\mathcal G^{(1)}|}{N_1} \\
&=
\frac{1}{N_1}
\sum_{i\in\mathcal G^{(1)}}1 \\
&\le
\frac{2P}{N_1}
\sum_{i\in\mathcal G^{(1)}}
\left\|
C_i^{(1)}\mathcal O^T-\mathcal C_i^{(1)}
\right\|_2^2 \\
&=
\frac{2P}{N_1}
\sum_{i\in\mathcal G^{(1)}}
\left\|
C_i^{(1)}-Z_i^{(1)}\mu\mathcal O
\right\|_2^2 \\
&\le
\frac{2P}{N_1}
\|Q^{(1)}-Z^{(1)}\mu\mathcal O\|_F^2 \\
&\le
\frac{8P}{N_1}
\|U^{(1)}-\mathcal U^{(1)}\mathcal O\|_F^2 \\
&\le
\frac{8P}{N_1}
\|U-\mathcal U\mathcal O\|_F^2 .
\end{align*}

Substituting the eigenspace perturbation bound obtained in Theorem~3, we conclude that, with probability at least \(1-\epsilon\),
\begin{align*}
\frac{|\mathcal M_1|}{N_1}
&\le
\frac{8P}{N_1}
\left(
\frac{
8\delta
\sqrt{3TK\log(8N/\epsilon)}
}
{\lambda_{TK}}
\right)^2 \\
&=
\frac{3\times 8^3\,\delta^2\,P\,TK\,
\log(8N/\epsilon)}
{N_1\lambda_{TK}^2}.
\end{align*}

Hence,
\[
\frac{|\mathcal M_1|}{N_1}
\le
c\,
\frac{
\delta^2\,P\,TK\,
\log(8N/\epsilon)
}
{N_1\lambda_{TK}^2},
\]
where
\[
c=3\times 8^3.
\]

An identical argument applies to nodes of type \(2\), thereby establishing the stated misclustering error bound.

\bibliographystyle{plainnat}   
\bibliography{reference}

@article{het_sengupta,
 ISSN = {10170405, 19968507},
 URL = {http://www.jstor.org/stable/24721222},
 author = {Srijan Sengupta and Yuguo Chen},
 journal = {Statistica Sinica},
 number = {3},
 pages = {1081--1106},
 publisher = {Institute of Statistical Science, Academia Sinica},
 title = {SPECTRAL CLUSTERING IN HETEROGENEOUS NETWORKS},
 urldate = {2025-08-27},
 volume = {25},
 year = {2015}
}

@article{davis_kahan_1970,
 ISSN = {00361429},
 URL = {http://www.jstor.org/stable/2949580},
 author = {Chandler Davis and W. M. Kahan},
 journal = {SIAM Journal on Numerical Analysis},
 number = {1},
 pages = {1--46},
 publisher = {Society for Industrial and Applied Mathematics},
 title = {The Rotation of Eigenvectors by a Perturbation. III},
 urldate = {2025-08-27},
 volume = {7},
 year = {1970}
}

@inproceedings{qin_rohe_2013,
author = {Qin, Tai and Rohe, Karl},
title = {Regularized spectral clustering under the Degree-Corrected Stochastic Blockmodel},
year = {2013},
publisher = {Curran Associates Inc.},
address = {Red Hook, NY, USA},
booktitle = {Proceedings of the 27th International Conference on Neural Information Processing Systems - Volume 2},
pages = {3120–3128},
numpages = {9},
location = {Lake Tahoe, Nevada},
series = {NIPS'13}
}

@InProceedings{chaudhuri12,
  title = 	 {Spectral Clustering of Graphs with General Degrees in the Extended Planted Partition Model},
  author = 	 {Chaudhuri, Kamalika and Chung, Fan and Tsiatas, Alexander},
  booktitle = 	 {Proceedings of the 25th Annual Conference on Learning Theory},
  pages = 	 {35.1--35.23},
  year = 	 {2012},
  editor = 	 {Mannor, Shie and Srebro, Nathan and Williamson, Robert C.},
  volume = 	 {23},
  series = 	 {Proceedings of Machine Learning Research},
  address = 	 {Edinburgh, Scotland},
  month = 	 {25--27 Jun},
  publisher =    {PMLR},
  url = 	 {https://proceedings.mlr.press/v23/chaudhuri12.html}
}

@article{goldenberg_survey,
author = {Goldenberg, Anna and Zheng, Alice X. and Fienberg, Stephen E. and Airoldi, Edoardo M.},
title = {A Survey of Statistical Network Models},
year = {2010},
issue_date = {February 2010},
publisher = {Now Publishers Inc.},
address = {Hanover, MA, USA},
volume = {2},
number = {2},
issn = {1935-8237},
url = {https://doi.org/10.1561/2200000005},
doi = {10.1561/2200000005},
journal = {Found. Trends Mach. Learn.},
month = feb,
pages = {129–233},
numpages = {105}
}

@Inbook{Sengupta_survey,
author="Sengupta, Srijan",
editor="Ghosal, Subhashis
and Roy, Anindya",
title="Statistical Network Analysis: Past, Present, and Future",
bookTitle="Frontiers of Statistics and Data Science",
year="2025",
publisher="Springer Nature Singapore",
address="Singapore",
pages="153--179",
isbn="978-981-96-0742-6",
doi="10.1007/978-981-96-0742-6_7",
url="https://doi.org/10.1007/978-981-96-0742-6_7"
}

@article{Handcock_07,
    author = {Handcock, Mark S. and Raftery, Adrian E. and Tantrum, Jeremy M.},
    title = {Model-Based Clustering for Social Networks},
    journal = {Journal of the Royal Statistical Society Series A: Statistics in Society},
    volume = {170},
    number = {2},
    pages = {301-354},
    year = {2007},
    month = {04},
    issn = {0964-1998},
    doi = {10.1111/j.1467-985X.2007.00471.x},
    url = {https://doi.org/10.1111/j.1467-985X.2007.00471.x},
    eprint = {https://academic.oup.com/jrsssa/article-pdf/170/2/301/49619054/jrsssa_170_2_301.pdf},
}

@article{Luxburg_sc_2007,
  title={A tutorial on spectral clustering},
  author={Ulrike von Luxburg},
  journal={Statistics and Computing},
  year={2007},
  volume={17},
  pages={395-416},
  url={https://api.semanticscholar.org/CorpusID:3264198}
}

@inproceedings{jordan_sc,
 author = {Ng, Andrew and Jordan, Michael and Weiss, Yair},
 booktitle = {Advances in Neural Information Processing Systems},
 editor = {T. Dietterich and S. Becker and Z. Ghahramani},
 pages = {},
 publisher = {MIT Press},
 title = {On Spectral Clustering: Analysis and an algorithm},
 url = {https://proceedings.neurips.cc/paper_files/paper/2001/file/801272ee79cfde7fa5960571fee36b9b-Paper.pdf},
 volume = {14},
 year = {2001}
}

@article{Hoff_02,
author = {Peter D Hoff and Adrian E Raftery and Mark S Handcock},
title = {Latent Space Approaches to Social Network Analysis},
journal = {Journal of the American Statistical Association},
volume = {97},
number = {460},
pages = {1090--1098},
year = {2002},
publisher = {Taylor \& Francis},
doi = {10.1198/016214502388618906},


URL = { 
    
        https://doi.org/10.1198/016214502388618906
    
    

},
eprint = { 
    
        https://doi.org/10.1198/016214502388618906
    
    

}

}

@article{sbm_nowicki,
author = {Krzysztof Nowicki and Tom A. B Snijders},
title = {Estimation and Prediction for Stochastic Blockstructures},
journal = {Journal of the American Statistical Association},
volume = {96},
number = {455},
pages = {1077--1087},
year = {2001},
publisher = {Taylor \& Francis},
doi = {10.1198/016214501753208735},

URL = {         https://doi.org/10.1198/016214501753208735
},
eprint = {  https://doi.org/10.1198/016214501753208735
}
}

@article{sbm_holland,
title = {Stochastic blockmodels: First steps},
journal = {Social Networks},
volume = {5},
number = {2},
pages = {109-137},
year = {1983},
issn = {0378-8733},
doi = {https://doi.org/10.1016/0378-8733(83)90021-7},
url = {https://www.sciencedirect.com/science/article/pii/0378873383900217},
author = {Paul W. Holland and Kathryn Blackmond Laskey and Samuel Leinhardt}
}

@article{rohe_chatterjee_2010,
author = {Rohe, Karl and Chatterjee, Sourav and Yu, B.},
year = {2010},
month = {07},
pages = {},
title = {Spectral clustering and the high-dimensional stochastic blockmodel},
volume = {39},
journal = {Annals of Statistics - ANN STATIST},
doi = {10.1214/11-AOS887}
}

@article{dcbm,
  title = {Stochastic blockmodels and community structure in networks},
  author = {Karrer, Brian and Newman, M. E. J.},
  journal = {Phys. Rev. E},
  volume = {83},
  issue = {1},
  pages = {016107},
  numpages = {10},
  year = {2011},
  month = {Jan},
  publisher = {American Physical Society},
  doi = {10.1103/PhysRevE.83.016107},
  url = {https://link.aps.org/doi/10.1103/PhysRevE.83.016107}
}

@article{pabm,
 ISSN = {13697412, 14679868},
 URL = {http://www.jstor.org/stable/44681844},
 author = {Srijan Sengupta and Yuguo Chen},
 journal = {Journal of the Royal Statistical Society. Series B (Statistical Methodology)},
 number = {2},
 pages = {365--386},
 publisher = {[Royal Statistical Society, Wiley]},
 title = {A block model for node popularity in networks with community structure},
 urldate = {2026-03-03},
 volume = {80},
 year = {2018}
}

@article{cov_newman,
author = {Newman, M. and Clauset, Aaron},
year = {2015},
month = {07},
pages = {},
title = {Structure and inference in annotated networks},
volume = {7},
journal = {Nature Communications},
doi = {10.1038/ncomms11863}
}

@article{Yan_sarkar_21,
author = {Bowei Yan and Purnamrita Sarkar},
title = {Covariate Regularized Community Detection in Sparse Graphs},
journal = {Journal of the American Statistical Association},
volume = {116},
number = {534},
pages = {734--745},
year = {2021},
publisher = {Taylor \& Francis},
doi = {10.1080/01621459.2019.1706541},
URL = { 
        https://doi.org/10.1080/01621459.2019.1706541
    
},
eprint = { 
        https://doi.org/10.1080/01621459.2019.1706541

}

}

@article{zhang_2016,
author = {Zhang, Yuan and Levina, Elizaveta},
year = {2016},
month = {01},
pages = {},
title = {Community detection in networks with node features},
volume = {10},
journal = {Electronic Journal of Statistics},
doi = {10.1214/16-EJS1206}
}

@article{yang_2021,
author = {Weng, Haolei and Feng, Yang},
year = {2021},
month = {10},
pages = {},
title = {Community detection with nodal information: Likelihood and its variational approximation},
volume = {11},
journal = {Stat},
doi = {10.1002/sta4.428}
}

@article{cov_sc_bink,
 ISSN = {00063444},
 URL = {http://www.jstor.org/stable/26363727},
 author = {N. Binkiewicz and J. T. Vogelstein and K. Rohe},
 journal = {Biometrika},
 number = {2},
 pages = {361--377},
 publisher = {[Oxford University Press, Biometrika Trust]},
 title = {Covariate-assisted spectral clustering},
 urldate = {2025-08-27},
 volume = {104},
 year = {2017}
}

@article{net_adj_cov_24,
    author = {Hu, Y and Wang, W},
    title = {Network-adjusted covariates for community detection},
    journal = {Biometrika},
    volume = {111},
    number = {4},
    pages = {1221-1240},
    year = {2024},
    month = {02},
    issn = {1464-3510},
    doi = {10.1093/biomet/asae011},
    url = {https://doi.org/10.1093/biomet/asae011},
    eprint = {https://academic.oup.com/biomet/article-pdf/111/4/1221/60783767/asae011.pdf},
}

@article{cov_dim_reduction_21,
    author = {Zhao, Junlong and Liu, Xiumin and Wang, Hansheng and Leng, Chenlei},
    title = {Dimension reduction for covariates in network data},
    journal = {Biometrika},
    volume = {109},
    number = {1},
    pages = {85-102},
    year = {2021},
    month = {02},
    issn = {1464-3510},
    doi = {10.1093/biomet/asab006},
    url = {https://doi.org/10.1093/biomet/asab006},
    eprint = {https://academic.oup.com/biomet/article-pdf/109/1/85/42362112/asab006.pdf},
}

@ARTICLE{profile_likelihood_cov,
  author={Chandna, Swati and Bagozzi, Benjamin E. and Chatterjee, Snigdhansu},
  journal={IEEE Transactions on Network Science and Engineering}, 
  title={Profile Least Squares Estimation in Networks With Covariates}, 
  year={2026},
  volume={13},
  number={},
  pages={1662-1675},
  doi={10.1109/TNSE.2025.3598705}}

@ARTICLE{sc_sbm_cov,
  author={Mu, Cong and Mele, Angelo and Hao, Lingxin and Cape, Joshua and Athreya, Avanti and Priebe, Carey E.},
  journal={IEEE Transactions on Network Science and Engineering}, 
  title={On Spectral Algorithms for Community Detection in Stochastic Blockmodel Graphs With Vertex Covariates}, 
  year={2022},
  volume={9},
  number={5},
  pages={3373-3384},
  doi={10.1109/TNSE.2022.3177708}}

@article{graphon_cov,
  title={Network Estimation via Graphon With Node Features},
  author={Yi Su and Raymond K. W. Wong and Thomas C.M. Lee},
  journal={IEEE Transactions on Network Science and Engineering},
  year={2018},
  volume={7},
  pages={2078-2089},
  url={https://api.semanticscholar.org/CorpusID:88522656}
}

@article{pairwise_cov_18,
  title={PCABM: Pairwise Covariates-Adjusted Block Model for Community Detection},
  author={Sihan Huang and Yang Feng},
  journal={Journal of the American Statistical Association},
  year={2018},
  volume={119},
  pages={2092 - 2104},
  url={https://api.semanticscholar.org/CorpusID:88518040}
}

@article{mult_cov_1,
author = {Shirong Xu and Yaoming Zhen and Junhui Wang},
title = {Covariate-Assisted Community Detection in Multi-Layer Networks},
journal = {Journal of Business \& Economic Statistics},
volume = {41},
number = {3},
pages = {915--926},
year = {2023},
publisher = {Taylor \& Francis},
doi = {10.1080/07350015.2022.2085726},


URL = { 
    
        https://doi.org/10.1080/07350015.2022.2085726
    
    

},
eprint = { 
    
        https://doi.org/10.1080/07350015.2022.2085726
}

}

@article{mult_cov_2,
author = {Zhao, Da and Wang, Wanjie and Li, Jialiang},
title = {Spectral clustering on aggregated multilayer networks with covariates},
year = {2025},
issue_date = {Aug 2025},
publisher = {Kluwer Academic Publishers},
address = {USA},
volume = {35},
number = {5},
issn = {0960-3174},
url = {https://doi.org/10.1007/s11222-025-10661-3},
doi = {10.1007/s11222-025-10661-3},
journal = {Statistics and Computing},
month = jun,
numpages = {15}
}

@article{voeten_2009,
  title={United Nations general assembly voting data},
  author={Voeten, Erik and Strezhnev, Anton and Bailey, Michael},
  journal={Harvard Dataverse},
  volume={28},
  year={2009}
}

@article{voeten_2017,
author = {Michael A. Bailey and Anton Strezhnev and Erik Voeten},
title ={Estimating Dynamic State Preferences from United Nations Voting Data},

journal = {Journal of Conflict Resolution},
volume = {61},
number = {2},
pages = {430-456},
year = {2017},
doi = {10.1177/0022002715595700},

URL = { 
    
        https://doi.org/10.1177/0022002715595700
    
    

},
eprint = { 
    
        https://doi.org/10.1177/0022002715595700
    
    

}
}

@article{voeten_2018,
    title={A Two-Dimensional Analysis of Seventy Years of United Nations Voting},
    author={Bailey, Michael and Voeten, Erik},
    url={http://dx.doi.org/10.2139/ssrn.3166115},
    doi={10.2139/ssrn.3166115},
    journal={SSRN Electronic Journal},
    publisher={Elsevier BV},
    year={2018},
    month={Jan},
    language={en},
}

@article{unga_ddpm,
author = {Qiushi Yu},
title = {Dynamic Dirichlet process mixture model for identifying voting coalitions in the United Nations General Assembly human rights roll call votes},
journal = {Journal of Applied Statistics},
volume = {49},
number = {12},
pages = {3002--3021},
year = {2022},
publisher = {Taylor \& Francis},
doi = {10.1080/02664763.2021.1931820},
note ={PMID: 36035611},
URL = {      https://doi.org/10.1080/02664763.2021.1931820    
},
eprint = { https://doi.org/10.1080/02664763.2021.1931820
}

}

@article{Gallop_2021,
  title={A network approach to measuring state preferences},
  author={Max Gallop and Shahryar Minhas},
  journal={Network Science},
  year={2021},
  volume={9},
  pages={135 - 152},
  url={https://api.semanticscholar.org/CorpusID:234182482}
}

@inproceedings{McSherry_2001,
author = {McSherry, F.},
title = {Spectral Partitioning of Random Graphs},
year = {2001},
isbn = {0769513905},
publisher = {IEEE Computer Society},
address = {USA},
booktitle = {Proceedings of the 42nd IEEE Symposium on Foundations of Computer Science},
pages = {529},
series = {FOCS '01}
}

@Manual{unvotes_rpkg,
    title = {unvotes: United Nations General Assembly Voting Data},
    author = {David Robinson},
    year = {2021},
    note = {R package version 0.3.0},
    url = {https://CRAN.R-project.org/package=unvotes},
  }

@Manual{wdi_rpkg,
    title = {WDI: World Development Indicators and Other World Bank Data},
    author = {Vincent Arel-Bundock},
    year = {2025},
    note = {R package version 2.7.9},
    url = {https://CRAN.R-project.org/package=WDI},
  }

@article{grdpg_cov,
  title={Spectral Inference for Large Stochastic Blockmodels With Nodal Covariates},
  author={Angelo Mele and Lingxin Hao and Joshua Cape and Carey E. Priebe},
  journal={SSRN Electronic Journal},
  year={2019},
  url={https://api.semanticscholar.org/CorpusID:201070815}
}

@article{country_classification,
author = {Harris, Dan and Moore, Mick and Schmitz, Hubert},
title = {Country Classifications for a Changing World},
journal = {IDS Working Papers},
volume = {2009},
number = {326},
pages = {01-48},
doi = {https://doi.org/10.1111/j.2040-0209.2009.00326\_2.x},
url = {https://onlinelibrary.wiley.com/doi/abs/10.1111/j.2040-0209.2009.00326_2.x},
eprint = {https://onlinelibrary.wiley.com/doi/pdf/10.1111/j.2040-0209.2009.00326_2.x},
year = {2009}
}

@techreport{russian_crisis_1,
  author      = {{European Bank for Reconstruction and Development}},
  title       = {Transition Report 1999: Ten Years of Transition --
                 Economic Transition in Central and Eastern Europe,
                 the Baltic States and the CIS},
  institution = {EBRD},
  year        = {1999},
  address     = {London},
  isbn        = {9781898802150},
  issn        = {1356-3424}
}

@TechReport{russian_crisis_2,
type={UCL SSEES Economics and Business working paper series},
institution={UCL School of Slavonic and East European Studies (SSEES)},
author={Venla SipilÃƒÂ¤},
title={The Russian triple crisis 1998: currency, finance and budget},
year={2002},
month={Mar},
number={17},
doi={None},
url={https://ideas.repec.org/p/see/wpaper/17.html},
}

\end{document}